\documentclass{aa}

\usepackage{graphicx}

\usepackage{txfonts}

\begin{document}

\title{Insights into the inner regions of the FU~Orionis disc}

\titlerunning{Light variations in FU~Ori}

\authorrunning{M. Siwak, M. Winiarski, W. Og{\l}oza}

\author
{Micha{\l} Siwak\inst{1}\and
Maciej Winiarski\inst{1}\and
Waldemar Og{\l}oza\inst{1}\and
Marek Dr{\'o}{\.z}d{\.z}\inst{1}\and
Stanis{\l}aw Zo{\l}a\inst{1,2}\and
Anthony F.\ J.\ Moffat\inst{3}\and
Grzegorz Stachowski\inst{1}\and
Slavek M.\ Rucinski\inst{4}\and
Chris Cameron\inst{5,6}\and
Jaymie M.\ Matthews\inst{7}\and
Werner W.\ Weiss\inst{8}\and
Rainer Kuschnig\inst{8}\and
Jason F.\ Rowe\inst{3}\and
David B.\ Guenther\inst{9}\and
Dimitar Sasselov\inst{10}
}

\institute{Mount Suhora Observatory, Krakow Pedagogical University,
ul.\ Podchorazych 2, 30-084 Krakow, Poland\\
\email{siwak@oa.uj.edu.pl}
\and
Astronomical Observatory, Jagiellonian University, ul. Orla 171, 
30-244 Krakow, Poland
\and
D\'{e}partment de Physique, Universit\'{e} 
de Montr\'{e}al, C.P.6128, Succursale: Centre-Ville,
Montr\'{e}al, QC, H3C~3J7, Canada
\and
Department of Astronomy and Astrophysics,
University of Toronto, 50 St.\ George St., Toronto,
Ontario, M5S~3H4, Canada
\and
Department of Mathematics, Physics \& Geology, Cape Breton University, 
1250 Grand Lake Road, Sydney,NS, B1P 6L2, Canada
\and
Canadian Coast Guard College, Dept. of Arts, Sciences, and Languages,
Sydney, Nova Scotia, B1R 2J6, Canada
\and
Department of Physics \& Astronomy, University of
British Columbia, 6224 Agricultural Road, Vancouver, B.C., V6T~1Z1, Canada
\and
Universit\"{a}t Wien, Institut f\"{u}r Astrophysik, 
T\"{u}rkenschanzstrasse 17, A-1180 Wien, Austria
\and
Institute for Computational Astrophysics, Department of Astronomy and Physics,
Saint Marys University, Halifax, N.S., B3H~3C3, Canada
\and
Harvard-Smithsonian Center for Astrophysics,
60 Garden Street, Cambridge, MA 02138, USA
}

\date{Received 09/05/2018; accepted 04/07/2018}

\abstract
{We investigate small-amplitude light variations in FU~Ori occurring in timescales of days and weeks.}
{We seek to determine the mechanisms that lead to these light changes.}
{The visual light curve of FU~Ori gathered by the {\it MOST} satellite continuously for 55 days 
in the 2013-2014 winter season and simultaneously obtained ground-based 
multi-colour data were compared with 
the results from a disc and star light synthesis model.
}
{Hotspots on the star are not responsible for the majority of observed light variations. 
Instead, we found that
the long periodic family of 10.5-11.4~d (presumably) quasi-periods showing light variations 
up to 0.07~mag may arise owing
to the rotational revolution of disc inhomogeneities located between 16-20~R$_{\sun}$. 
The same distance is obtained by assuming that these light variations arise because of a purely 
Keplerian revolution of these inhomogeneities for a stellar mass of 0.7~M$_{\sun}$. 
The short-periodic ($\sim 3-1.38$~d) small amplitude ($\sim 0.01$~mag) light variations show a clear sign 
of period shortening, similar to what was discovered in the first {\it MOST} observations of FU~Ori. 
Our data indicate that these short-periodic oscillations may arise because of changing 
visibility of plasma tongues (not included in our model), 
revolving in the magnetospheric gap and/or likely related hotspots as well.
}
{
Results obtained for the long-periodic 10-11~d family of light variations appear to be roughly in line 
with the colour-period relation, which assumes that longer periods are produced by more external 
and cooler parts of the disc.
Coordinated 
observations in a broad spectral 
range are still necessary to fully understand 
the nature of the short-periodic 1-3~d family of light variations and their period changes.
}


\maketitle

\section{Introduction}

\label{intro}

FU~Orioni-type stars (FUors) were already recognised as classical 
T~Tauri-type stars (CTTS) undergoing a phase of enhanced disc brightness by \citet{herbig77}. 
The light outburst is due to an increased mass accretion rate 
from $10^{-11}-10^{-7}$~M$_{\sun}$~yr$^{-1}$ typical in CTTS, 
up to  $10^{-5}-10^{-4}$~M$_{\sun}$~yr$^{-1}$ 
in FUors \citep{hartmann85,hartmann96}.
During the FUor phase, the emission-line rich visual spectrum of the formerly quiet CTTS is dominated by the absorption features produced in the inner accretion disc. 
Its inner parts radiate as the photosphere of an F-G supergiant star, while the slightly colder 
and more distant parts of the disc produce a K-M type supergiant spectrum that can be observed 
in the near-infrared \citep{kenyon88}. 
In these circumstances disc radiation dominates stellar radiation (usually an early-M or 
late-K dwarf) by 100 -- 1000 times. 
This makes FUors well suited for inner disc variability studies in visual bands 
in early stages of star formation, presumably during the first 0.3~Myr of evolution 
when discs are gravitationally unstable \citep{hartmann98,liu16}.

In 1937, FU~Ori , which is the prototype of FUors, increased its brightness from $15.5\pm0.5$ to 
$9.7\pm0.1$~mag \citep{clarke05} in the photographic system, whose effective wavelength $\lambda$ 
is similar to that of the Johnson $B$ filter. 
The star still remains in a high state of brightness with 
only 1.3~mag decrease noticed until 2014 in the $B$ filter (see Section~\ref{multi-saao}). 
Our target was briefly described in our first paper of this series \citep{siwak13} and 
also in \citet{powell12}  and \citet{audard14}; 
the latter
authors presented a detailed review regarding eruptive young stars, both FUors and 
EX~Lupi-type stars (EXors), emerging from observations obtained in a broad spectral range.

Most of the photometric and spectroscopic historical papers 
concern the variability of FU~Ori occurring on timescales of 
months to years (see e.g. \citealt{clarke05}). 
In the pre-space telescope era nightly 
breaks and non-uniform weather patterns imposed severe limitations 
on light variability studies on a daily scale, occurring on top of the light plateau 
caused by enhanced accretion. 
The most in-depth though indirect study of this subject was made by \citet{kenyon2000}. 
After a careful consideration of several hypotheses the authors concluded 
that the disc likely shows flickering, occurring 
with a characteristic timescale of predominantly about one~day, and that variations of the colour 
indices obtained in the Johnson filters point to the region extending between 
the stellar photosphere and the radius where the disc temperature reaches its maximum 
as the dominant source of the variability.
We note that \citet{kolotilov85} and \citet{ibragimov93} were also unable to find any stable periods 
in the data sets gathered during the 1980s. 
Instead the authors found quasi-periodic oscillations (QPOs) that appeared to evolve 
from 18.35 to 9.19~d in the course of the single 1984-1985 observing season. 
The 9.2~d QPO was also visible during 1987-1988 along with possible 9.8 and 11.4~d QPOs.

Having the advantage of contiguous monitoring of stars for a few  weeks 
from space with the {\it MOST} satellite and with such photometric 
precision unavailable from the ground, we decided to use this space telescope for 
a direct search and characterisation of short-period, small-amplitude light variations 
in FU~Ori during winter 2010-2011. 
Our first results \citep{siwak13} can be summarised as follows:\newline
The light curve itself and its Fourier and wavelet transform spectra reveal 
two major QPOs. The first has a larger amplitude and 
possibly changes its period in the range 8 and 9 days. The second, of a smaller amplitude, is apparently 
time coherent, and was
initially observed at about 2.4 days, but drifted down to 2.2 days, 
well before the end of the run. 
The longer variation was less securely defined in the 28-day-long run 
to conclude that this was really a time-coherent QPO. 
Light changes occurring on the timescale of $\leq1$~d, predicted 
by the Monte Carlo model of \citet{kenyon2000} to dominate in the light curve, 
were observed in the light curve for a very limited time only. 
Moreover, \citet{zhu07} investigated the spectral energy distribution in FU~Ori 
and argued against the putative boundary layer extending towards the stellar 
photosphere (see Sec.~5.1 of their paper), i.e. the `energy release zone' 
\citep{luybarskii1997} proposed to be responsible for the disc flickering 
by \citet{kenyon2000}. 
In these circumstances, assuming a stellar mass of 0.3~M$_{\sun}$ \citep{zhu07}, we indisputably 
interpreted that variations observed by {\it MOST} 
were produced by the hot plasma condensations that develop in the magneto-rotationally 
unstable inner parts of the disc at distances of 
about 12 and 5~R$_{\sun}$, respectively. 
The period shortening may be interpreted as spiralling-in or inward 
drifts in the inner disc. 
Assuming Keplerian rotation of the disc, we also temporarily proposed that the shortest 
observed period of $2.2\pm0.1$~d may define the inner edge of the disc 
at $4.8\pm0.2$~R$_{\sun}$, which agrees with the $5.5^{+2.9}_{-1.8}$~R$_{\sun}$ result from 
the interferometric observations of \citet{malbet05}.\newline 
Our two-colour Str{\"o}mgren $v$ and $b$ filter ground-based observations 
of FU~Ori at the Mount Suhora Observatory ({\it MSO}) substantially confirmed 
the $(B-V)$ versus $V$ relation obtained by \citet{kenyon2000}. 
However, we obtained a slightly redder colour index of the 
dominant 8-day variation, which we tentatively interpreted 
as due to a more outward location of the inhomogeneity  causing 
this quasi-periodicity. 
This would agree with the interpretation of the colour-period 
relationship through different 
locations of the dominant variable flux with longer periods 
produced by more external and cooler parts of the disc. 
We decided to examine this relationship based on new {\it MOST} data collected 
as long as technically possible, and simultaneously obtained multi-colour 
observations from the ground during the 2013-2014 observing season. 

We describe the {\it MOST} satellite and ground-based multi-colour observations 
of FU~Ori obtained at the {\it MSO} 
and at the South African Astronomical Observatory ({\it SAAO}) in Section~\ref{observations}. 
Results from the data analyses are presented in Section~\ref{results}.  
We discuss the obtained results in the context of 
recent theories proposed for CTTS and FUors in Section~\ref{discussion} and summarise in Section~\ref{summary}.

\section{Observations}

\label{observations}

\subsection{MOST observations}

\label{MOSTobs}

The pre-launch characteristics of the {\it MOST\/} satellite mission are described in \citet{WM2003} and the initial post-launch performance in \citet{M2004}. The satellite observes in one broadband filter covering the spectral range from 370 nm to 750~nm with effective wavelength similar to that of the Johnson $V$ filter.

\begin{figure*}

  \centering

  \includegraphics[width=0.85\linewidth]{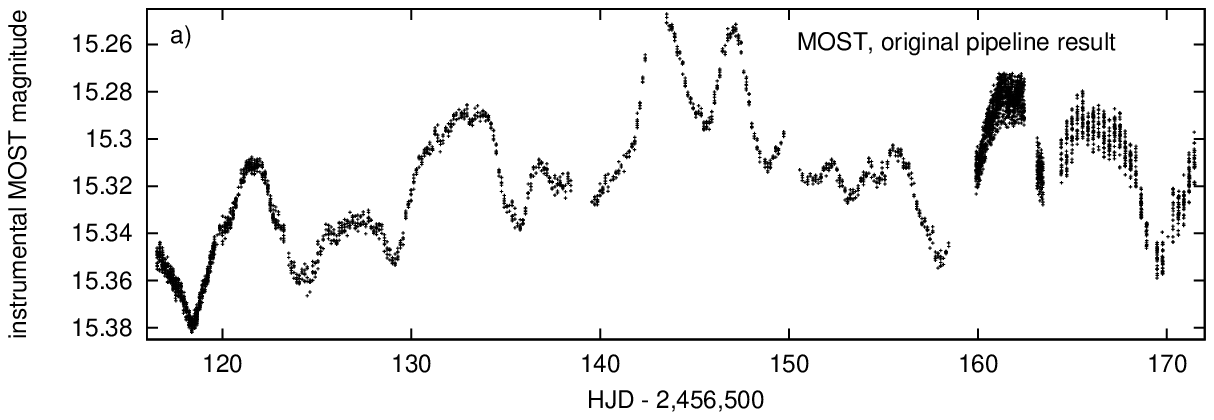}

  \includegraphics[width=0.85\linewidth]{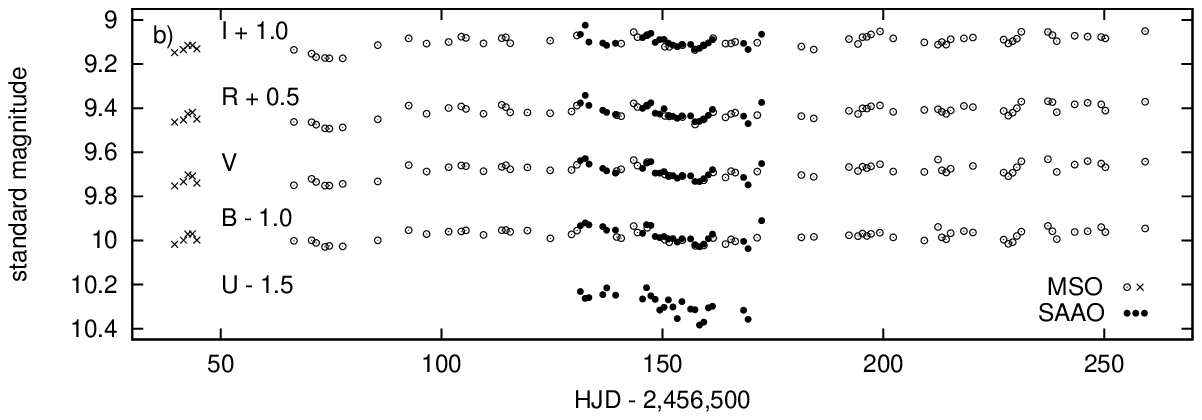}

  \includegraphics[width=0.85\linewidth]{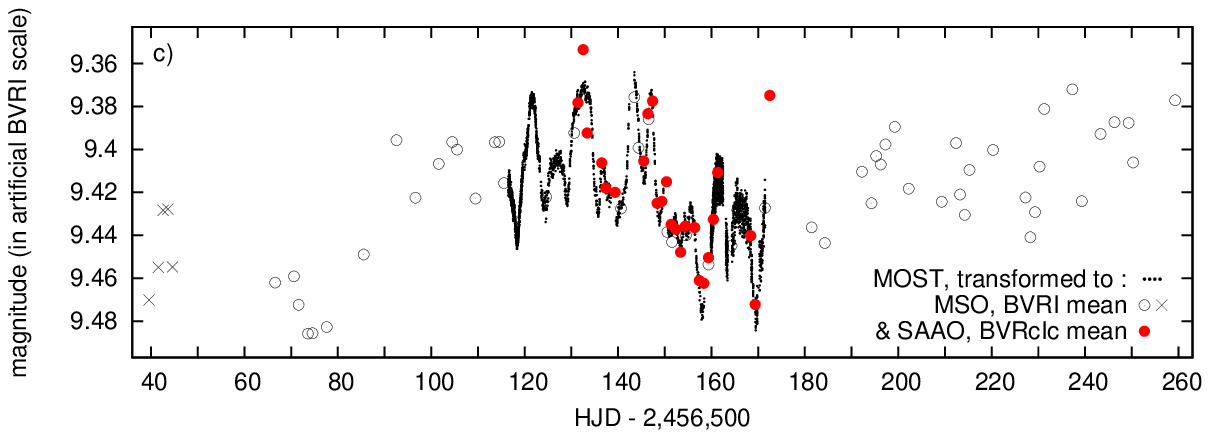}

  \includegraphics[width=0.85\linewidth]{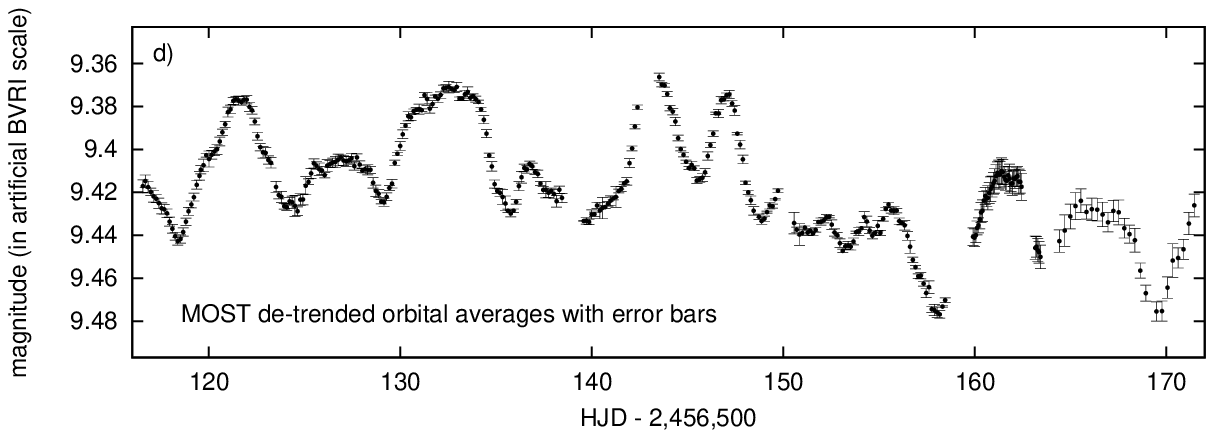}
  \caption{First panel shows the 2013-2014 {\it MOST} light curve 
           of FU~Ori provided by the pipeline. 
           The standardised {\it SAAO} {\it UBVR$_c$I$_c$} (filled circles) 
           and aligned {\it MSO} {\it BVRI} light curves 
           (open circles and crosses) are shown in the second panel. 
           The third panel shows averaged {\it SAAO} {\it BVR$_c$I$_c$} 
           (filled circles) and {\it MSO} {\it BVRI} 
           (open circles) observations used for correction of the instrumental 
           trend in the original {\it MOST} data, as described in Sec.~\ref{MOSTobs}. 
           The de-trended {\it MOST} data are plotted in the same panel. 
           The fourth panel shows the de-trended {\it MOST} data in the form 
           of orbital averages with error bars ($\sigma$).
           }
\label{Fig.dat1}
\end{figure*}

The observations of FU~Ori were made in the direct imaging data acquisition mode of the satellite. A run of length 54.93 days started on November 20, 2013 and lasted until January 15, 2014 (HJD=2\,456\,616.5408 -- 2\,456\,671.4662). Because of the slow temporal changes in the light curve noticed during the first {\it MOST} run in 2010-2011, and to permit alternate, multi-object observations, the star was observed during every second satellite orbit, 
i.e. with a typical cadence of 202.87 minutes. 
The individual exposures were 60~s long and typically ten to a few dozen exposures per orbit were obtained. The photometry (Figure~\ref{Fig.dat1}a) was extracted from the raw data and de-correlated with known instrumental effects using the dedicated pipeline introduced by \cite{rowe06}. In spite of this advanced process, a significant systematic trend is visible in the primary and secondary targets; this trend is possibly owing to changing sensitivity of the electronic system. This affected all stars observed in the field, but affected each star in a slightly different way, preventing their use in removing the FU Ori light-curve trend. To eliminate the problem, we used our $BVR_cRI_cI$ filter data for FU~Ori obtained simultaneously at the {\it SAAO} and {\it MSO} (see in Figure~\ref{Fig.dat1}b, and next two Sections for details). In this process, the $U$-filter data had to be abandoned owing to their significantly higher scatter. The eight-degree polynomial fitted to the differences between the {\it MOST} and the $BVR_cRI_cI$ averaged data (matching the bandwidth of the {\it MOST} satellite) was then subtracted from the {\it MOST} pipeline result giving variations as shown in Figure~\ref{Fig.dat1}c; the magnitude level is arbitrary in this panel. Finally, we calculated mean-orbital data points, as shown in Figure~\ref{Fig.dat1}d. Their median error ($\sigma$) is 0.0019~mag 
in the full range  0.0002-0.0078~mag.
One can notice a few major gaps in the data acquisition 
at HJD-2\,456\,500=138.5-139.6, 
142.4-143.5, 149.7-150.6, 158.5-159.9, 162.4-163.1 
and 163.4-164.4 owing to observations of other, time-critical targets. 
These gaps were short in comparison to the variability timescales 
observed in this star and do not affect our conclusions.


\subsection{{ Observations by \it MSO uvyBVRI} filters}

\label{multi-mso}

In order to search for long-period light variations, which is not possible using {\it MOST} because of the technical limitation of a single run to about 55~days, we conducted time variability monitoring of FU~Ori at the {\it MSO}. We started our observations on September 3, 2013, and continued the monitoring for 220 days, i.e.\ until April 11, 2014, when the altitude of the star amounted to only 30~deg at civil dusk. During this interval, useful data were obtained on 69 nights, out of which 20 coincided with the satellite run.

We used three different photometric systems on the 0.6~m telescope. During the first five nights of September, we used a {\it SBIG ST10XME} CCD camera equipped with the Johnson-Morgan {\it BVRI} filters provided by {\it SBIG}. The data obtained with this instrument are indicated as crosses in Figure~\ref{Fig.dat1}b. On September 30, 2013, the {\it Apogee Alta U42} CCD camera, equipped with a different set of wide band {\it BVRI} Johnson (Bessell) filters manufactured by {\it Custom Scientific}, was installed on the telescope and the data are indicated by open circles. We decided to use only this uniform data set for further analyses, which limited the monitoring time to 194 days. Additionally, for 24 nights between  November 12, 2013 and January 13, 2014 (18 nights during the {\it MOST} run), we observed the star with Str{\"o}mgren {\it uvy} filters.

On each clear night typically 10-30 single exposures per filter for FU~Ori were obtained. All frames were {\it dark} and {\it flat-field} calibrated in a standard way within the {\small \sc MIDAS} package \citep{W1991}. The correction for {\it bias} was accomplished using the same exposure for {\it dark} as for science frames. The photometric data were extracted using the {\small \sc C-Munipack} programme \citep{Motl11} using the {\small \sc DAOPHOT~II} package \citep{stet87}. A constant aperture of 12 pixels (corresponding to 13.44~arcsec on the sky for a scale of 1.12~arcsec per pixel) was used for FU~Ori and the comparison stars.To avoid possible errors associated with the arc-like nebula surrounding FU~Ori, relatively large annuli (from 20 to 30 pixels, i.e.\ 20-34~arcsec) were used for the sky background calculations.

The differential photometry used a mean comparison star that  was formed from two stars (Table~\ref{Tab.comp}), following \citet{siwak13}, both apparently stable to about 0.01~mag in all filters throughout the entire {\it MSO} run.

We corrected our data for dominant colour extinction effects using the mean coefficients for this site. Since FU~Ori turned out to be essentially constant during nightly monitoring sessions of about 20--30~minutes, nightly averages were calculated. The data were left in the  instrumental system, but were later manually aligned with the {\it BV$R_cI_c$} standardised {\it SAAO} light curves (see below, and also in Fig.~\ref{Fig.dat1}b).

\subsection{{Observations by \it SAAO UBVR$_c$I$_c$} filters }

\label{multi-saao}


\begin{table}
\caption{Standard Johnson-Cousins $V$-filter magnitudes and colour indices of FU~Ori 
and its comparison stars evaluated using Equation~1, during the single 
night of January 11-12, 2014 (see Table~\ref{Tab.coeff} and Sec.~\ref{saao_system}).
The errors in parentheses account only for photometric noise ; the systematic 1-2\% uncertainty of zero points $C_{ft}$ (see Table~\ref{Tab.coeff}) 
has not been taken into account. 
The same comparison stars formed the mean comparison star 
for the {\it MSO} observations.}
\begin{tabular}{c l c c}

\hline

         &       FU~Ori & GSC~00714-00203 & GSC~00715-00188 \\ \hline

$V$      &    9.748(5)  & 10.566(5)       &  10.827(6)      \\

$U-B$    &    0.842(21) &  1.079(14)      &   1.327(14)     \\

$B-V$    &    1.291(10) &  0.779(38)      &   1.258(85)     \\ 

$V-R_c$  &    0.778(10) &  0.592(13)      &   0.719(13)     \\

$R_c-I_c$&    0.839(8)  &  0.620(12)      &   0.735(15)     \\ \hline

\end{tabular}

\label{Tab.comp}

\end{table}


FU~Ori was observed at the {\it SAAO} on 25 nights from December 5, 2013 to January 14, 2014 using  the 0.5 m {\it Boller} \& {\it Chivens} telescope equipped with well-known for its reliability single-channel {\it Modular Photometer}. This photometer was equipped with a Hamamatsu R943-02 GaAs photomultiplier and a set of {\it UBVR$_c$I$_c$} Johnson-Cousins filters (see in Figure~\ref{Fig.dat1}b). For all observations, a 25~arcsec aperture was utilised; thereby a part of the associated arc-like nebulae light (however, negligibly small) was also recorded. The differential photometry of FU~Ori utilised the same two comparison stars as those used at the {\it MSO} (Tab.~\ref{Tab.comp}). Performing {\it all-sky} absolute photometry, commonly practised at the {\it SAAO} with this instrument, was impossible owing to a large number of not fully photometric nights during the first two weeks of the run.
We measured the target ({\it var}), comparison stars ({\it comp1, comp2}), and sky background ({\it sky})  in the sequence of {\it sky-comp1-comp2-sky-var-sky-comp2-comp1-sky-comp2-comp1-sky-var-sky-comp1-comp2-sky}, repeated two or three times.  During the moonless nights (except for those with a high airglow activity) the {\it sky} sampling rate was reduced since the background level remained stable for about an hour. The rate of {\it sky} monitoring was considerably increased during the rising and setting of the Moon. The two full Moon passages through the field during the run forced us to discard observations of FU~Ori during that time owing to strong background gradients, especially in the $UB$ filters, and to rely entirely on simultaneous CCD observations at the {\it MSO}, which have the advantage that the sky level is individually calculated from pixels surrounding each star.

\subsubsection{Determination of colour equations for the {\it SAAO} system}

\label{saao_system}


\begin{table}

\caption{Colour equation coefficients determined for the {\it SAAO} 
$UBVR_cI_c$ system during the night of  January 11-12, 2014. 
The appropriate values of $CI$ for each star are given in Table~\ref{Tab.comp}.}  



\begin{tabular}{c c c c c c}

\hline

$ft$       & $k_{ft}$ & $\beta_{ft}$ & $\mu_{ft}$    & $C_{ft}$        & $CI$      \\ \hline

$U$        & 0.62  & -0.052    &-0.0307(38) & 18.8811(192) & $U-B$     \\    

$B$        & 0.27  & -0.031    & 0.0508(16) & 19.5517(129) & $B-V$     \\ 

$V$        & 0.13  & -0.010    & 0.0212(33) & 19.8781(137) & $B-V$     \\ 

$R_c$      & 0.10  & -0.006    & 0.0077(64) & 19.8270(128) & $V-R_c$   \\  

$I_c$      & 0.07  & -0.008    & 0.0024(55) & 19.3665(101) & $R_c-I_c$ \\ \hline

\end{tabular}

\label{Tab.coeff}

\end{table}


We observed a set of eight standard stars from the {\it E400} region \citep{menzies89} in the night of January 11-12, 2014, in excellent photometric conditions. Transformation equations to standard magnitudes $m_{ft}^{std}$ for each  filter $ft\in\{U,B,V,R_c,I_c\}$ took the following form:
\begin{equation}
m_{ft}^{std}=m_{ft}^{obs} - (k_{ft} + \beta_{ft} \times CI) \times X(z) + \mu_{ft} \times CI + C_{ft},
\end{equation}
where $m_{ft}^{obs}$ is the observed (instrumental) magnitude calculated from the {\it dead-time} and {\it sky-level} corrected counts, $k_{ft}$ is the mean {\it SAAO} differential extinction, $\beta_{ft}$ is the colour extinction evaluated for {\it SAAO} from \cite{fukugita96} and Mt.\ Palomar sites, $\mu_{ft}$ is the system transformation coefficient, the constant $C_{ft}$ is the zero point in magnitude, and $CI$ is the colour index defined as the magnitude difference of neighbouring filters. We present the obtained values and respective $CI$ terms for every filter in Table~\ref{Tab.coeff}, while the resulting magnitudes of the comparison stars are given in Table~\ref{Tab.comp}. This enabled the absolute calibration of the FU~Ori {\it UBVR$_c$I$_c$} light curves, as show in Figure~\ref{Fig.dat1}b. We did not derive colour equations for the {\it MSO} system; instead, the $BVRI$ light curves were aligned to the {\it SAAO} $BVR_cI_c$ data, using observations obtained simultaneously at both observatories. We estimate the accuracy of this procedure to be about 0.002~mag.

\subsubsection{Determination of colour indices from the {\it SAAO} data}

\label{CIsaao}

Using single channel photometry, colour indices can be calculated in two ways:
first,~through  subtraction of the two light curves obtained with the use of comparison stars (as for the {\it MSO} data); or \newline
second,~using the variable star only, relating its {\it sky}- and {\it dead-time}-corrected measurements in two colours and using colour equations determined above in Sec.~\ref{saao_system}. The first method involved too large scatter into the final results owing to the accumulation of noise from the comparison stars. This is particularly the case when these stars are fainter than the target and the sequence of measurements given in Section~\ref{multi-saao} was not executed in the fully photometric conditions that prevailed during the first two weeks of our run. For that reason, we calculated the colour indices using the second method. For example, to obtain $V-I_c$ colour indices, the flux ratios between the $V$-filter data points and the reference 1-3 degree polynomial (obtained from a fit to $I_c$-filter points) were transformed to the magnitude scale, corrected for differential and colour extinctions, and then transformed to the standard system using Equation~1 and coefficients listed in Table~\ref{Tab.coeff}. We plot the final results in Figure~\ref{Fig.rez1}.

\subsection{Spectroscopic {\it SAAO} observations}

\label{spec-saao}

In the evening of March 11, 2017, we obtained a few low-resolution spectra of FU~Ori at the {\it SAAO} using the {\it SpUpNIC -- Spectrograph Upgrade-Newly Improved Cassegrain} \citep{crause16}, mounted on the 1.9 m {\it Radcliffe} telescope. Grating~6 was used to cover the wavelength range from 3904~\AA ~ to 6650~\AA ~with a resolution of 1.35~\AA~pix$^{-1}$. Two spectroscopic standard stars LTT~2415 and LTT~3864 were observed immediately after our target using the same slit width of 3.59~arcsec. The spectra were extracted and then wavelength- and flux-calibrated within the {\small \sc IRAF} package. Only these FU~Ori and standard-star spectra, for which no telescope-guiding errors were noticed during the 120~sec-long integrations, were selected for further analyses.

\section{Results of data analysis}

\label{results}

\subsection{General description of variability}

\label{general_descr}


 \begin{figure}

  \includegraphics[width=1.0\linewidth]{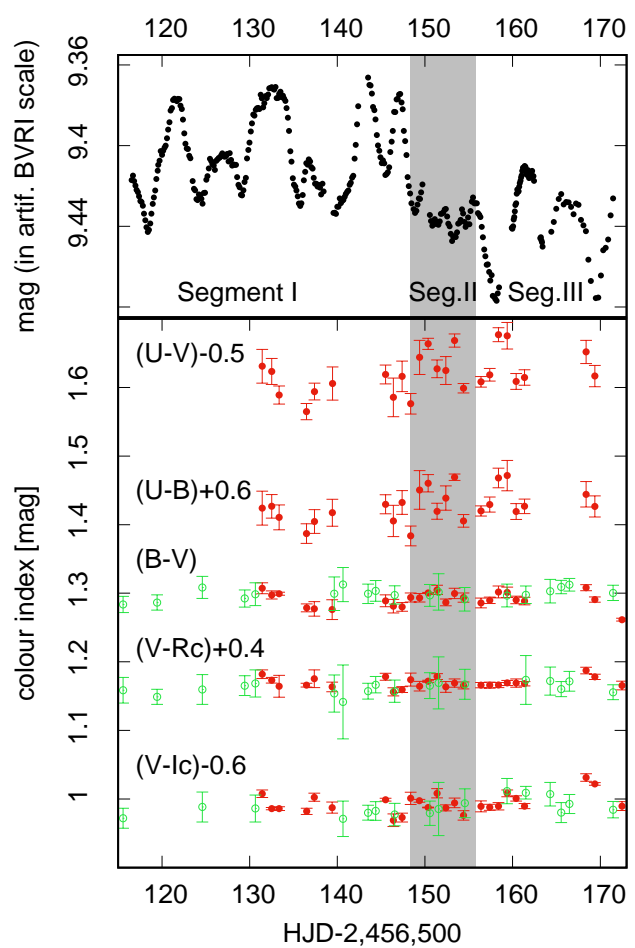}

  \caption{New {\it MOST} light curve of FU~Ori (top panel) 
  and the ground-based $U-V$, $U-B$, $B-V$, $V-R_c$ and $V-I_c$ colour indices 
  (lower panel) in standard magnitudes, with arbitrary offsets indicated. 
  As mentioned in the text, the {\it MSO} data (open circles) 
  were aligned to the standardised {\it SAAO} data (filled circles) 
  with an accuracy of 0.002~mag. 
  The grey-shaded area determines the approximate 
  boundaries of {\it Segment~II} data, as defined 
  in Sect.~\ref{results}. 
  The {\it Segment~I} data are shown on the left, 
  while the {\it Segment~III} data are shown on the right.} 

  \label{Fig.rez1}

 \end{figure}


The 2013-2014 {\it MOST} light curve (Fig.~\ref{Fig.dat1}d) appears to consist of three segments, each defined by a characteristic pattern of the FU~Ori variability. These patterns exist for some time and then disappear. The three segments, indicated in Figure~\ref{Fig.rez1}, have the following characteristics:
\begin{enumerate}

\item We define as {\it Segment~I} the part  from the beginning of the {\it MOST} observations (at least) until HJD$\approx$2\,456\,647.8, which is dominated by consecutive peaks of different heights, i.e. 0.07 and 0.02~mag. Careful insight into the ground-based averaged data (Fig~\ref{Fig.dat1}c) reveals that this variability pattern began in fact about 20 days earlier, at HJD$\approx$2\,456\,595, with the appearance of the broad local light maximum. The ground-based data also suggest that the height of the maximum observed by {\it MOST} as the smaller one
(i.e. of 0.02~mag), could initially be higher.\newline
This variability pattern is very similar to what was observed by {\it MOST} in IM~Lupi, which is visible at an inclination angle of 60~deg and where two stable polar hotspots played a major role in the observed light modulations \citep{siwak16}. This conclusion was inferred by utilising the numerical results of \citet{romanowa04} and \citet{kulkarni08} for stars accreting in a stable regime, which  more likely is the case for CTTS with low accretion rates.  We note that stable polar hotspots were also found in EX~Lupi (the prototype of EXors) during a quiescent phase using spectroscopic data \citep{sicilia-aguilar15}. For geometrical reasons one may think that the same mechanism operates in FU~Ori, which most likely is observed at an inclination of 55~degrees \citep{malbet05}. Additionally, the characteristic re-appearance time for the observed peaks, both larger ($\sim0.07$~mag) and smaller ($\sim0.02$~mag) appears to be about 10-11 days, which is of the same order as the 7.2-7.6~d rotational period of IM~Lupi, and 7.41~d in EX~Lupi \citep{sipos09}. However, despite large observational errors, the $U-V$, $U-B$ and $B-V$ colour indices appear to be systematically slightly redder when the star is brighter, while the $V-R_c$ and $V-I_c$ light curves appear to be stable, as shown in Figure~\ref{Fig.rez1} for the {\it Segment~I} data. This seems to contradict the hotspot hypothesis and we  return to this issue later in this paper.
\newline We note that the last maximum in {\it Segment~I} centred at HJD$-2\,456\,500\approx146$ is double-peaked. Starting from this place the present light variability pattern appears to smoothly transform into the next, which is defined below.

\item {\it Segment~II} of the light curve consists of small-amplitude sine-like variations, showing a period shortening of each successive oscillation. It is indicated by a grey-shaded area in Figure~\ref{Fig.rez1}. The sine-like wave lasted for at least eight days and ceased at HJD$-2\,456\,500\approx156$, when its initial period of about three~days shortened to $1.38\pm0.04$~d. The colour indices appear to be stable but this is not a very significant statement given the small 0.005-0.01~mag amplitude of these oscillations, which is comparable to the measurement errors of the ground-based data.

\item {\it Segment~III} of the light curve begins after HJD$-2\,456\,500\approx156$, when the star brightness started to drop by about 0.05~mag (until the first minimum at HJD$-2\,456\,500\approx158.1$) and then it rose by 0.06-0.07~mag in the {\it MOST} magnitude system. The symmetry of these light changes and a second, similar event at HJD$-2\,456\,500\approx169.5$, appearing 11.4~d after the first, suggest their similar origin. Unfortunately, the {\it MOST} monitoring finished  soon after the second deep light minimum event owing to technical limitations on the satellite run length. Furthermore, termination of our {\it SAAO} run and poor weather conditions over the {\it MSO} at the same time, shortened the photometric monitoring of this feature from the ground. Therefore, the question of whether we observed an initiation of a new QPO remains open. Its putative 11.4~d period might suggest the same physical mechanism that operated during {\it Segment~I}, especially that also a secondary light drop occurred at HJD$-2\,456\,500\approx163$. Interestingly, all but the $U-V$ and $U-B$ colour indices in {\it Segment~III} appear to be stable during the first deep minimum, but during the second, at HJD$-2,456,500\approx169.5$, two slightly redder points in the $V-R_c$ and $V-I_c$ light curves were found in the {\it SAAO} data.

\end{enumerate}


  \begin{figure}

   \includegraphics[width=1.0\linewidth]{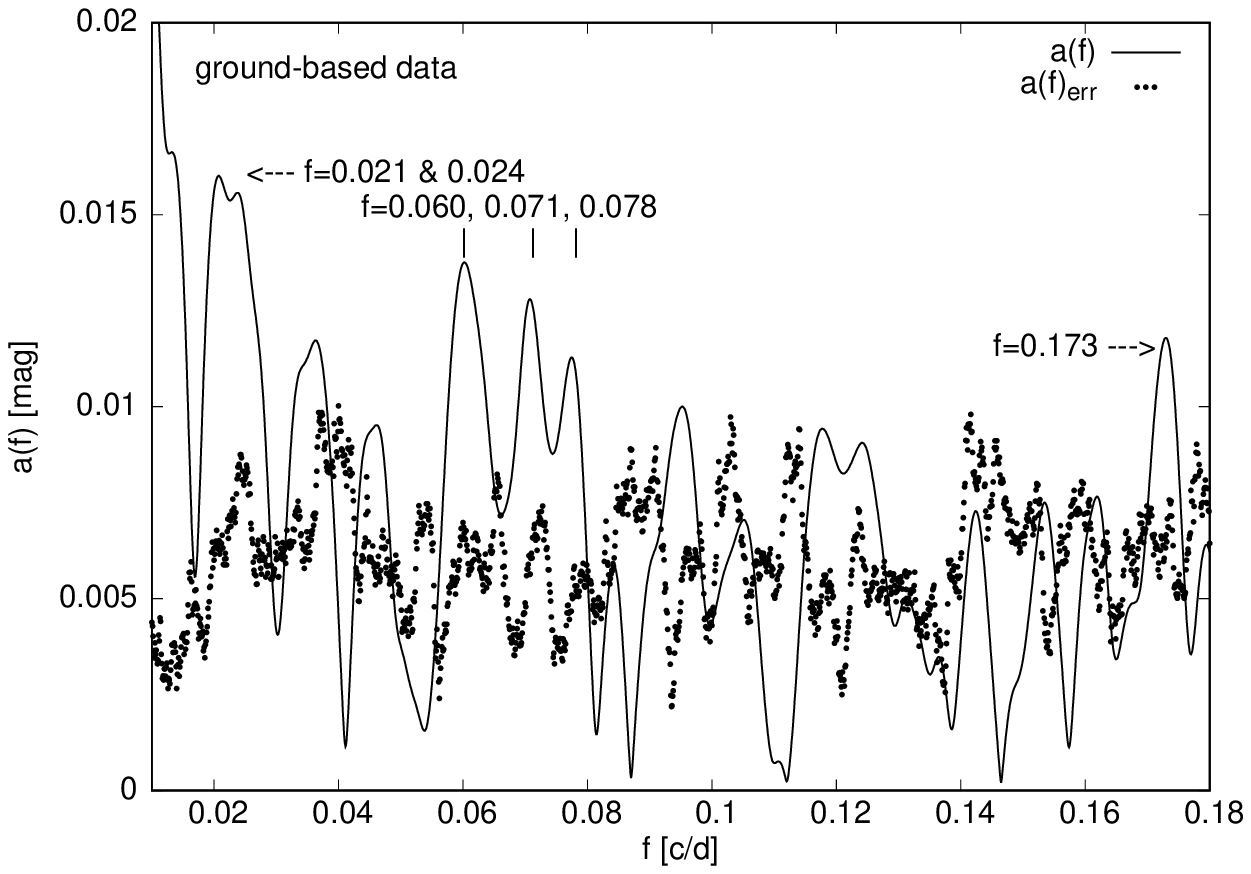}

   \includegraphics[width=1.0\linewidth]{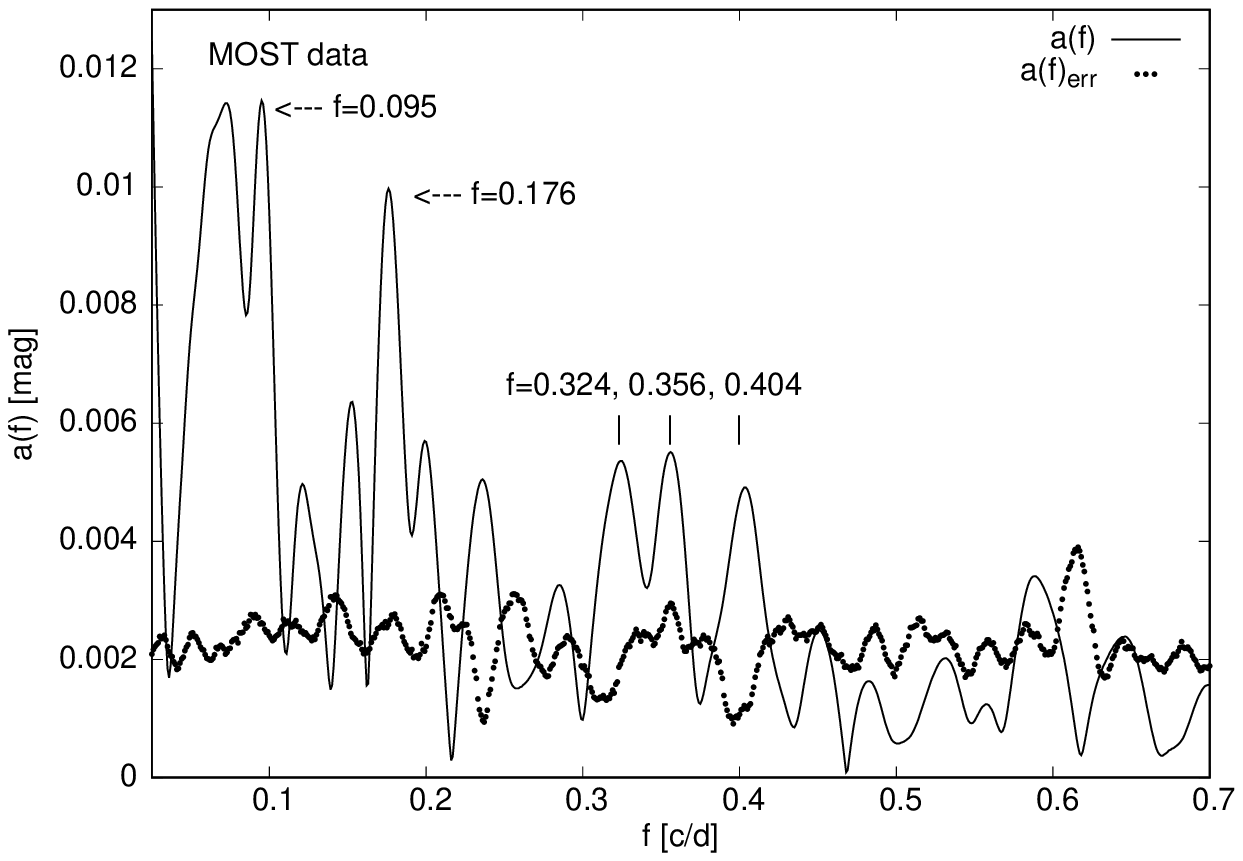}

  \caption{Results of Fourier analysis of ground-based (the upper panel) 
  and {\it MOST} (the lower panel) data in 
  form of the amplitude $a(f)$ vs. the frequency ($f$) are shown as
  continuous lines. The amplitude errors, determined from
  bootstrap sampling, are shown as dots.}

  \label{Fig.rez2}

 \end{figure}


\subsection{Frequency analysis of the FU~Ori data}

\label{MOSTfr}

We performed frequency analyses of two data sets: the first utilised the long-term {\it MSO} $BVRI$ and {\it SAAO} $BVR_cI_c$ nightly-averaged data points, i.e. those that were previously used as a fiducial comparison star for the {\it MOST} data trend removal;  the second utilised the {\it MOST\/} mean-orbital data points. 

We used the procedure previously developed by \citet{ruc08}. The Fourier analysis was done by consecutive, in the frequency $f$ space with a step of $\Delta f = 0.001$, least-squares fits of expressions of the form $l(f ) = c_0 (f) + c_1 (f) cos[2\pi (t - t_0) f ] + c_2 (f ) sin [2\pi (t - t_0) f ]$. The amplitude $a(f)$ for each frequency was found as the modulus of the periodic component, $a(f)=\sqrt{c_1^2 (f) + c_2^2(f)}$. The bootstrap sampling technique permitted evaluation of mean standard errors of the amplitudes from the spread of the coefficients $a_i$.

The Fourier spectra of ground-based data (the upper panel in Figure~\ref{Fig.rez2}) shows two families of peaks above the noise level (of about 0.007~mag) in the ranges  $f=0.021-0.024$~c~d$^{-1}$ (48-42~d) and $0.060-0.078$~c~d$^{-1}$ (16.6-12.9~d). Since the significance of the wide peaks is low, we have been unable to draw any firm conclusions on the existence of long-period QPOs in the combined {\it MSO} and {\it SAAO} data set.\newline
The {\it MOST} data (lower panel in Figure~\ref{Fig.rez2}) reveal three families of periods above the noise level (of about 0.0025~mag): $f=0.095$~c~d$^{-1}$ (10.5~d), 0.176~c~d$^{-1}$ (5.7~d) and $0.324-0.404$~c~d$^{-1}$ (3.09-2.48~d). The period 5.7~d corresponds roughly to one-half of the dominant third periodicity in {\it Segments~I} and {\it III} of the light curve; we note that a similar peak is also visible at $f=0.173$~c~d$^{-1}$ in the ground-based data.  
We note an absence of 2.5-1.4~d peaks expected from the sine-like QPO observed in {\it Segment~II} of the light curve, although this can be explained by a continuous period change of this wave train, as described in point~2 of Section~\ref{general_descr}.

The peaks in the amplitude-frequency spectra appear to scale as a flicker-noise ($a(f)\sim1/\sqrt f$) and this fact was also noticed in our first paper about FU~Ori. This may suggest that the observed variability may be intrinsic to the disc, for example a consequence of instabilities in the mass transfer leading to light variations either of quasi-periodic or irregular nature, typical for flickering \citep{luybarskii1997}. The same flicker-noise character is visible in the amplitude-frequency spectra of TW~Hya \citep{ruc08,siwak11,siwak14, siwak18} and RU~Lup \citep{siwak16}, whose variability is due to changing visibility of hotspots produced during moderately stable and unstable accretion regimes \citep{kulkarni09,blinova16}.


 \begin{figure}

  \centering

  \includegraphics[width=1.0\linewidth]{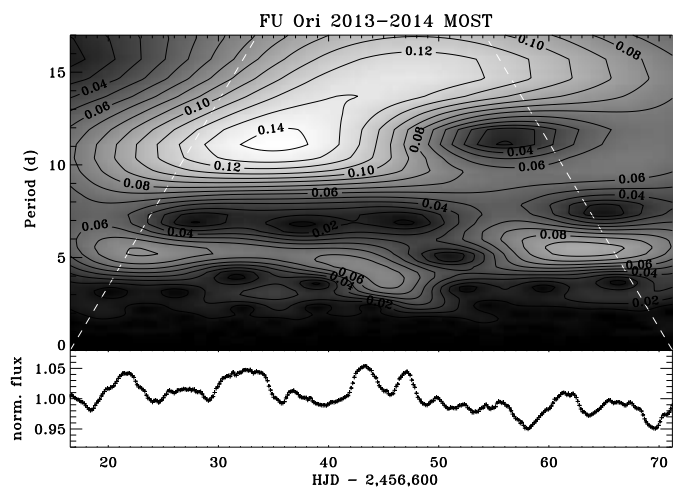}

 \caption{Wavelet spectrum computed from the 2013-2014 {\it MOST} data. 
 Edge effects are present beyond the white broken lines. 
}

 \label{Fig.rez3}

 \end{figure}



 \begin{figure*}

  \centering

  \includegraphics[width=0.48\linewidth]{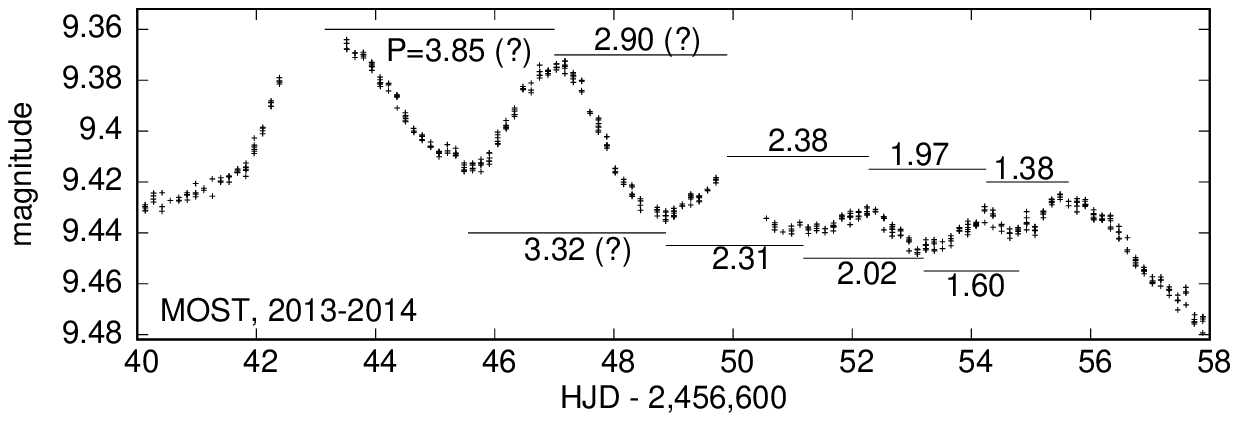}
  \includegraphics[width=0.48\linewidth]{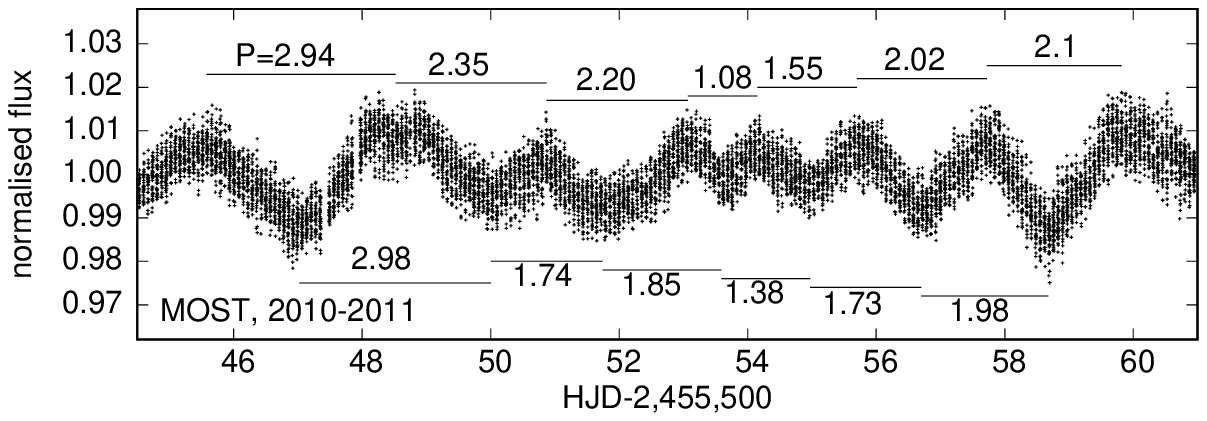}

 \caption{Short-periodic fragments of 2013 and 2010 {\it MOST} 
          light curves. In both cases maxima and minima were 
          localised by eye to about 0.02~d, which leads to the 0.04~d uncertainty of each indicated temporal period $P$ value. Modulations seen at the beginning 
          are marked by ``?'' as they occurred prior the official start of {\it Segment~II}.
          }

 \label{Fig.rez3a}

 \end{figure*}


\subsection{Wavelet analysis of the {\it MOST} FU~Ori data}

\label{MOSTwav}

To obtain a uniform data sampling required for the wavelet analysis of the star, we interpolated with splines the 344 mean-orbital data points into a grid of 389 equally spaced points at 0.14088~day intervals. We present the results obtained with the Morlet-6 wavelet in Figure~\ref{Fig.rez3}; it shows the spectrum for the relevant period range up to 17~days. The re-sampled light curve is plotted directly below the wavelet spectrum for clarity. The spectrum confirms primary characteristics of light curve segments, as defined in Section~\ref{general_descr}. The main reservation is that the 11.4~d modulation during {\it Segment~III} is not present in the spectrum; instead we see a 5.5~d modulation; this discrepancy is most likely due to the finite length of the run. Similarly, we also observe a false $\sim 5$~d periodicity during {\it Segment~I}, which is roughly half of the major 10.5-11~d double-peaked quasi-periodicity.

To better illustrate the period shortening in {\it Segment~II} and in the 2010 light curve, 
in Figure~\ref{Fig.rez3a} we additionally show fragments of both available {\it MOST} light 
curves with intervals $P$ between consecutive minima and maxima of the sine-like waves. 
The new analysis reveals that the short-periodic oscillation seen in the 2010--2011 light curve is more complex than that which appeared in our first paper \citep{siwak13}. We now see a 1.08~d single wave that appeared in the middle of the wave train. It could be either due to two independent, overlapping wave trains of similar periods (the maximum at HJD$-2\,455\,500\approx48.5$ may be double-peaked) or a single event in the disc 
or on the star. According to our previous interpretation \citep{siwak13}, the entire wave train showed period shortening from 2.4 to 2.2~d and this estimate was only based on the blurred wavelet spectrum. Currently, the light curve shows the period shortening from about 2.8 to 2.0-2.1~d, with some perturbations at about HJD$-2\,455\,500\approx53-56$, as noticed above. It is also possible that the period shortening took place from 2.8 to 1.08~d, and later started to increase to 2.1~d.

We shortly conclude that these short-periodic sine-like oscillations were so far clearly observed in precise space-based light curves only for a limited time and they always showed clear instances of period changes, usually shortening. This suggests that they were not produced owing to the appearance of many independent flickering events in the disc, but were driven by a mechanism leading to coherent light variations.

\subsection{Colour-magnitude relations from the ground-based data}

\label{CIdiagrams}

To investigate the colour-magnitude relations for FU~Ori, it is more convenient to utilise the Johnson $V$-filter standardised magnitudes, as obtained at {\it SAAO} and {\it MSO}. The effective wavelength of the $V$ filter is similar to that of the {\it MOST} broadband filter and lies roughly in the middle of the investigated wavelength range. Although internally more accurate, the {\it MOST} magnitudes are linked to the $BVRI$ data through the de-trending operation, as described in Section~\ref{MOSTobs}. The colour indices that we consider are $U-V$, $B-V$, $V-R_c$, $V-R$, $V-I_c$ and $V-I$. Similar diagrams were also prepared for the {\it MSO} observations obtained in the Str{\"o}mgren {\it uvy} filters, which were left in the instrumental system. Because of the similarity of the effective wavelengths, the $\Delta y - \Delta(u-y)$ and $\Delta y - \Delta(v-y)$ diagrams are closely related with the $V-(U-V)$ and $V-(B-V)$ colour-magnitude diagrams.


\begin{figure*}

\centering

\includegraphics[width=0.33\linewidth]{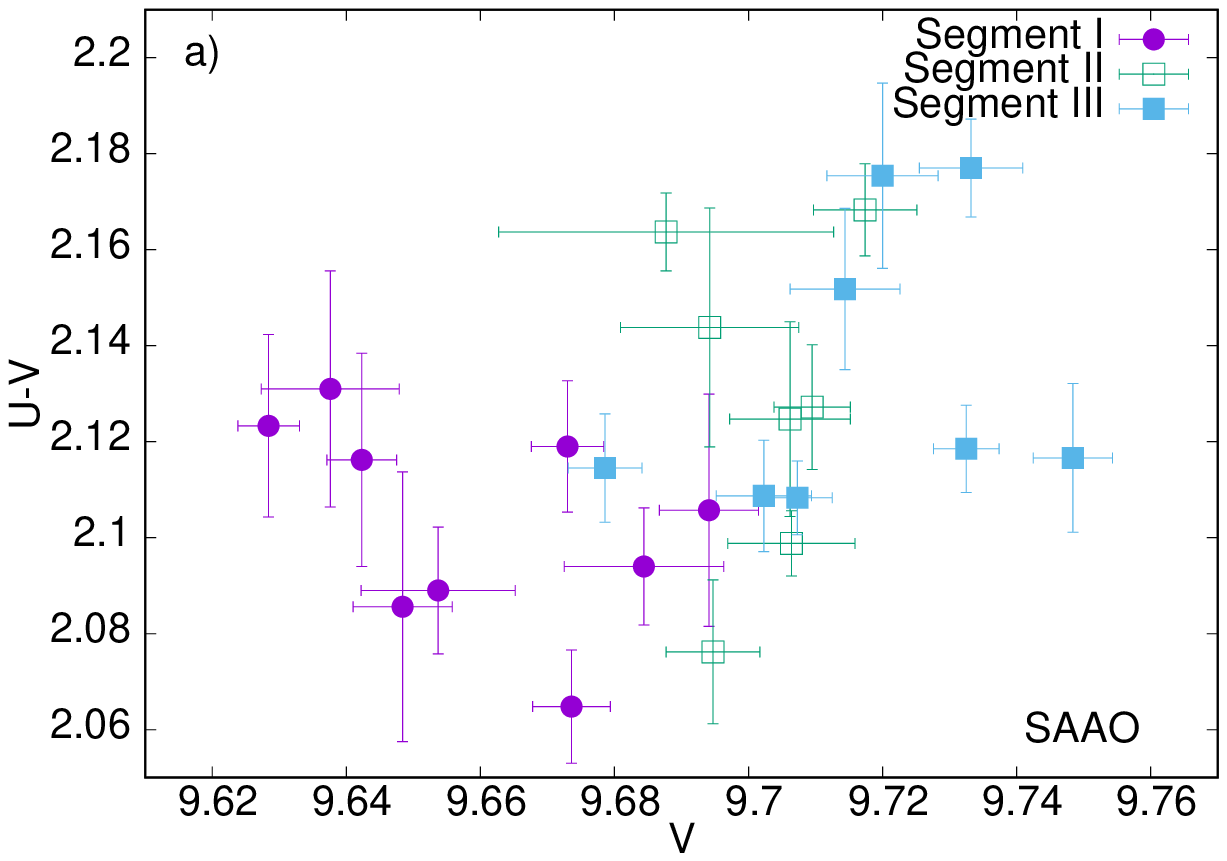}
\includegraphics[width=0.33\linewidth]{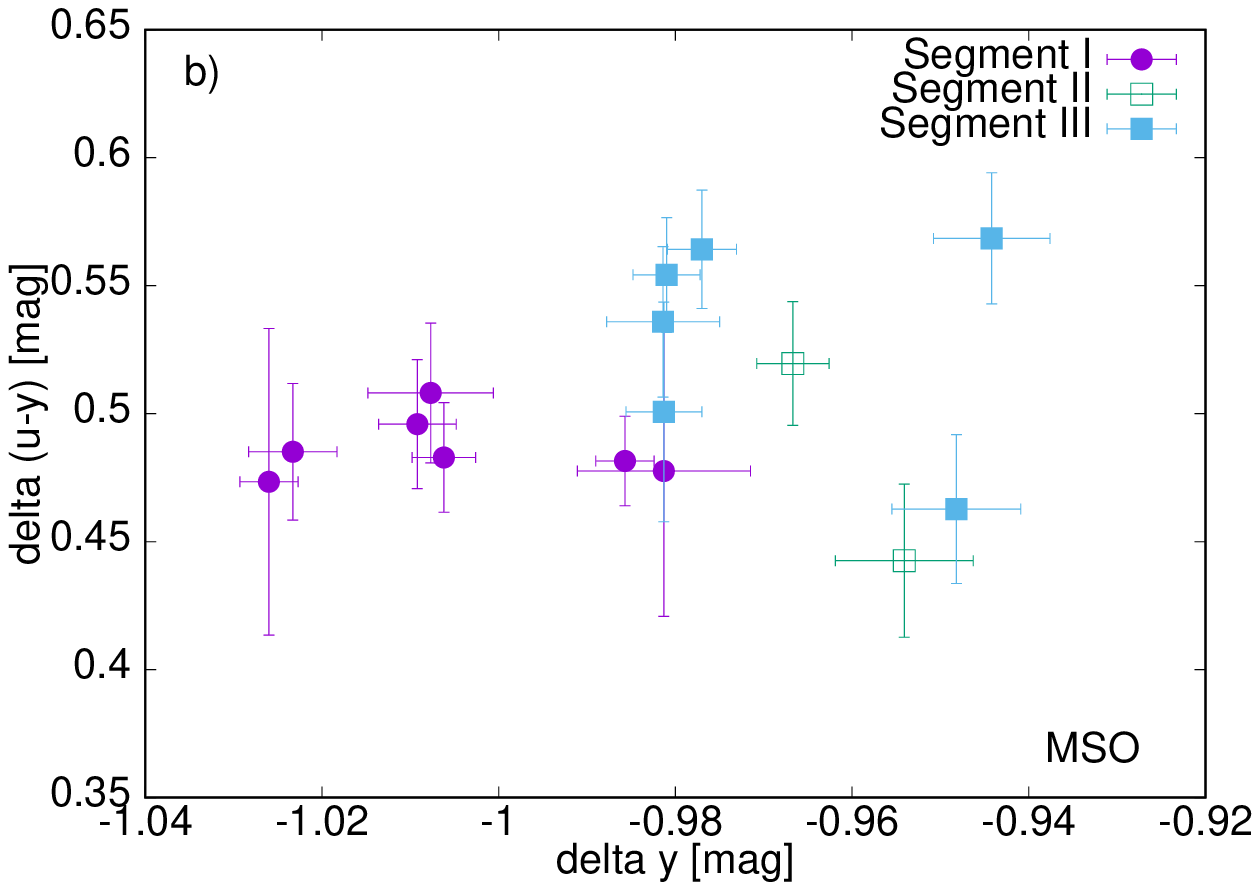}
\includegraphics[width=0.33\linewidth]{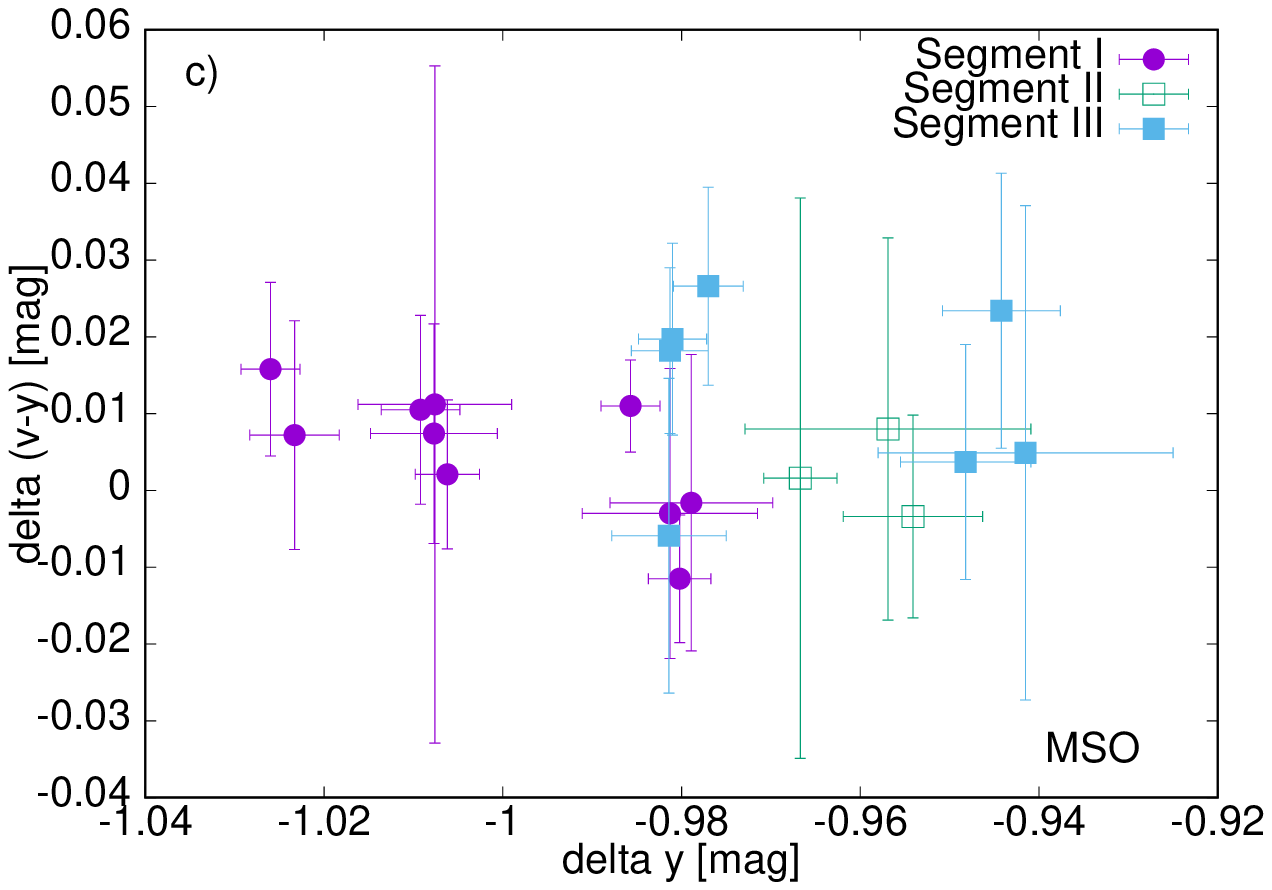}
\includegraphics[width=0.33\linewidth]{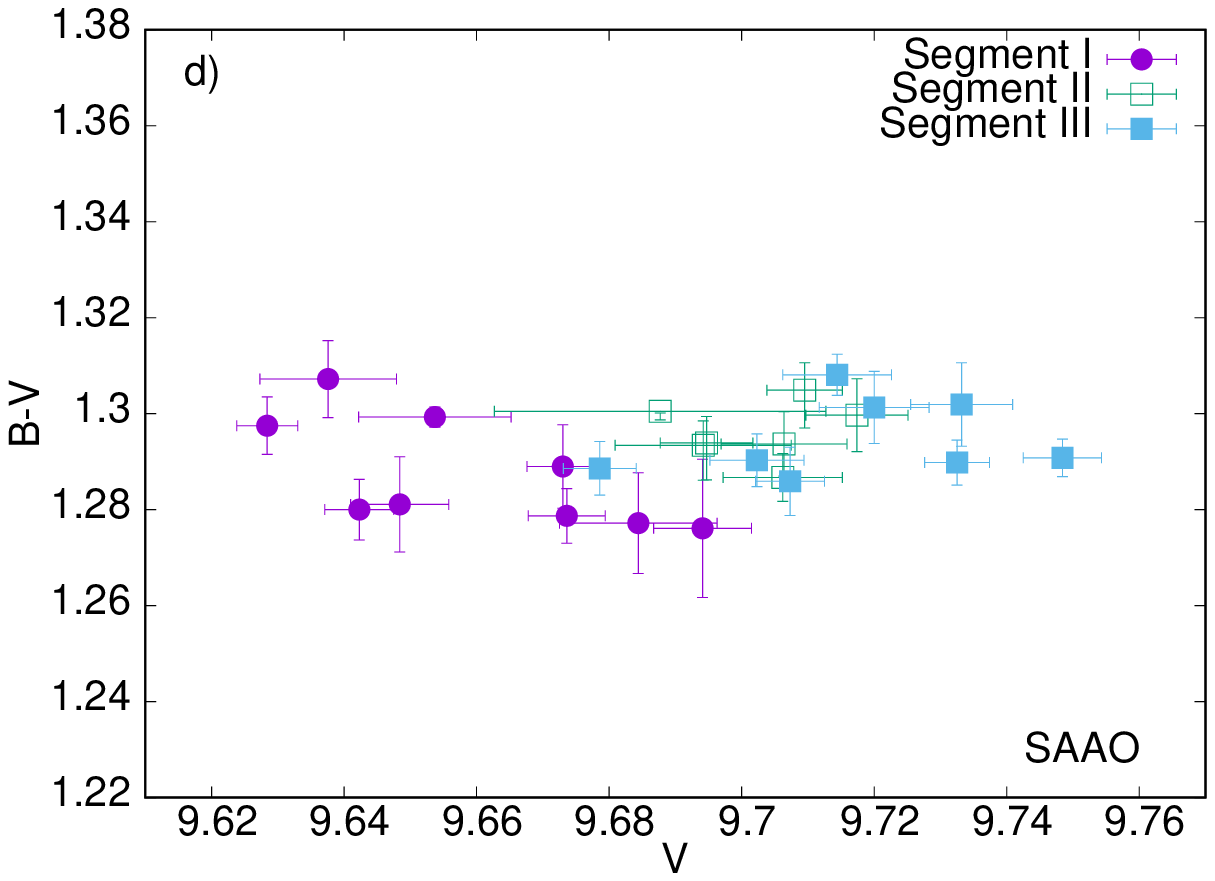}
\includegraphics[width=0.33\linewidth]{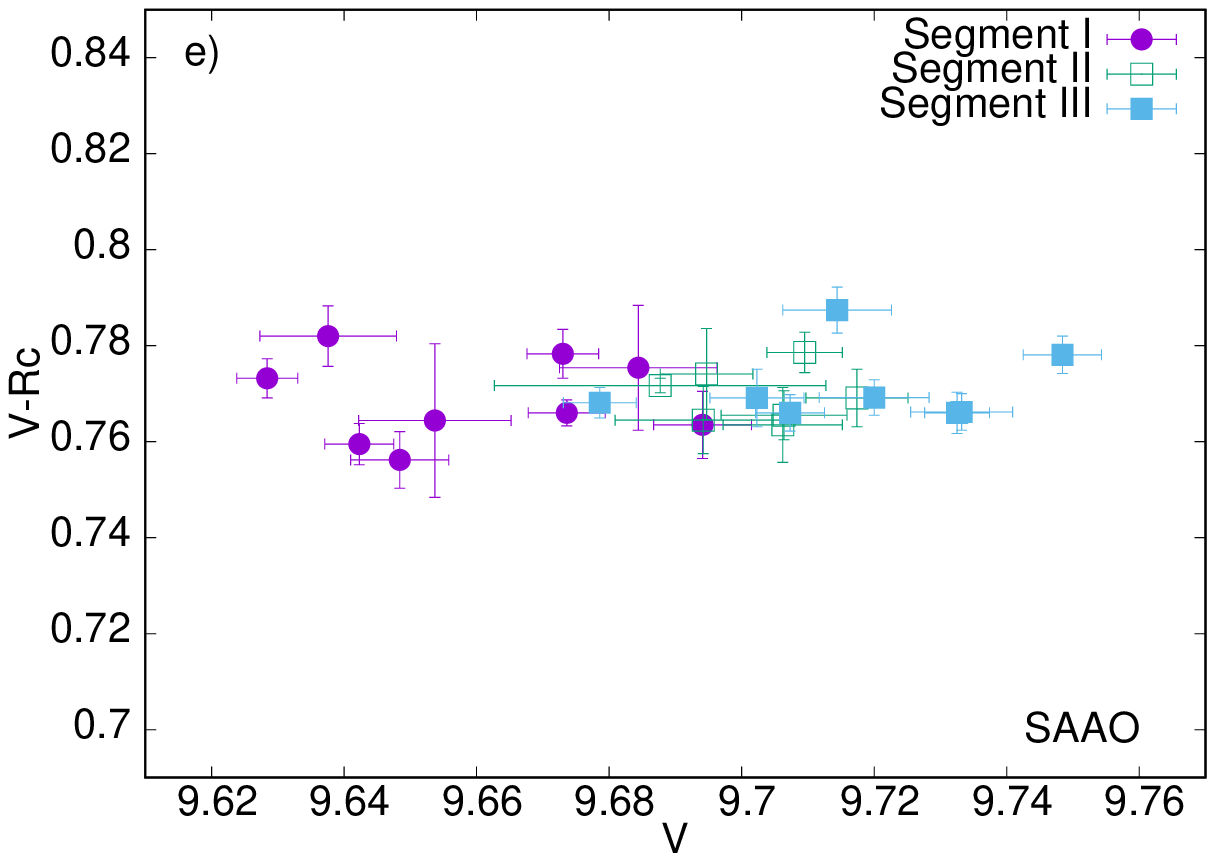}
\includegraphics[width=0.33\linewidth]{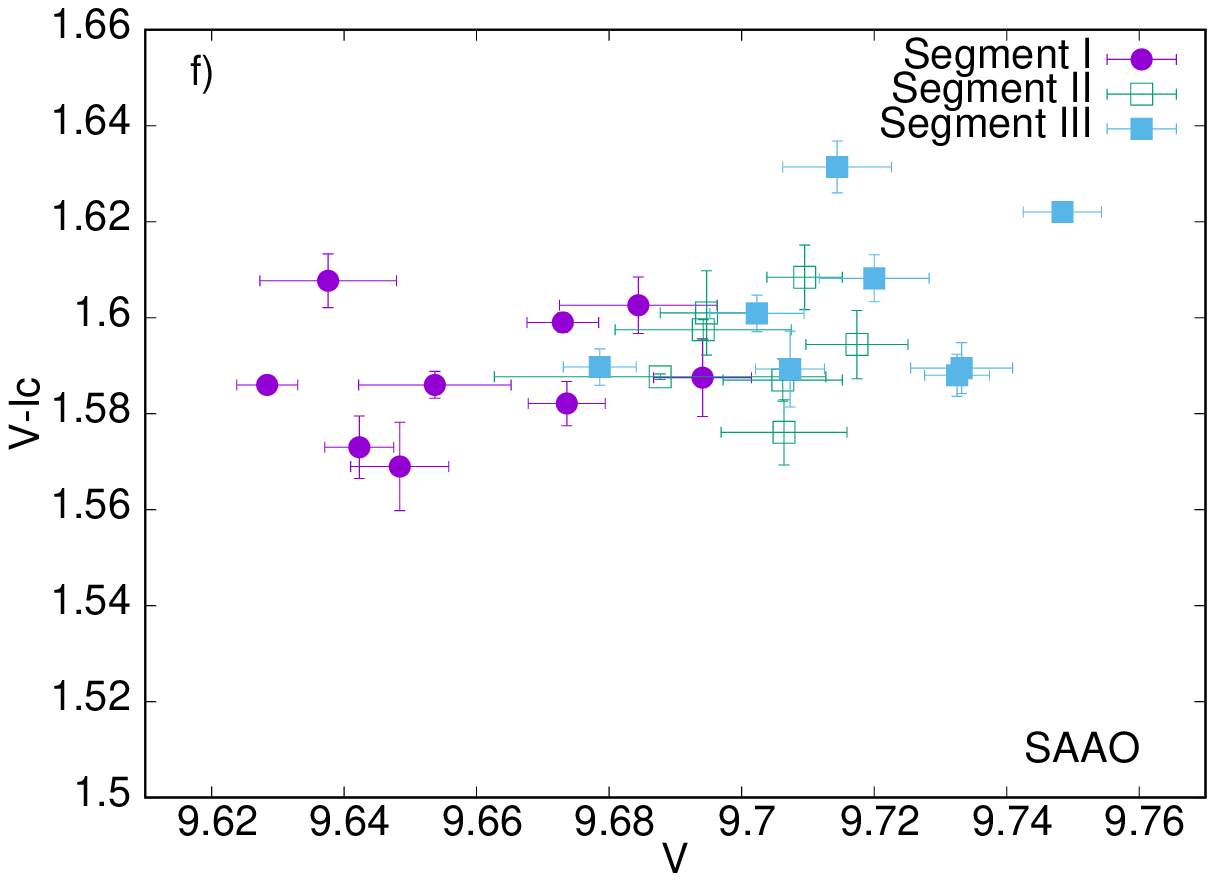}
\includegraphics[width=0.33\linewidth]{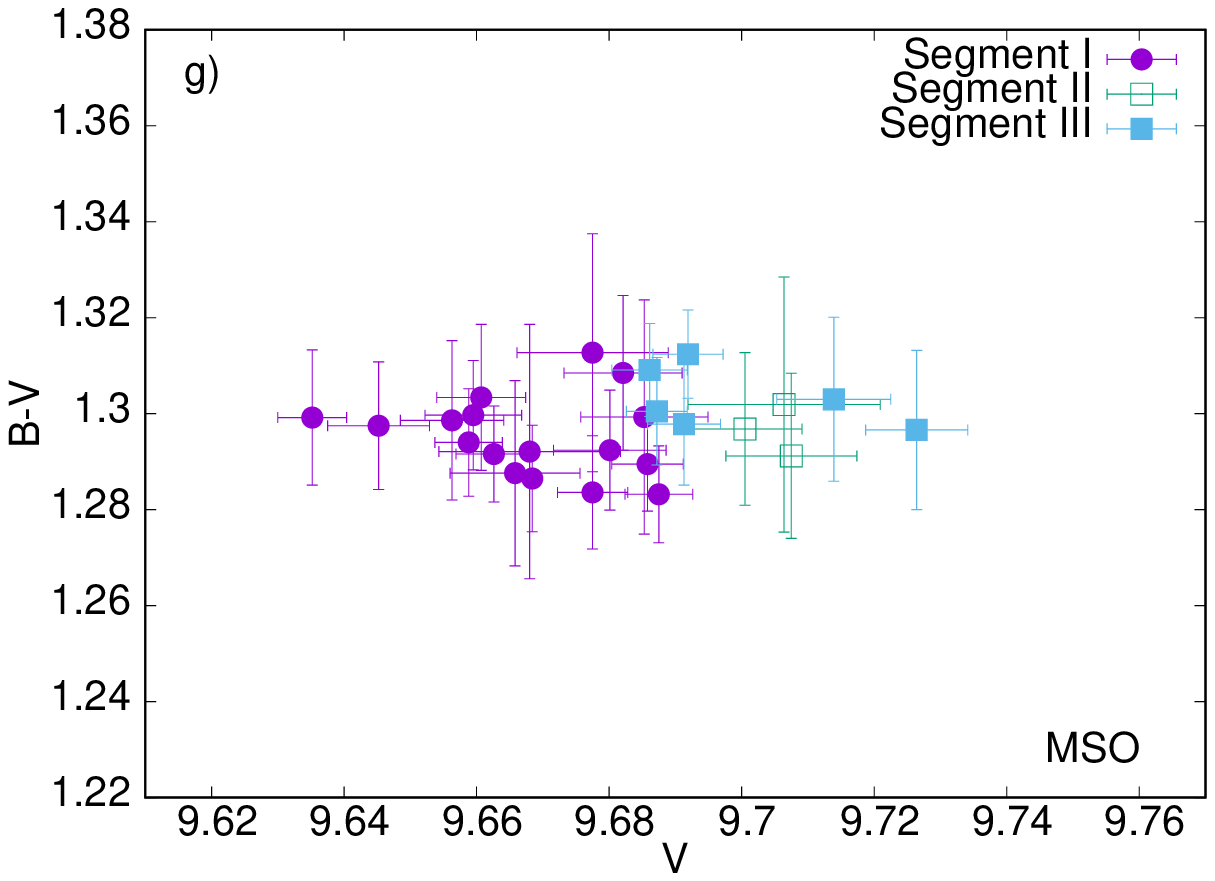}
\includegraphics[width=0.33\linewidth]{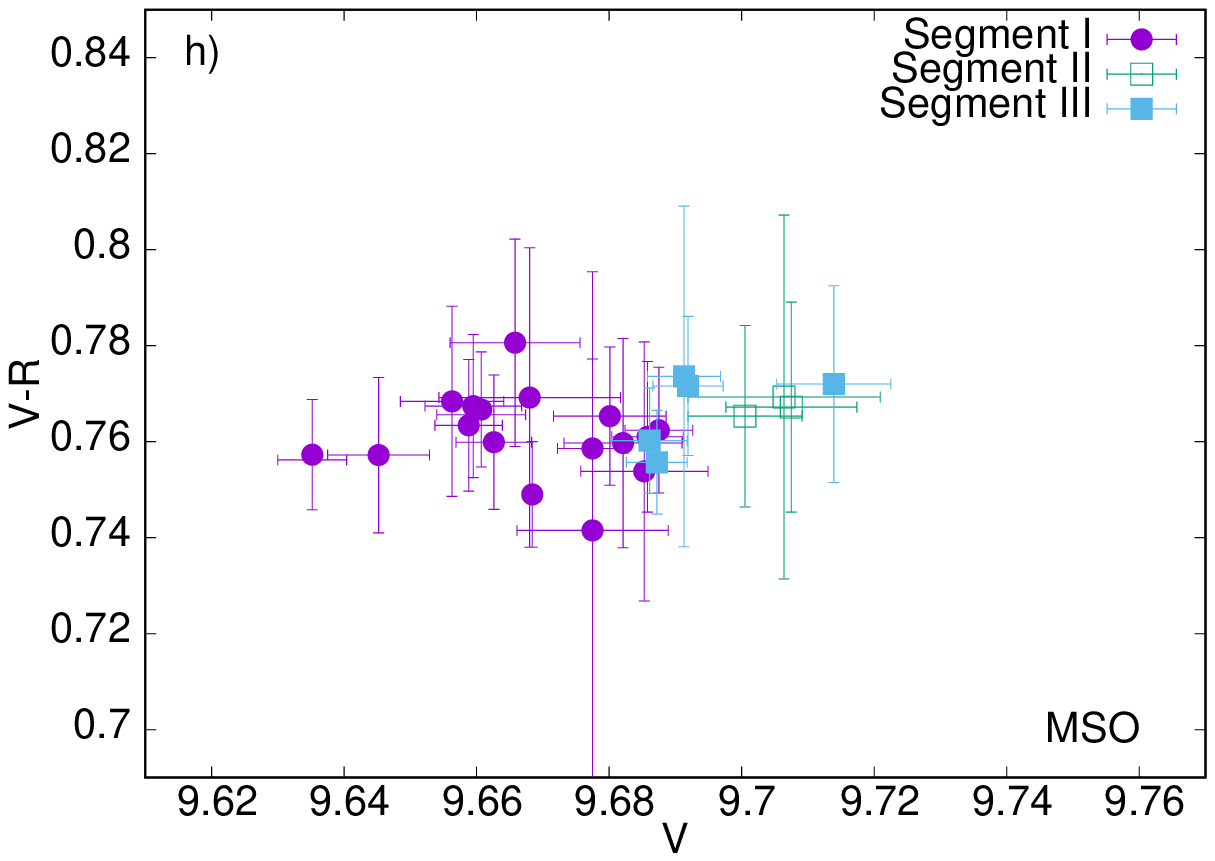}
\includegraphics[width=0.33\linewidth]{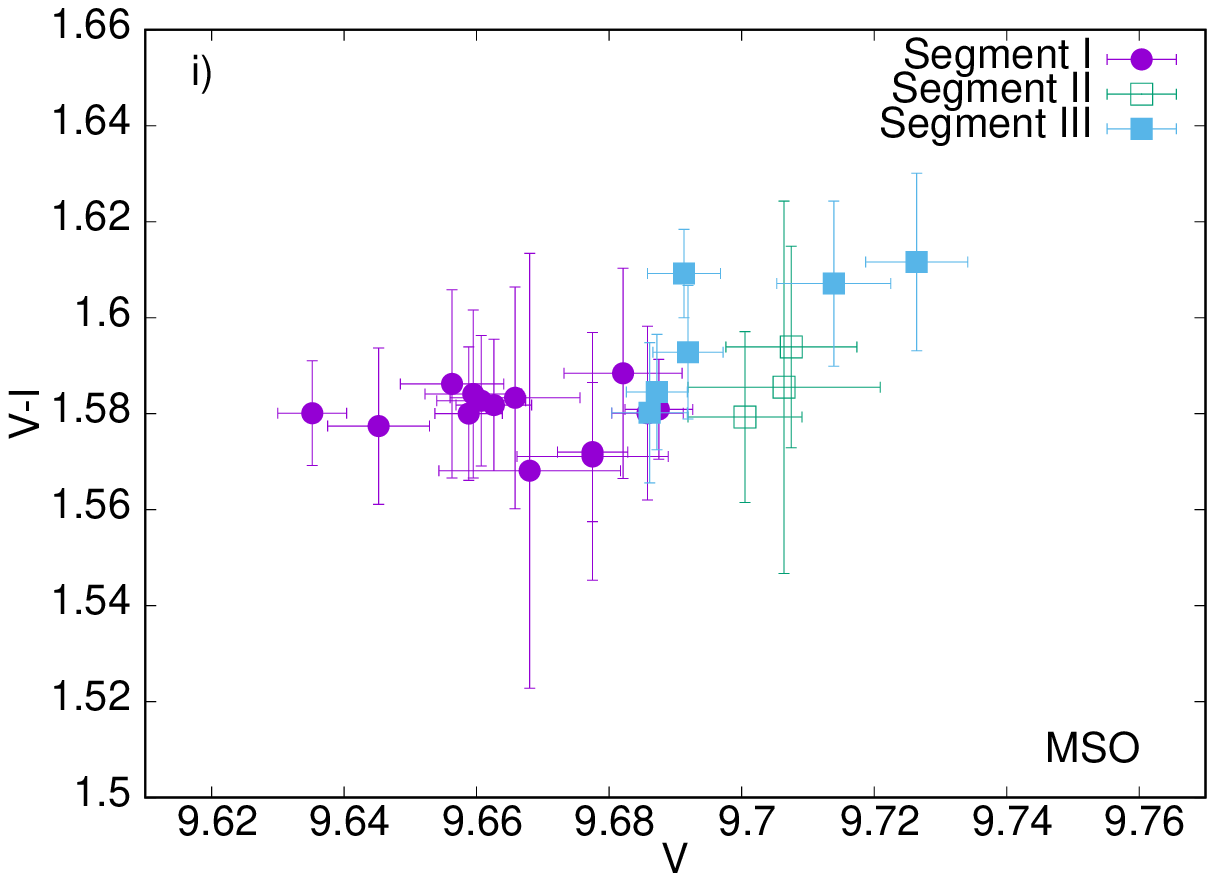}

\caption{Colour-magnitude diagrams for FU~Ori prepared from data obtained 
during {\it Segments I}, {\it II,} and {\it III}, as defined in Section~\ref{general_descr}. 
}

\label{Fig.rez4}

\end{figure*}

\begin{figure*}

\centering

\includegraphics[width=0.33\linewidth]{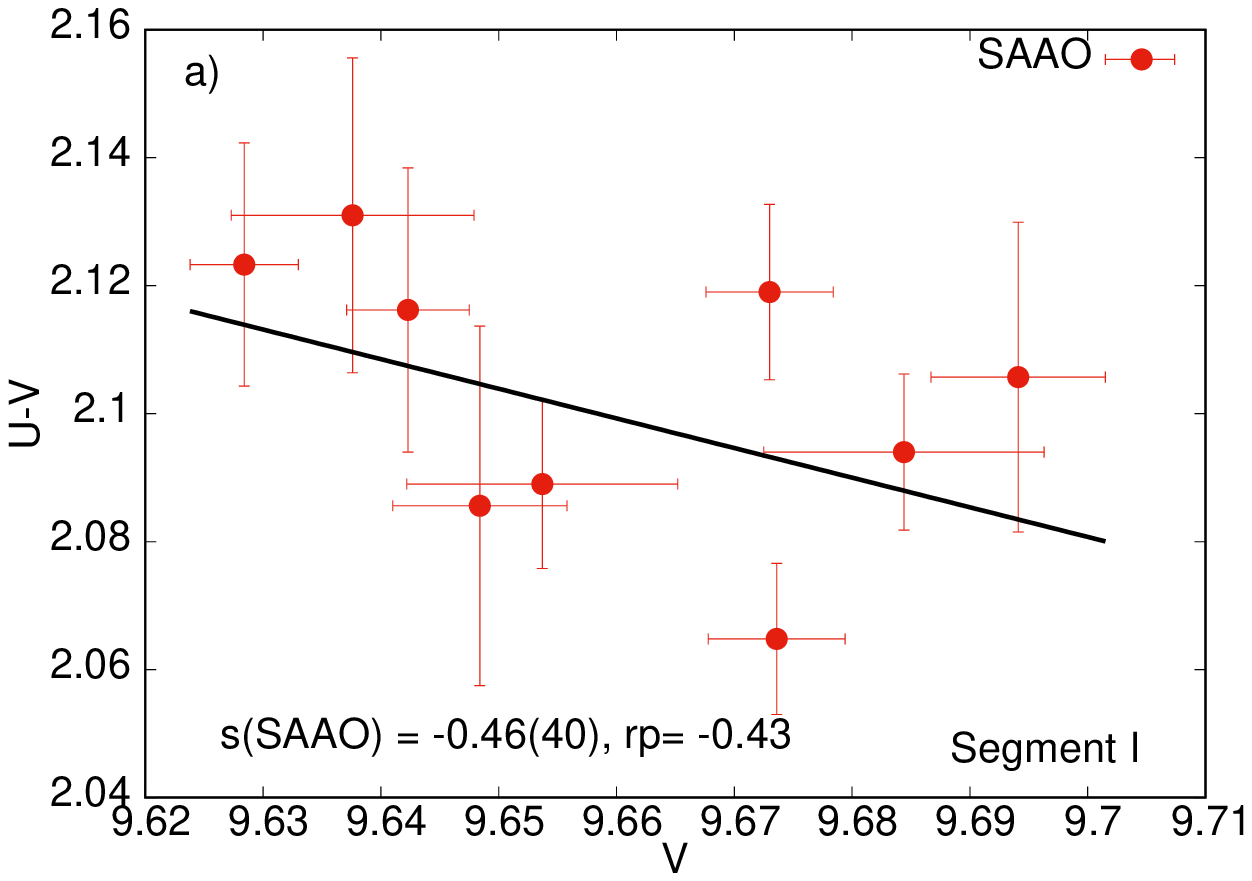}
\includegraphics[width=0.33\linewidth]{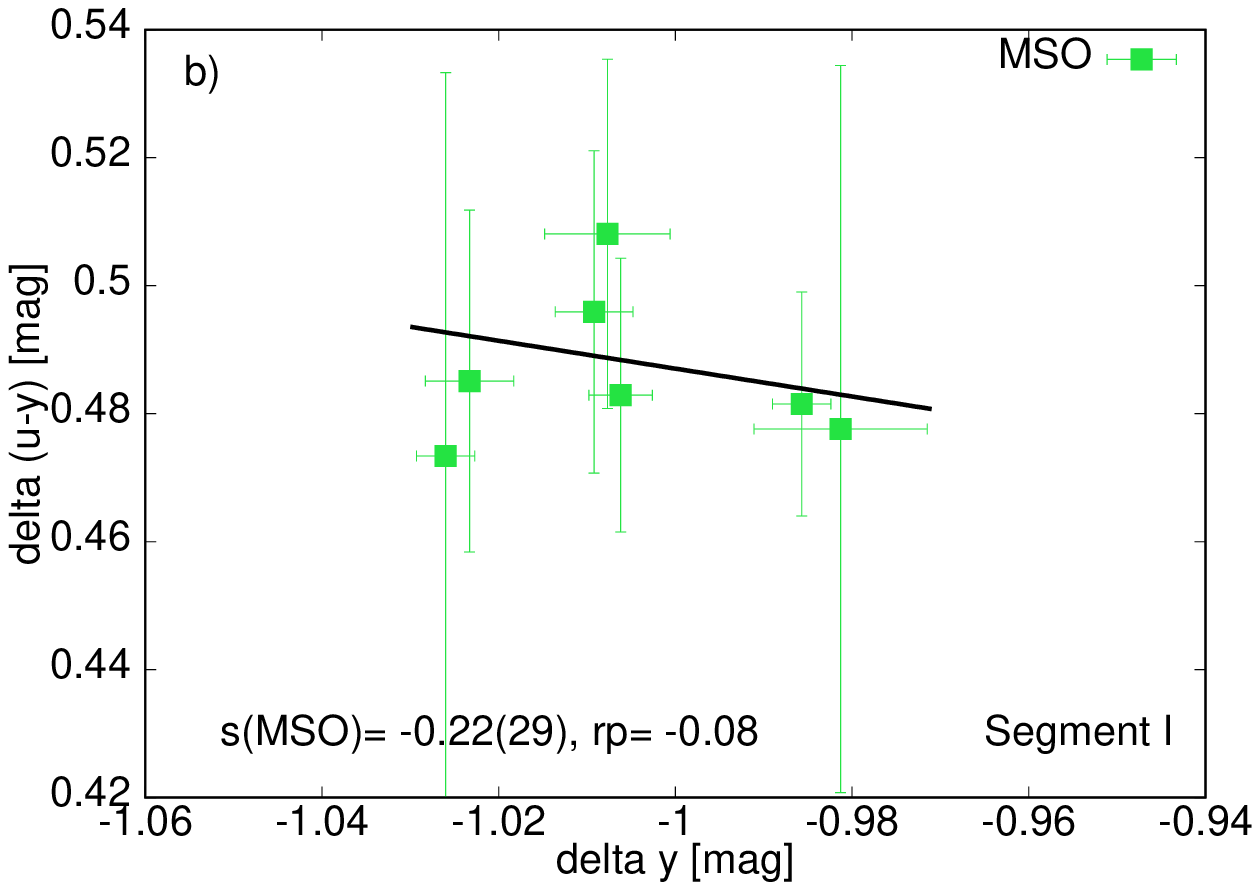}
\includegraphics[width=0.33\linewidth]{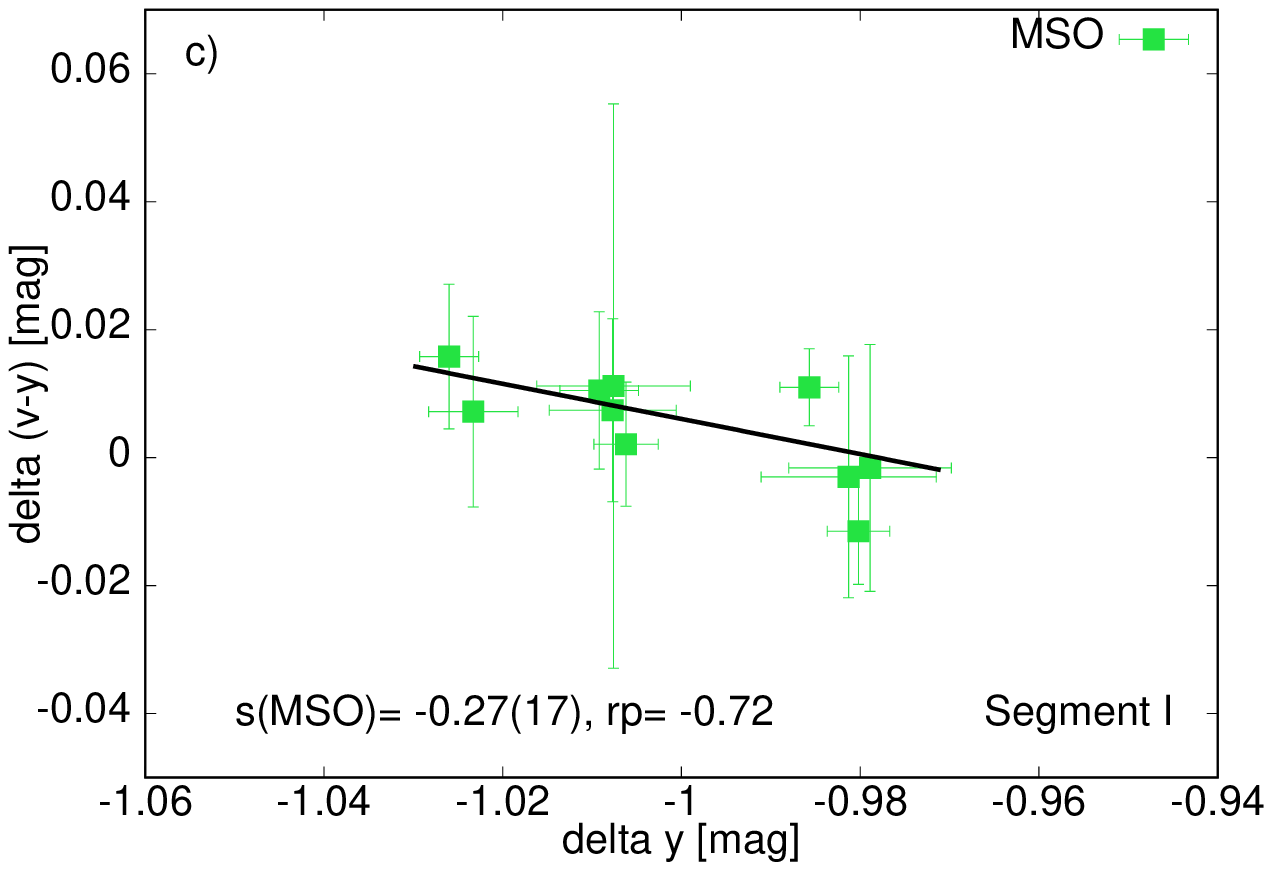}
\includegraphics[width=0.33\linewidth]{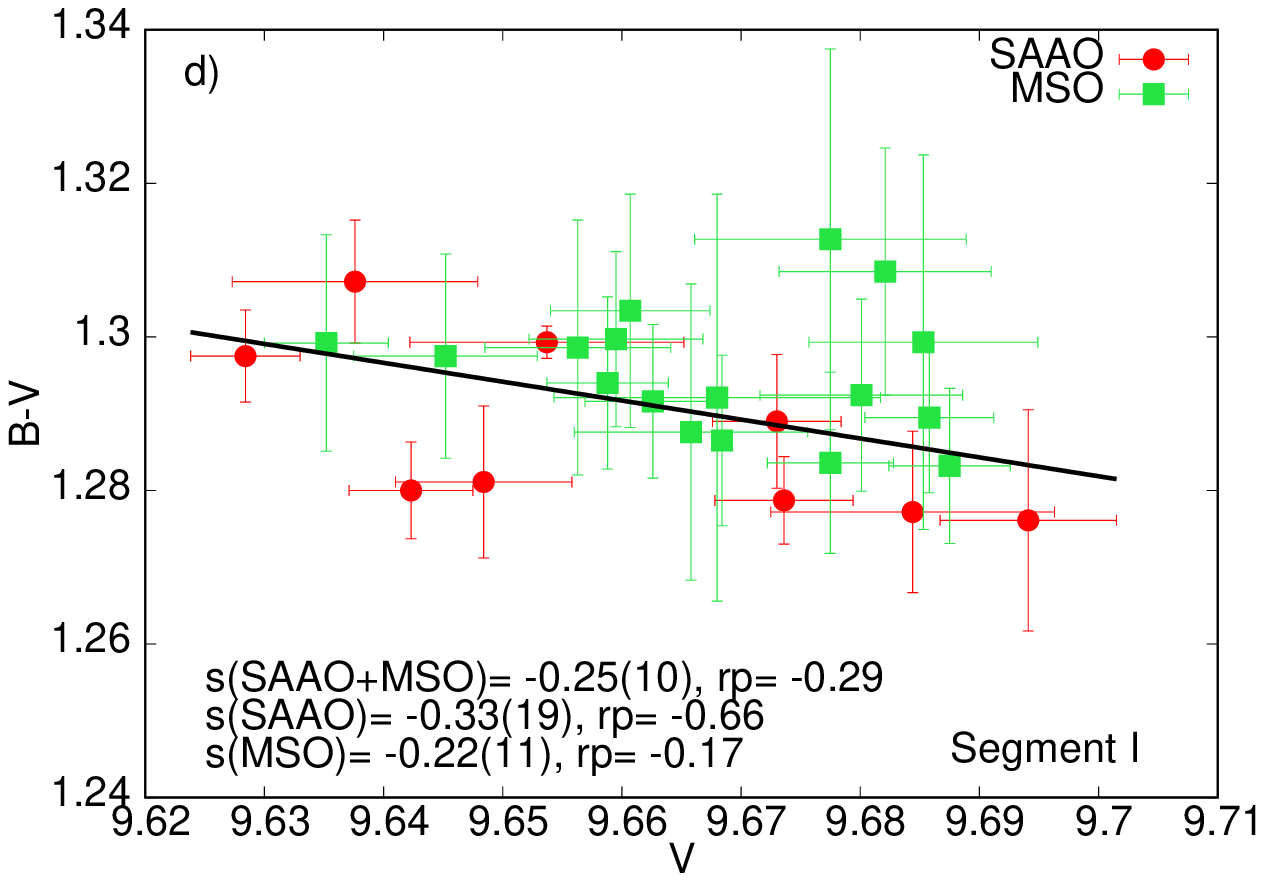}
\includegraphics[width=0.33\linewidth]{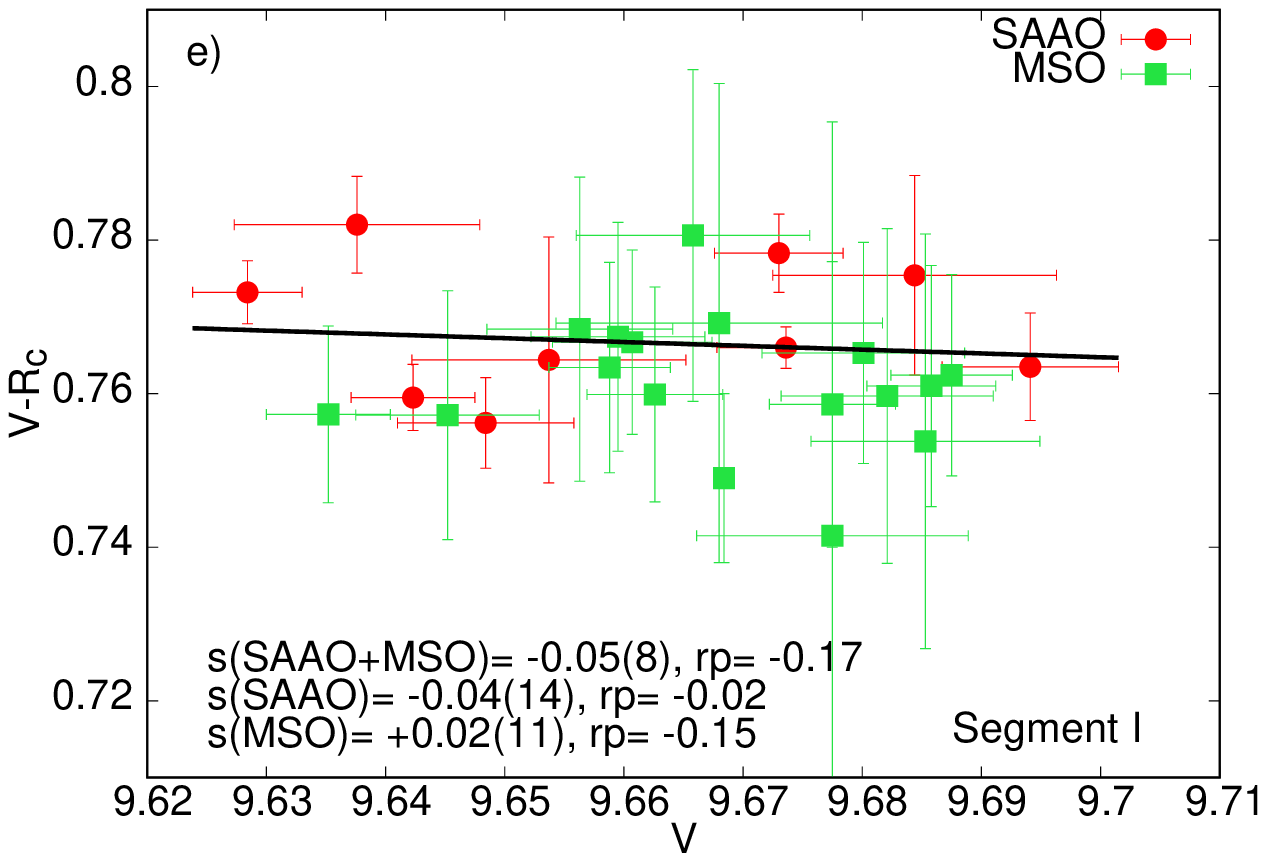}
\includegraphics[width=0.33\linewidth]{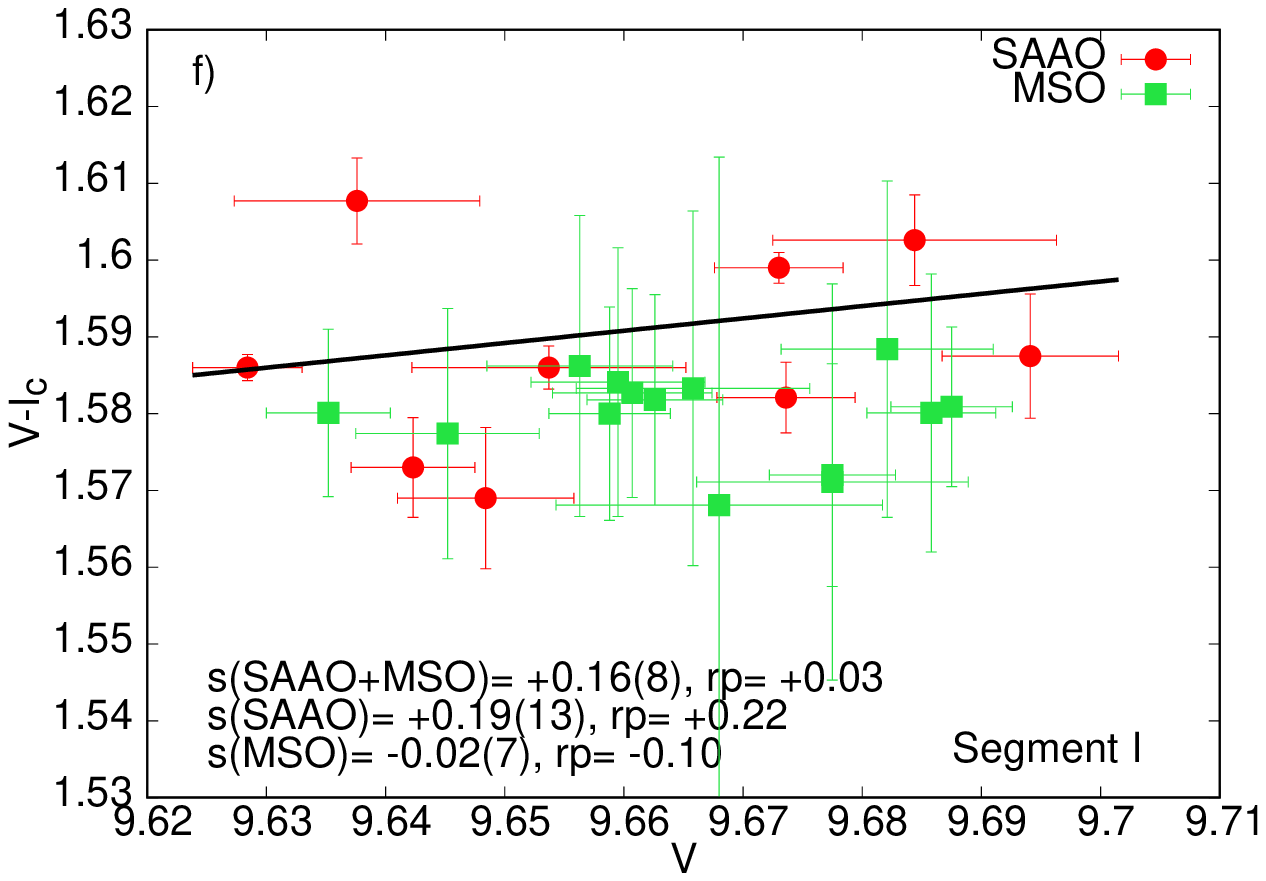}
\includegraphics[width=0.33\linewidth]{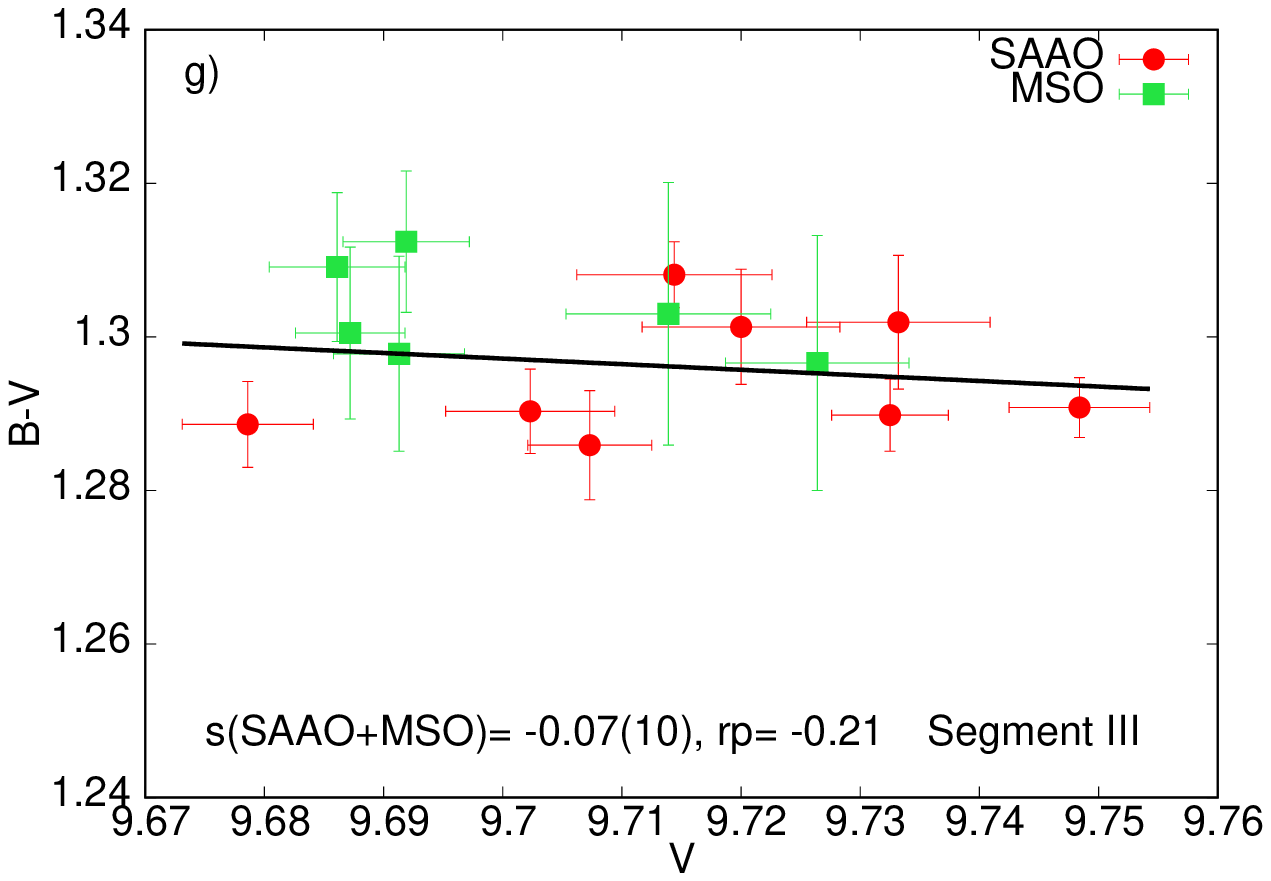}
\includegraphics[width=0.33\linewidth]{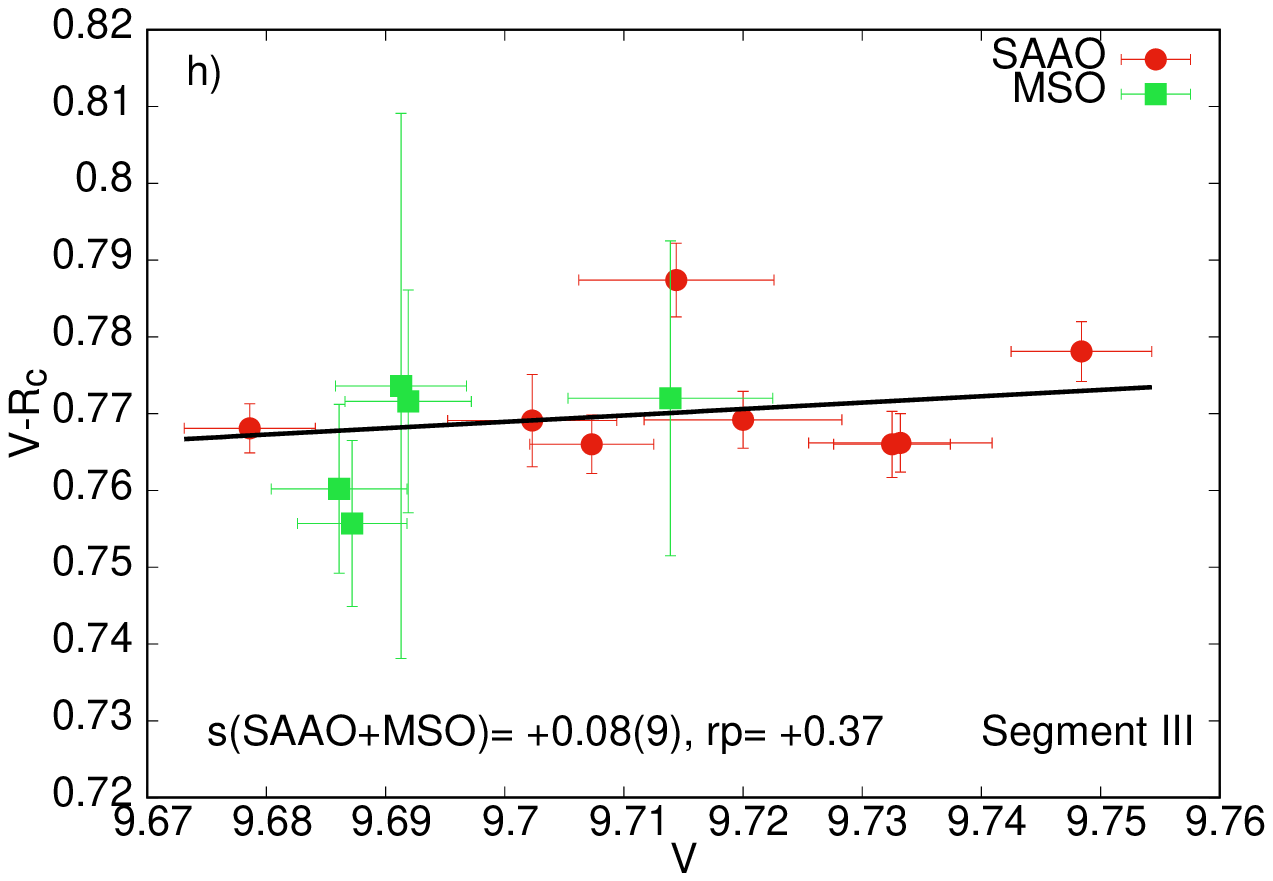}
\includegraphics[width=0.33\linewidth]{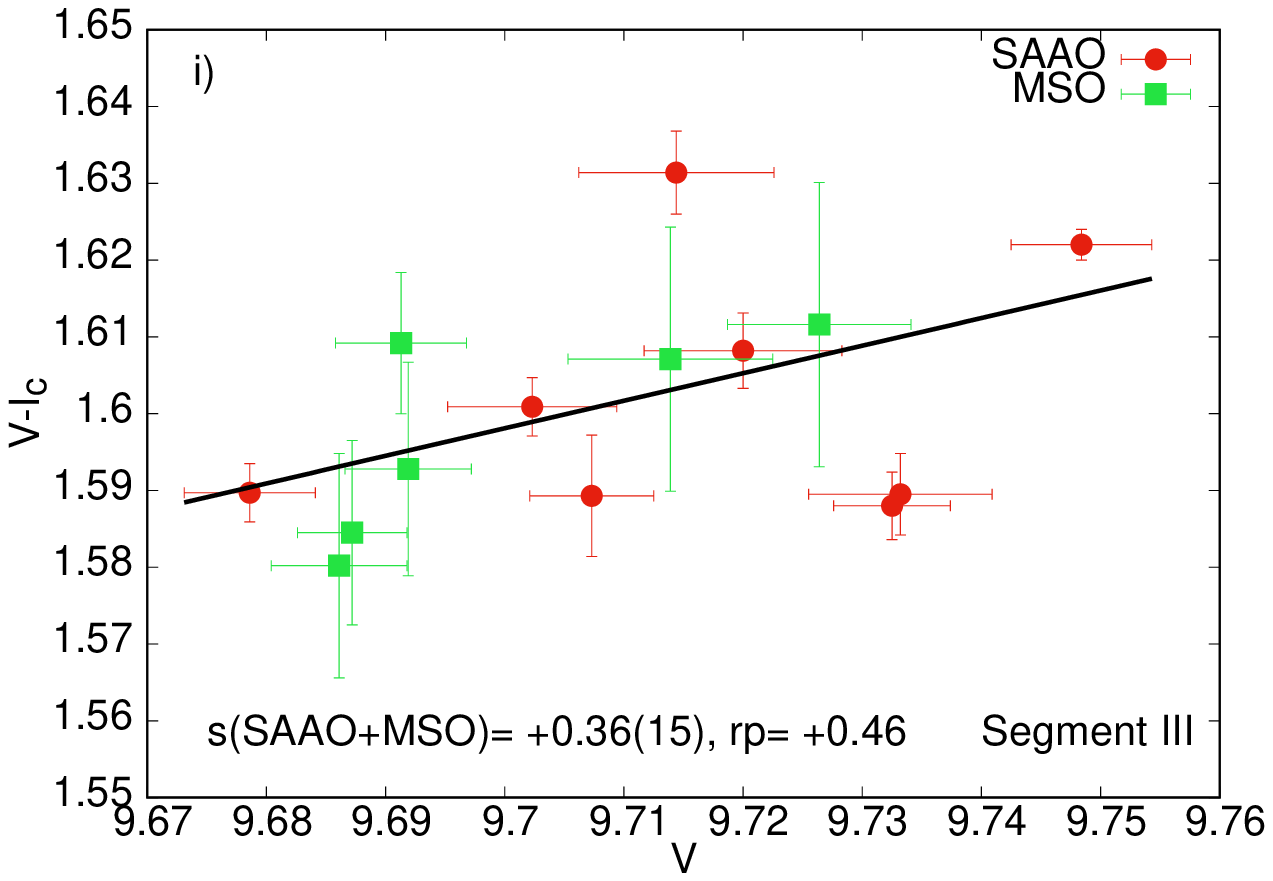}

\caption{Colour-magnitude diagrams for FU~Ori specifically prepared 
for the first and the third {\it Segment}, as defined in Section~\ref{general_descr}. 
A linear least-squares weighted
fit is shown and the numerical values of slopes 
with asymptotic standard error(s) in parentheses are given. 
Pearson correlation coefficients are also shown. 
} 

\label{Fig.rez5}

\end{figure*}

We do not show colour-magnitude diagrams prepared with the use of all data points as they contain a mix of effects from various light variability patterns that were observed at both observatories through different time spans. As announced in Section~\ref{intro}, to investigate colour-period relation we need to focus on colour-magnitude diagrams prepared for the pre-defined light curve segments; only such an approach allows us to make a link between variability patterns well defined in the {\it MOST} light curve and variability of their colour indices as a function of $V$ and $y$-filter brightness. We show colour-magnitude diagrams constructed from {\it SAAO} (Figure~\ref{Fig.rez4}a,d,e,f) and (separately) from {\it MSO} data (Figure~\ref{Fig.rez4}b,c,g,h,i) to highlight distinct properties of the three individual segments; these distinct properties are best visible in $V-(U-V)$, $V-(B-V)$ and in corresponding $\Delta y-\Delta(u-y)$ and $\Delta y-\Delta(v-y)$ diagrams, which show that especially the points obtained during {\it Segment~I}  occupy separate regions.
Whenever an unambiguous fit of a linear function (weighted by the $V$ or $y$ filter and colour index errors) was possible, we give numerical values of the slope coefficients $s$ and Pearson's correlation coefficients $rp$ both for combined and separate {\it MSO} and {\it SAAO} data sets (Figure~\ref{Fig.rez5}). Although a limited number of data points cause the significance of each separate fit to be low, we stress that we obtained the same tendencies for each of the three independent photometric systems; only the slope for the $V-(V-I)$ relation constructed from {\it MSO} data is negative (Fig.~\ref{Fig.rez5}f), but this is mostly due to its non-uniform coverage with data points. We also note that the slopes obtained from our data are in qualitative accordance with the slopes obtained from the by far much more numerous multi-season data set by \citet{kenyon2000}. These authors found -0.40(14) for the $V-(U-B)$, -0.12(2) for the $V-(B-V),$ and 0.15(2) 
for the $V-(V-R)$ relations.

Figures~\ref{Fig.rez5}a,b,c,d prepared for {\it Segment I} of the light curve show evidence for negative slopes in the $V-(U-V)$, $\Delta y-\Delta(u-y)$, $\Delta y-\Delta(v-y)$ and $V-(B-V)$ diagrams. This tendency is seen both for the {\it SAAO} and for the {\it MSO} data. The $V-(V-R_c)$ and $V-(V-I_c)$ diagrams (Figures~\ref{Fig.rez5}e,f) show tendencies for zero and positive slopes, respectively. \ We note that the ground-based observations obtained at the {\it MSO} in two photometric systems cover the full range of {\it Segment~I} variability, while at  {\it SAAO} ground-based observations were only collected during the second half.

\begin{figure*}

\centering

\includegraphics[width=0.33\linewidth]{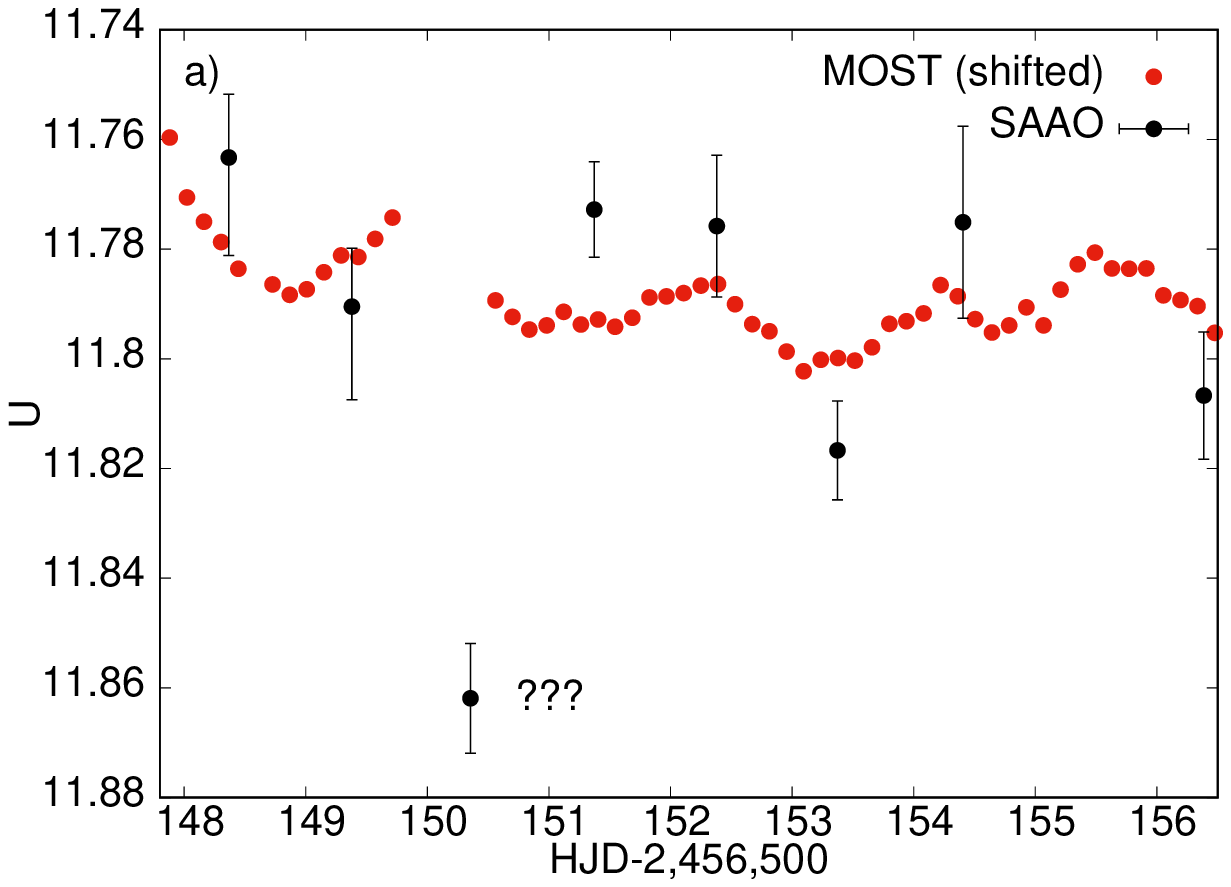}
\includegraphics[width=0.33\linewidth]{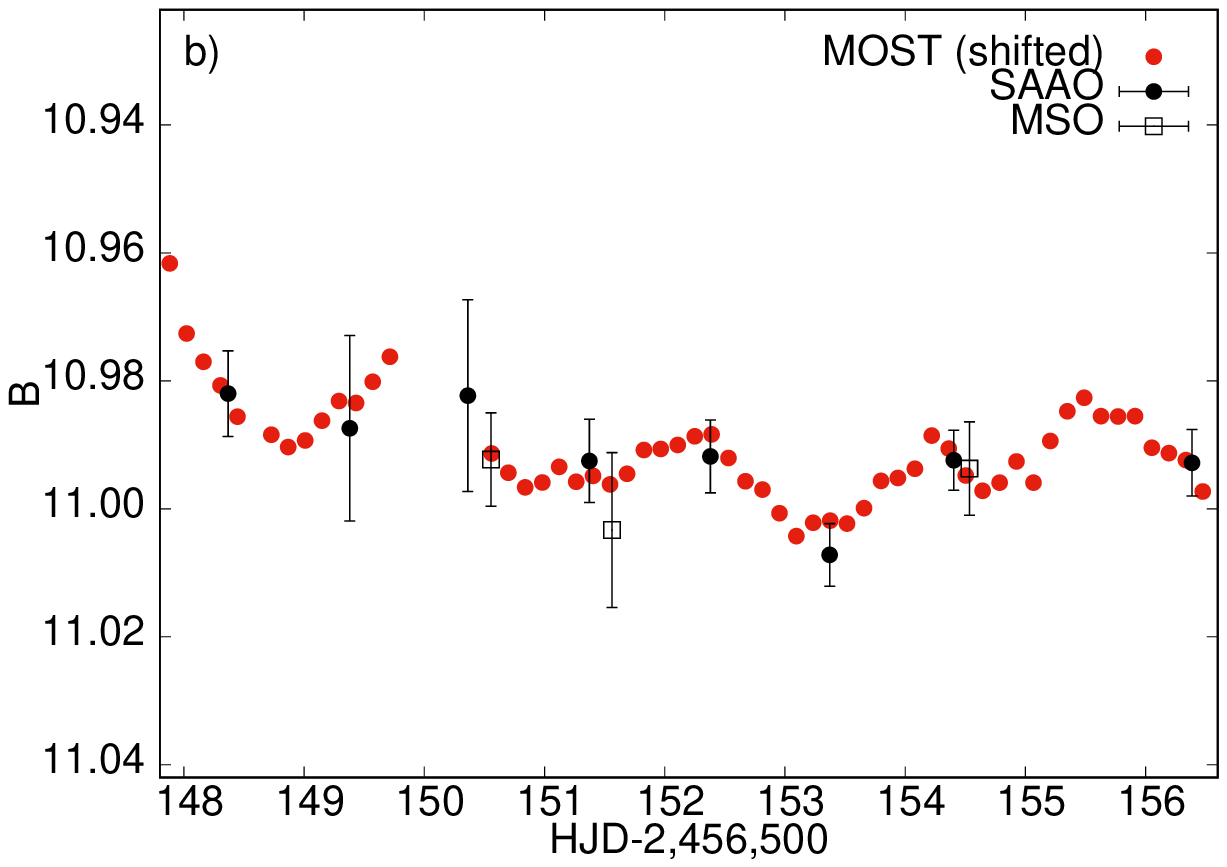}
\includegraphics[width=0.33\linewidth]{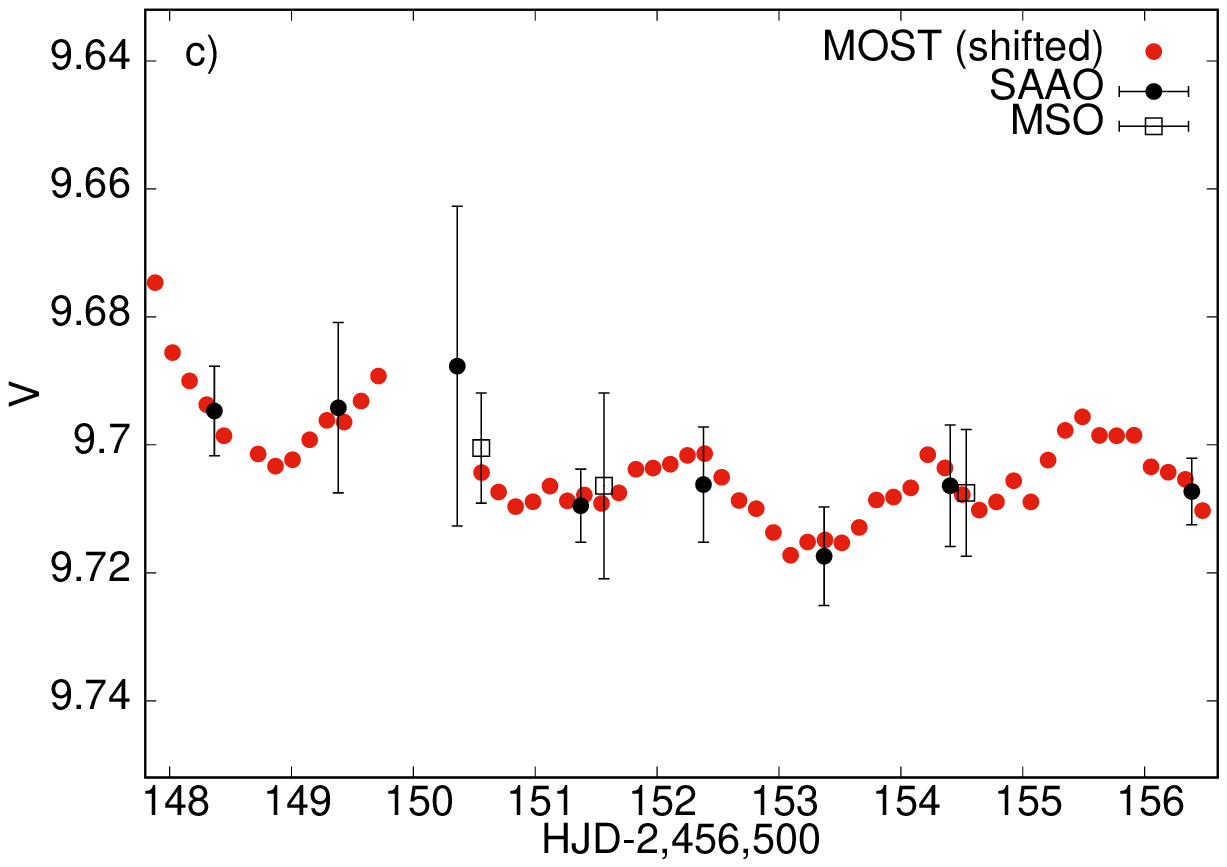}
\includegraphics[width=0.33\linewidth]{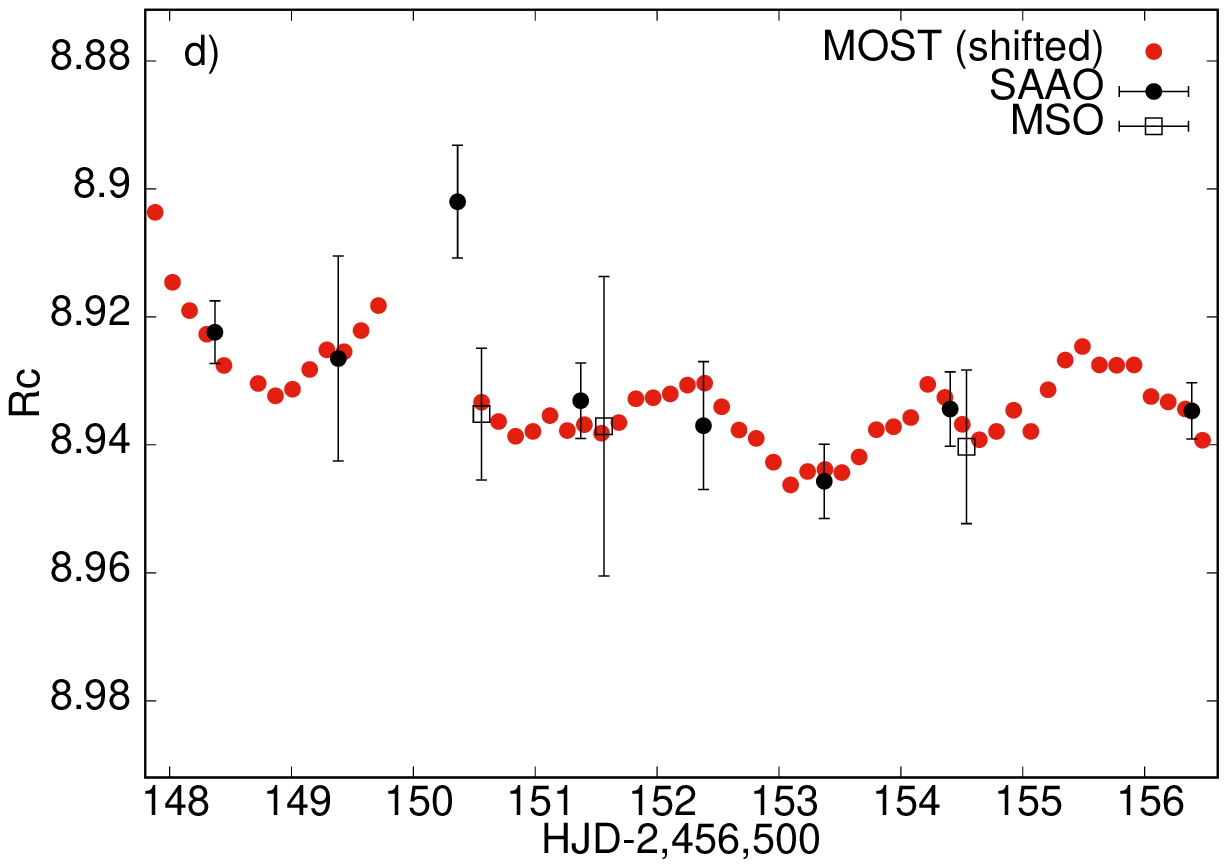}
\includegraphics[width=0.33\linewidth]{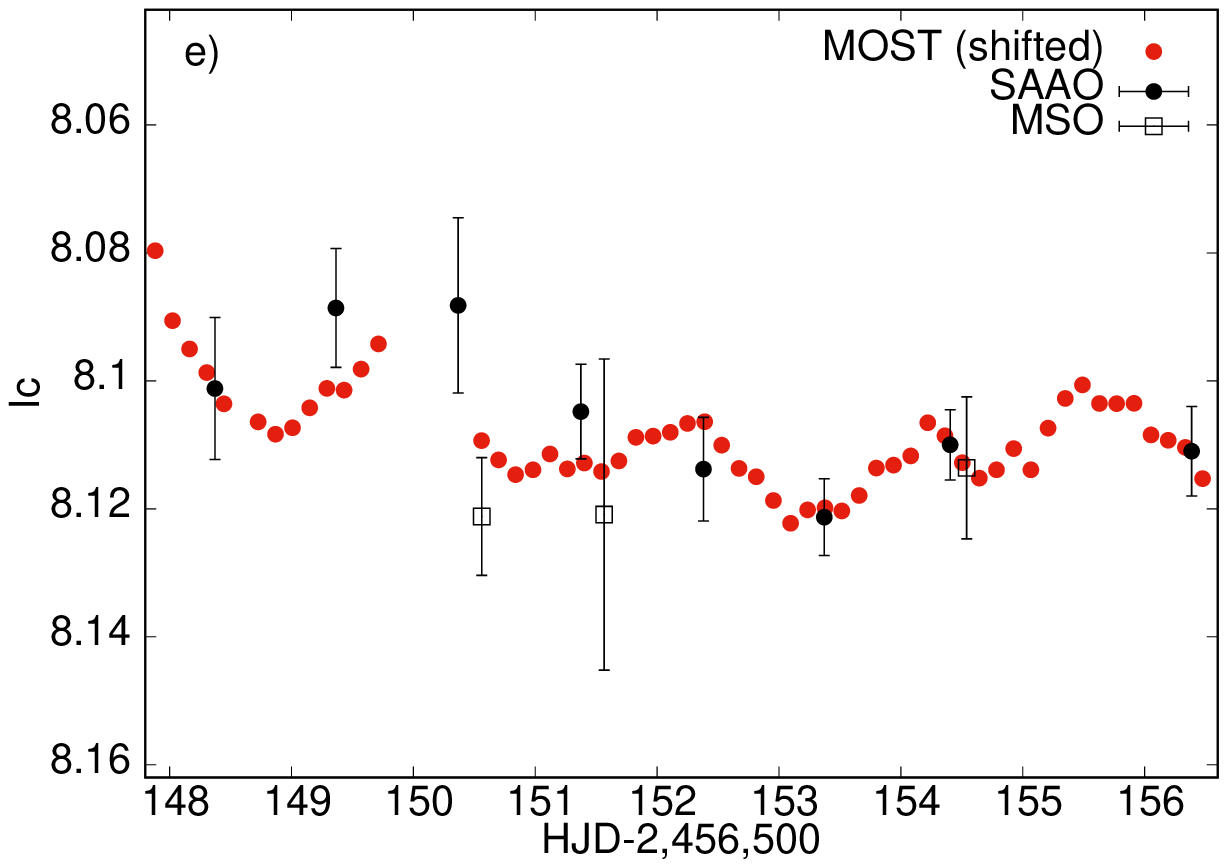}
\includegraphics[width=0.33\linewidth]{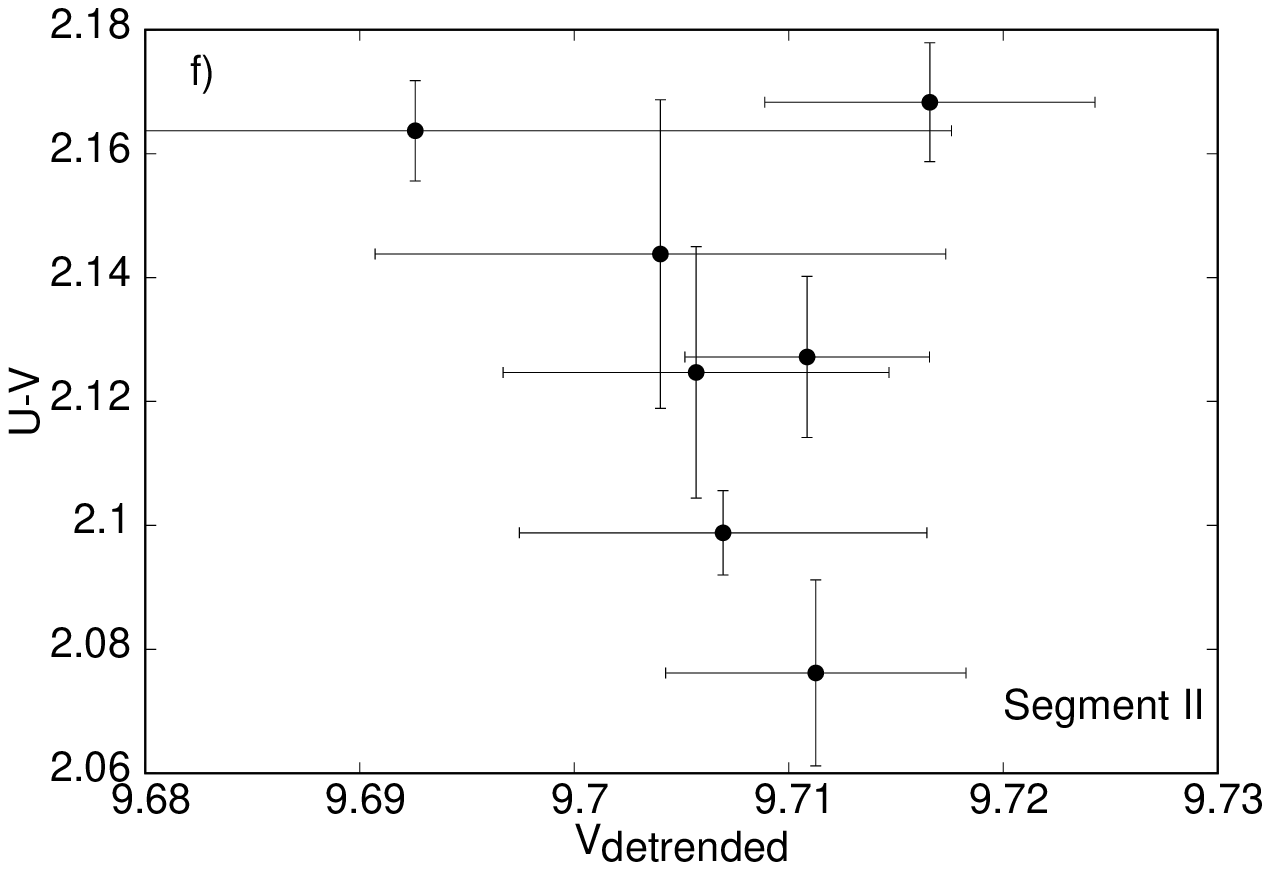}

\caption{Comparison of {\it Segment~II} of the {\it MOST} light curve 
with $UBVR_cI_cRI$ data obtained at ground-based observatories. 
The last panel shows $V-(U-V)$ diagram for this segment; the diagram indicates on larger variability amplitude in $U$- than in $V$ filter.}
\label{Fig.rez6}

\end{figure*}

The colour-magnitude diagrams for {\it Segment~II} do not show any dependency (and therefore they are not shown here). Although we made an attempt to remove the downward trend in $V$-filter data visible through this time interval by a parabolic fit, the large errors of the few ground-based measurements, in comparison with the mean amplitude of the sine-like variation of only 0.0055~mag (in the {\it MOST} system), exclude the possibility of  finding any significant trends. We can only state that amplitudes in $BVR_cI_cRI$ filters were roughly constant. We arrived at this conclusion just by looking in Figures~\ref{Fig.rez6}b,c,d,e, where we plot the arbitrarily shifted {\it MOST} and ground-based light curves. We note that the $U$-filter data show the signature of about a twice larger amplitude than observed by {\it MOST} and in the remaining Johnson filters (Fig.~\ref{Fig.rez6}a,f).

The colour-magnitude diagrams for {\it Segment~III} (Fig.~\ref{Fig.rez5}g,h,i) show similar relations as those for {\it Segment~I}. We consider the results obtained by the least-square fits to the combined data sets only because of the very limited number of measurements in each separate sample. The scatter in the $V-(U-V)$ diagram (Fig.~\ref{Fig.rez4}a) is very large and we were unable to determine any relation. The data points in the related $\Delta y-\Delta (u-y)$ and $\Delta y-\Delta (v-y)$ diagrams are slightly less scattered (Fig.~\ref{Fig.rez4}b,c). They do not seem to contradict the statement that slopes in {\it Segment~III} show similar wavelength dependency as in the {\it Segment~I} diagrams.


\subsection{Search for origins of light variations 
using a disc and star light synthesis model}

\label{model}

Apart from the phenomenological considerations presented above, we decided to construct a simple disc and star light synthesis model to pinpoint the source of light variations in FU~Ori analytically. 
Our model (see in Appendix~\ref{model_app} and \ref{valid} for full description), 
is similar to that constructed by \citet{zhu07}. The major difference is that we use 
{\it PHOENIX} library of spectral intensities for ordinary supergiants \citep{husser13}.
By means of our model we calculated synthetic amplitudes in Johnson filters for all light curve 
{\it Segments}, assuming that either hotspots on the star or disc inhomogeneities 
are responsible for observed light variations.

\subsubsection{Testing the accretion hotspot scenario}  

\label{spotmodel}

According to the idea of \citet{kenyon2000} and \citet{audard14}, flux modulations induced by the changing visibility of hotspots on the star could potentially be noticed despite the prevailing disc light and lead to quasi-periodic variability, with amplitudes suppressed from 1-2~mag (as for CTTS), to a few hundredths of a magnitude (as for FU~Ori). 
Though such a mechanism was excluded for past FU~Ori data by \citet{kenyon2000}, it may occasionally operate in this star even though the magnetosphere 
in FU~Ori is heavily compressed and much smaller than in typical CTTS \citep{konigl11}. 
To check the hotspot scenario, we calculated synthetic amplitudes caused by rotation of a spotted star for each of the Johnson-Cousins filters to compare the results with our observations. We considered possible values of the stellar radius ($R^{\star}$) in the range 1.5-2.0~R$_{\sun}$ along with several values (3500-4000~K) of effective temperatures of the stellar photosphere ($T_{eff}^{\star}$), but finally we decided to fix the parameters at $T_{eff}^{\star}=4000$~K and $R^{\star}=2$~R$_{\sun}$.\newline
The hotspot was assumed to lie at 50~deg latitude to let the whole spot hide behind the star and lead to the variability shape observed in {\it Segment~I}. The hotspot {\it banana-shape} was approximated in our model by a spherical rectangle with width of $\sim8$~deg in latitude and $\sim60$~deg in longitude, as suggested by the three-dimensional~magnetohydrodynamical numerical simulations of \citet{kulkarni13}. We stress that detailed values of these parameters, and whether the shape and position 
is typical for stable or unstable accretion regimes considered by the authors, do not impact 
our results in a significant way. We performed computations with typical hotspot temperatures in the range 7000-12000~K using corresponding {\it PHOENIX} intensities, 
although a more detailed treatment should also include emission lines calculated 
by \cite{dodin18}. For each hotspot temperature, a corresponding set of linear limb darkening coefficients for Johnson filters was applied \citep{diaz-cordoves95,claret95}.

Synthetic amplitudes of light variations for three selected hotspot temperature values $T_{spot}$ are presented in Table~\ref{Tab.hotspot}. A strong wavelength-amplitude dependency is obvious for all considered cases. The amplitudes are highest for the $U$ filter and decrease rapidly as the wavelength increases. This is in conflict with our observations (Fig.~\ref{Fig.dat1}b, Fig.~\ref{Fig.rez1} and 
Fig.~\ref{Fig.rez4}), which show similar amplitudes for almost all segments. Moreover, more detailed analysis revealed smaller amplitudes observed in ultraviolet and blue bands 
during {\it Segments~I} and {\it III} (Fig.~\ref{Fig.rez5}a,b,c,d,g). For this reason, we can firmly state that rotation of the stellar surface with hotspots on the photosphere is not responsible for the longer family of light variations observed in 2013-2014. 
This finding is also true for the eight~day event observed in the first {\it MOST} light curve of FU~Ori, where its {\it MOST} filter amplitude was found to be larger than that measured in the {\it MSO} Str{\"o}mgren $vb$ filters \citep{siwak13}.\newline 
The result obtained above may also suggest three other possibilities. 
First, hotspots are fairly uniformly 
distributed on the stellar surface, second they are not always formed on the star, 
and third their effective temperatures are only slightly larger than the effective 
temperature of the stellar photosphere. 
We think that the second possibility may be true for FU~Ori. During {\it Segment~II} the $U$-filter 
amplitude appears to be twice as large as in the remaining filters. Although the amplitudes in $BVR_cI_c$ filters seem to be very similar, we state that these data are not accurate enough to exclude the hotspot scenario for this segment with full certainty; the significant elongation in y-axis seen in the $V-(U-V)$ colour-magnitude diagram (Fig.~\ref{Fig.rez6}f) may suggest the presence of hot radiation sources. 
We note that similar behaviour was also found in TW~Hya. During March 9, 2016 some short-term 
hotspots appeared on the star as a consequence of inhomogeneous accretion. 
Whilst the $V-(B-V)$ relation remained stable over the entire night, the corresponding 
$V-(U-V)$ colour-magnitude diagram showed two separate relationships (see in Fig.~10 in \citealt{siwak18}).


\begin{table}

\caption{Synthetic amplitudes in magnitudes predicted for the rotating spotted stellar 
surface for Johnson-Cousins filters for three selected hotspot temperatures $T_{spot}$.
}

\begin{tabular}{c c c c} 

\hline

$T_{spot}=$  &7000~K &10000~K &12000~K \\ \hline

$\Delta U$~[mag]   & 0.037   & 0.159   & 0.294   \\    

$\Delta B$~[mag]   & 0.021   & 0.084   & 0.126   \\ 

$\Delta V$~[mag]   & 0.010   & 0.031   & 0.045   \\ 

$\Delta R_c$~[mag] & 0.005   & 0.012   & 0.018   \\  

$\Delta I_c$~[mag] & 0.001   & 0.003   & 0.005   \\ \hline

\end{tabular}

\label{Tab.hotspot}

\end{table}


\subsubsection{Testing the disc inhomogeneity scenario}

\label{discmodel}

In the first paper of this series we proposed that given the visibility inclination of 55~deg, some surface and/or disc temperature inhomogeneities, which appear from the interactions of stellar magnetosphere with the disc plasma and then disappear within the disc dynamical timescale, may cause quasi-periodic flux modulations, as they revolve around the star. We also claimed that variation of their colour indices versus {\it MOST} (or e.g. $V$-filter) magnitude may depend on sizes and locations of the inhomogeneities in the disc and lead to the colour-period relation.

We propose to approximate these inhomogeneities by structures similar to spiral arms or rings, recently imaged in protoplanetary discs of young stars by {\it ALMA} \citep{perez16}, {\it VLT-SPHERE} \citep{benisty15,stolker16,avenhaus18}, and {\it Gemini-GPI} and {\it Magellan-MagAO} \citep{follette17}. We believe that two different inhomogeneities on opposite sides of the inner disc, or a single inhomogeneity seen either behind or in front of the star, may lead to the double-peaked light 
features of various amplitudes, as observed in {\it Segment~I} of the 2013-2014 {\it MOST} light curve. If this is the case, in the flat surface disc model presented in this work, the longitudinal flux distribution 
of disc annuli disturbed by such inhomogeneities could be parameterised by local declines and increases 
of its temperature with respect to the value $T_{eff}(R)$, predicted 
for a steadily accreting disc, and given by equation~\ref{teff-r}
\footnote{One can question the approach adopted in this model because the  approximation of the disc inhomogeneities by $\Delta T$ entails the use of $I_{\lambda}$, whose values are not necessarily appropriate for the disturbed plasma. }. 
Illumination of such disc inhomogeneities by the flux emerging from the innermost disc 
and even the central star may additionally increase their brightness contrast. Let us consider the revolution of one inhomogeneity around the star, as our multi-colour data are sensitive only to variability caused by the higher amplitude light modulation in {\it Segment~I};
such a structure can be approximated assuming that the second half of the disturbed disc ring is brighter than the first half. This situation can be parameterised with the use of dimensionless factor 
$\Delta T=|T(R)-T_{eff}(R)|/T_{eff}(R)$ as follows:
\begin{equation}
\label{inhom}
T(R)= \left\{ \begin{array}{ccc} 
T_{eff}(R) & \mbox{for} & R_{inn} \leqslant R < R_{pert}^{inn}, \\  
                      & & 0 \leqslant \varphi < 2\pi \\                  
(1 + \Delta T)\times T_{eff}(R) & \mbox{for} & R_{pert}^{inn} \leqslant R \leqslant R_{pert}^{out},\\
                     & & 0 \leqslant \varphi < \pi \\
(1 - \Delta T)\times T_{eff}(R) & \mbox{for} & R_{pert}^{inn} \leqslant R \leqslant R_{pert}^{out},\\ 
                     & & \pi \leqslant \varphi < 2\pi\\
 T_{eff}(R) & \mbox{for} & R > R_{pert}^{out},\\
                     & & 0 \leqslant \varphi < 2\pi,\\
\end{array}\right. 
\end{equation}
where $R_{pert}^{inn}$ and $R_{pert}^{out}$ define the inner and outer radius of a disc ring, in which the real effective temperatures $T(R)$ deviate by $\Delta T$ from these predicted 
by Equation~\ref{teff-r}, 
while $\pi$ is an {\it a priori} chosen azimuthal angle $\varphi$, where the temperature deviation sign changes. Rotation of the disc inhomogeneity around the star is controlled by variable phase, i.e. $\varphi\rightarrow\varphi+\Delta\varphi$. By estimating the size of the inhomogeneous disc area (contained between $R_{pert}^{inn}$ and $R_{pert}^{out}$) for a set of small or moderate $\Delta T$, it would be possible to obtain the observed amplitude in $V$ filter and colour index variations with respect to the $V$-filter synthetic magnitudes, i.e. consistent with all colour-magnitude diagrams.

\begin{figure*}

\centering

\includegraphics[width=0.48\linewidth]{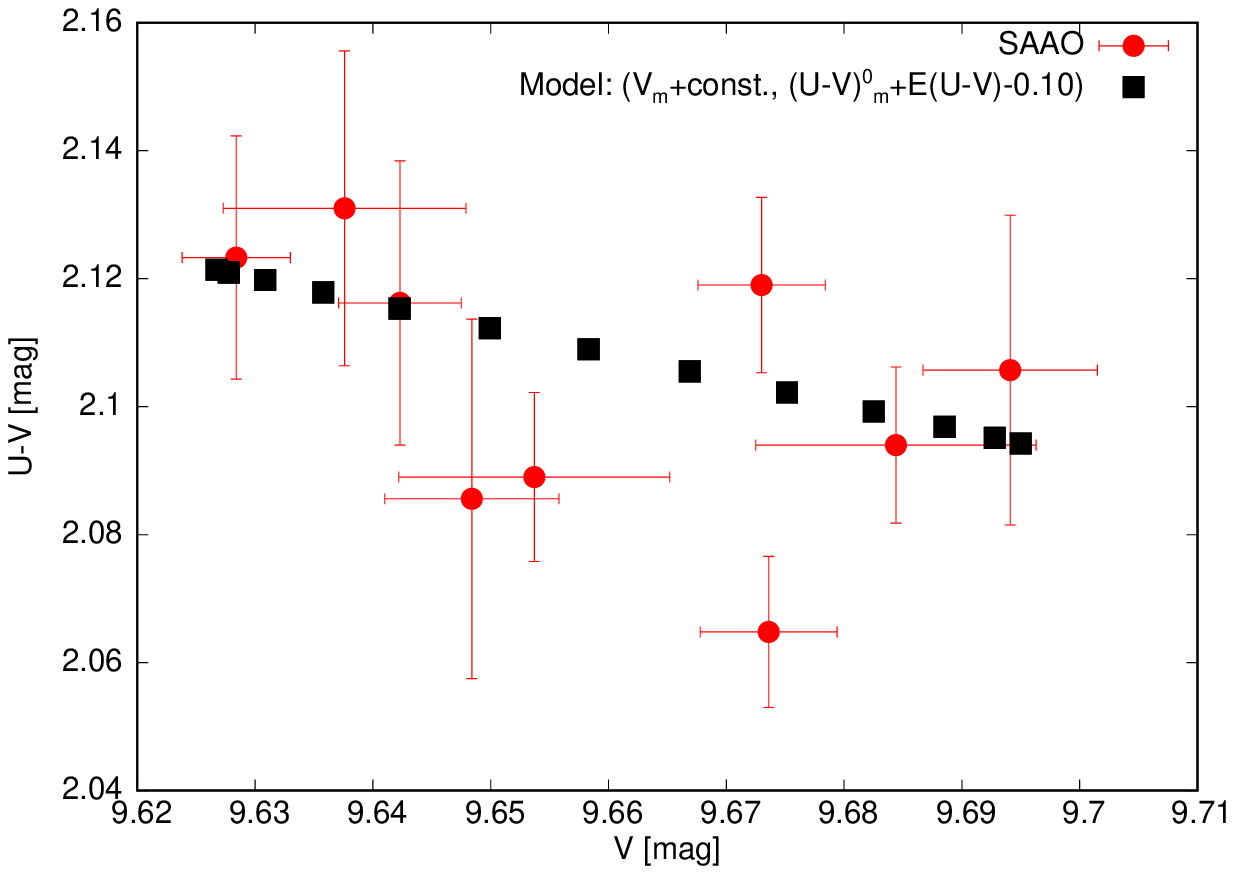}
\includegraphics[width=0.48\linewidth]{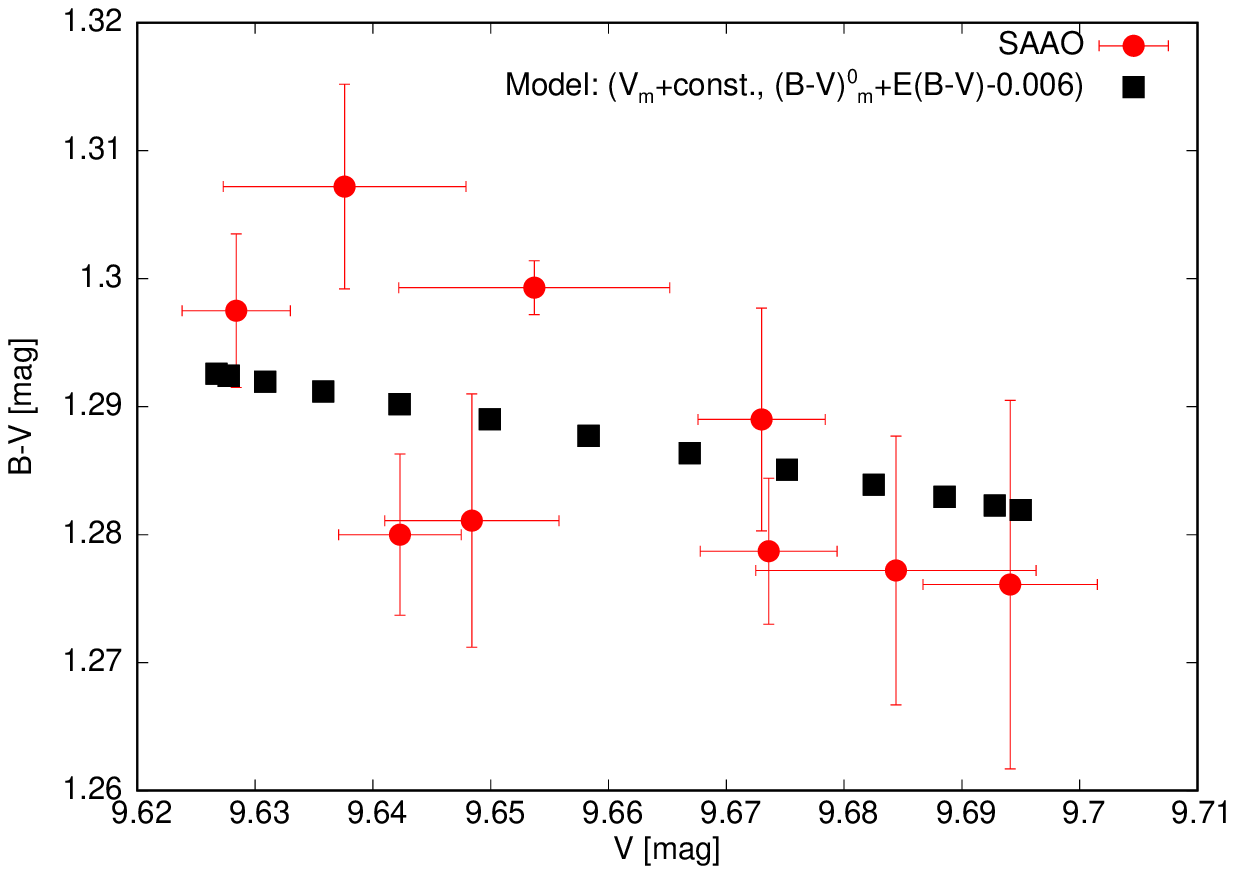}
\includegraphics[width=0.48\linewidth]{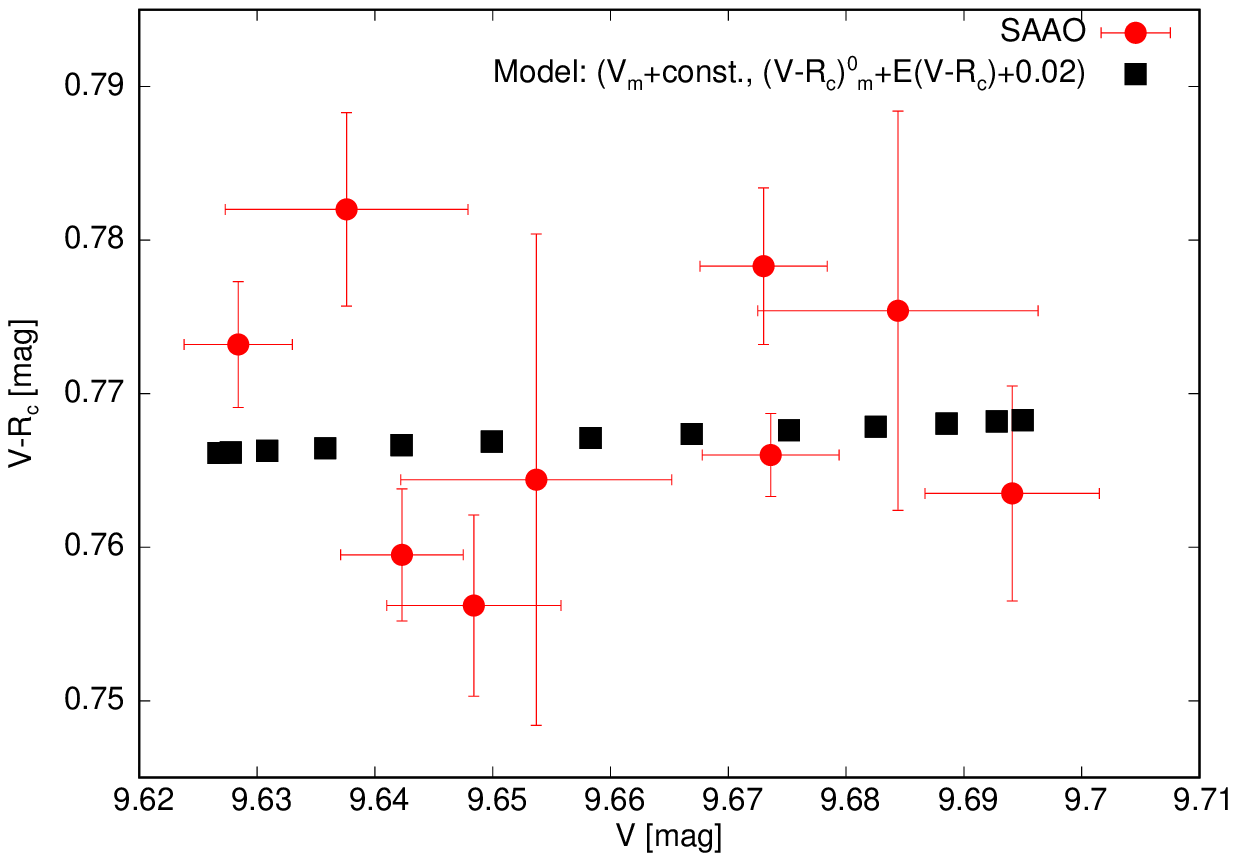}
\includegraphics[width=0.48\linewidth]{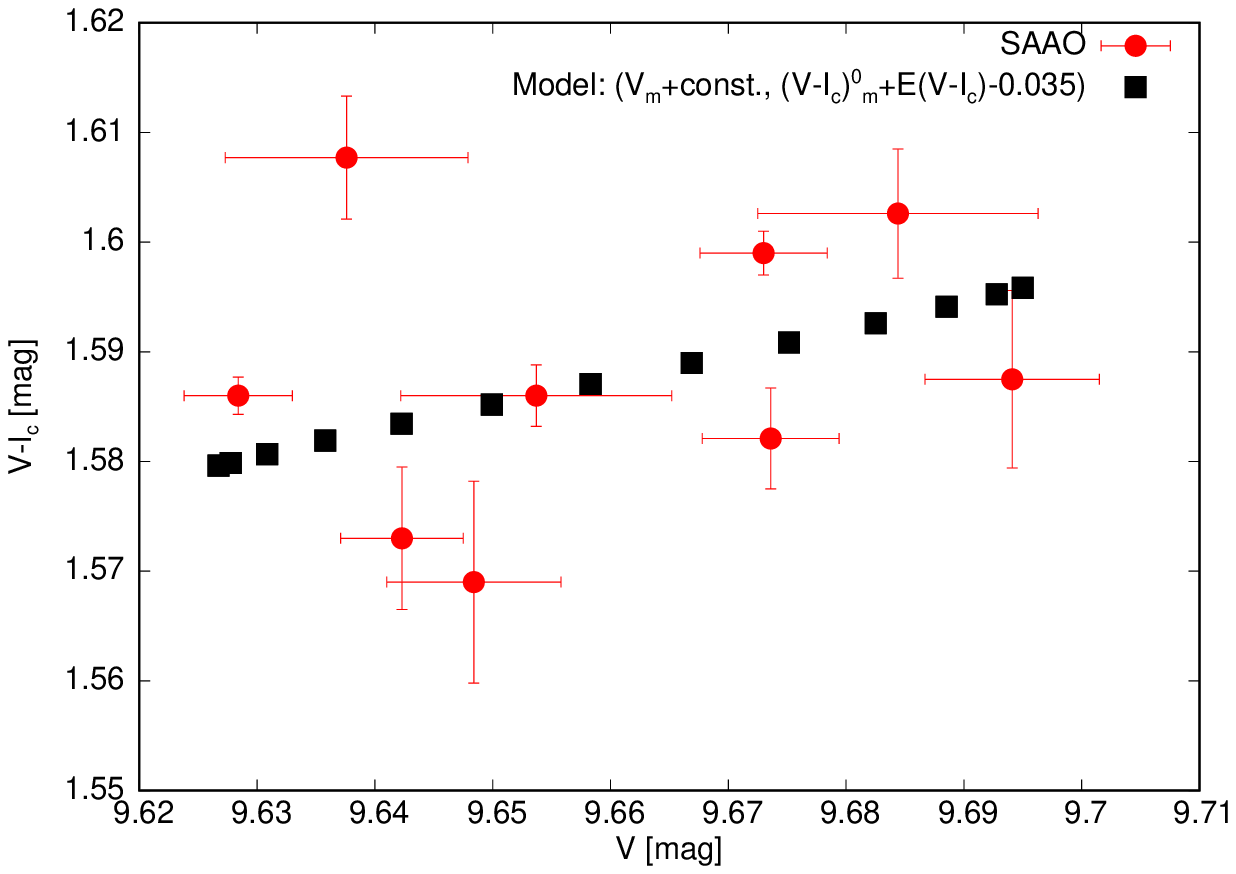}

\caption{Comparison of observed (circles, the {\it SAAO} data only) 
and synthetic (squares) colour-magnitude diagrams for {\it Segment~I}.}

\label{Fig.rez7}

\end{figure*}

\begin{figure*}

\centering

\includegraphics[width=0.48\linewidth]{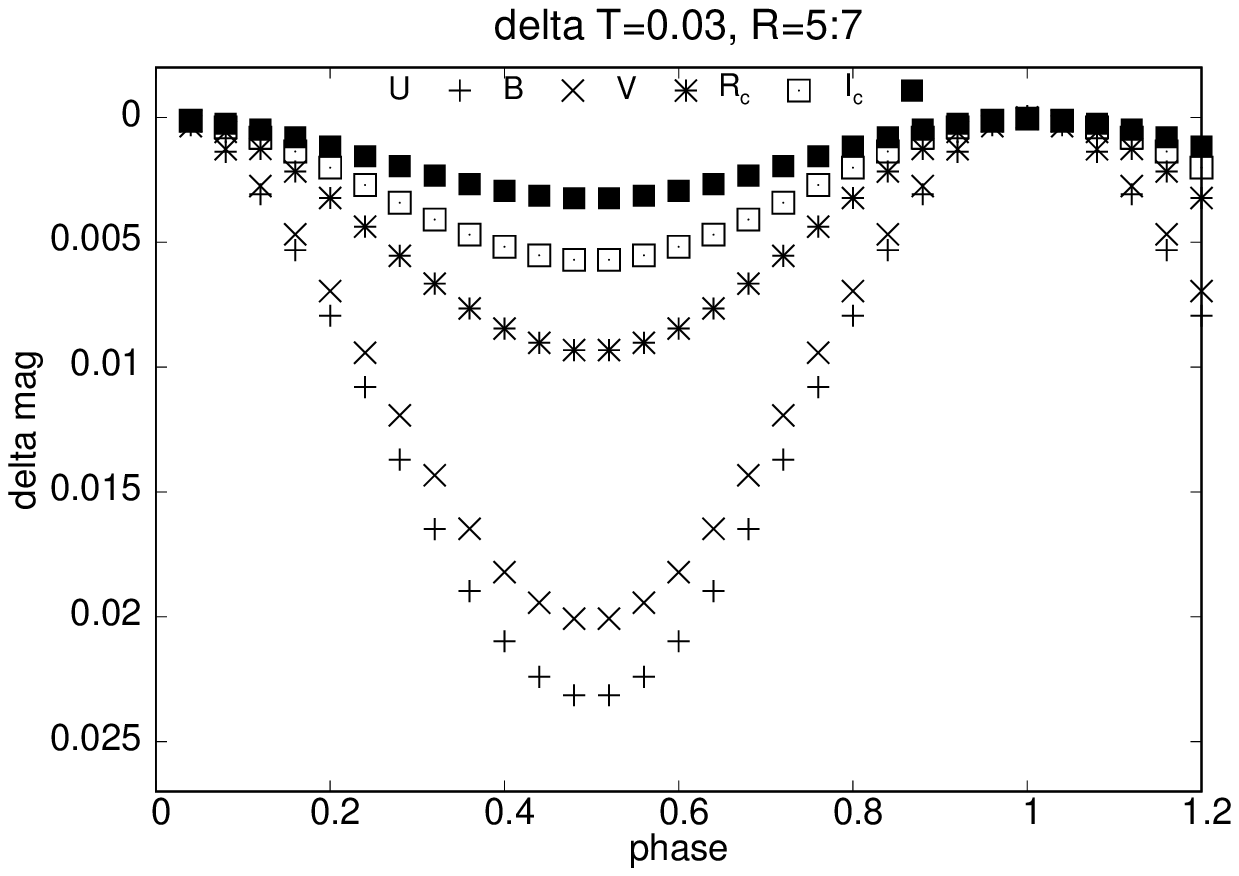}
\includegraphics[width=0.48\linewidth]{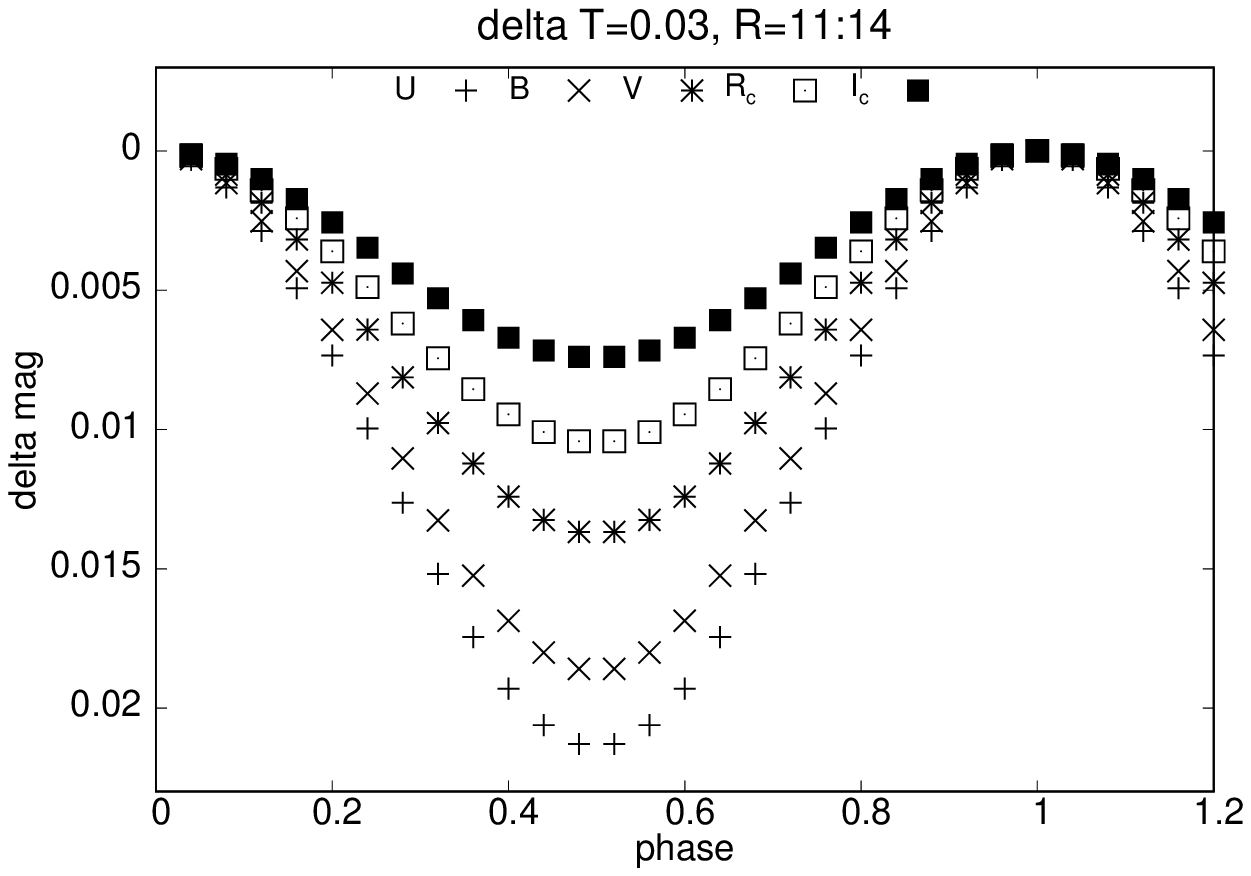}
\includegraphics[width=0.48\linewidth]{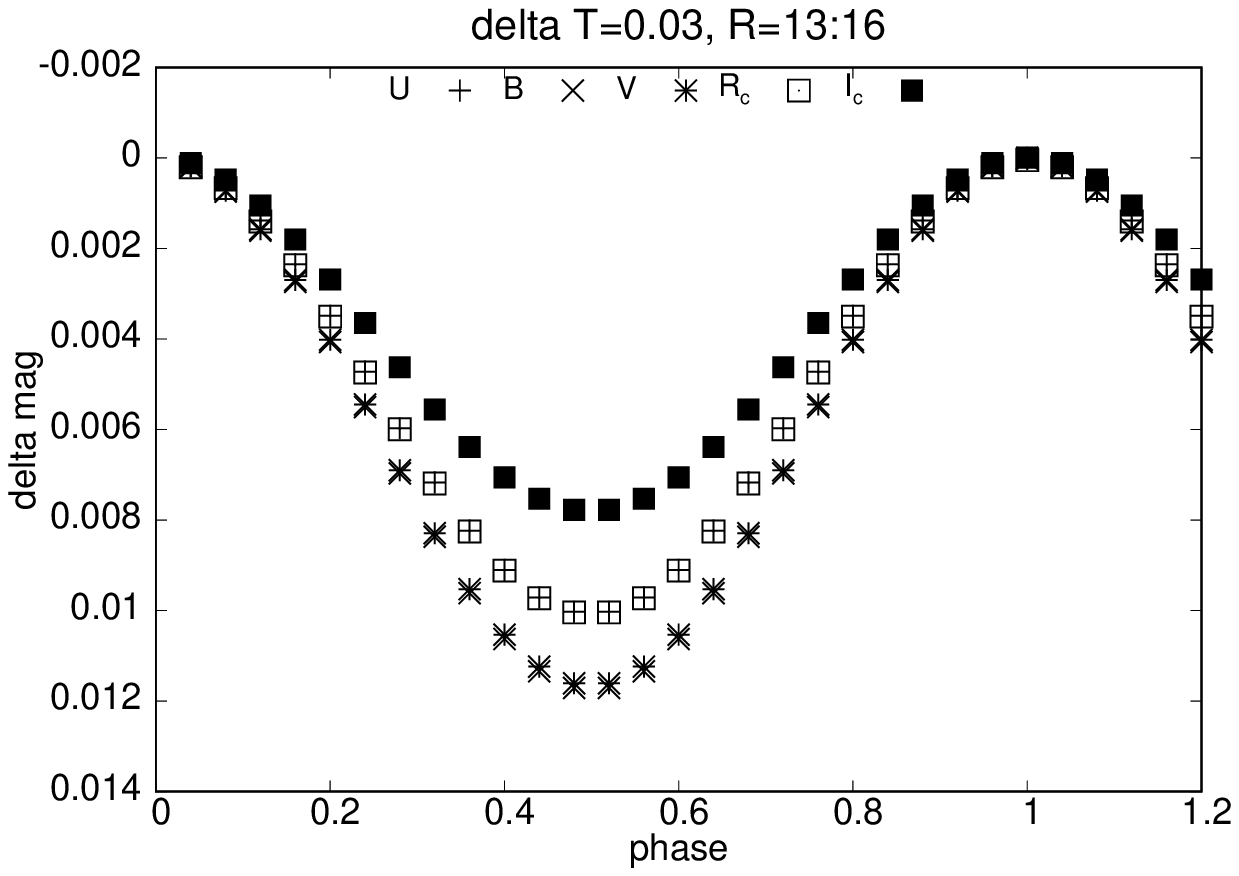} 
\includegraphics[width=0.48\linewidth]{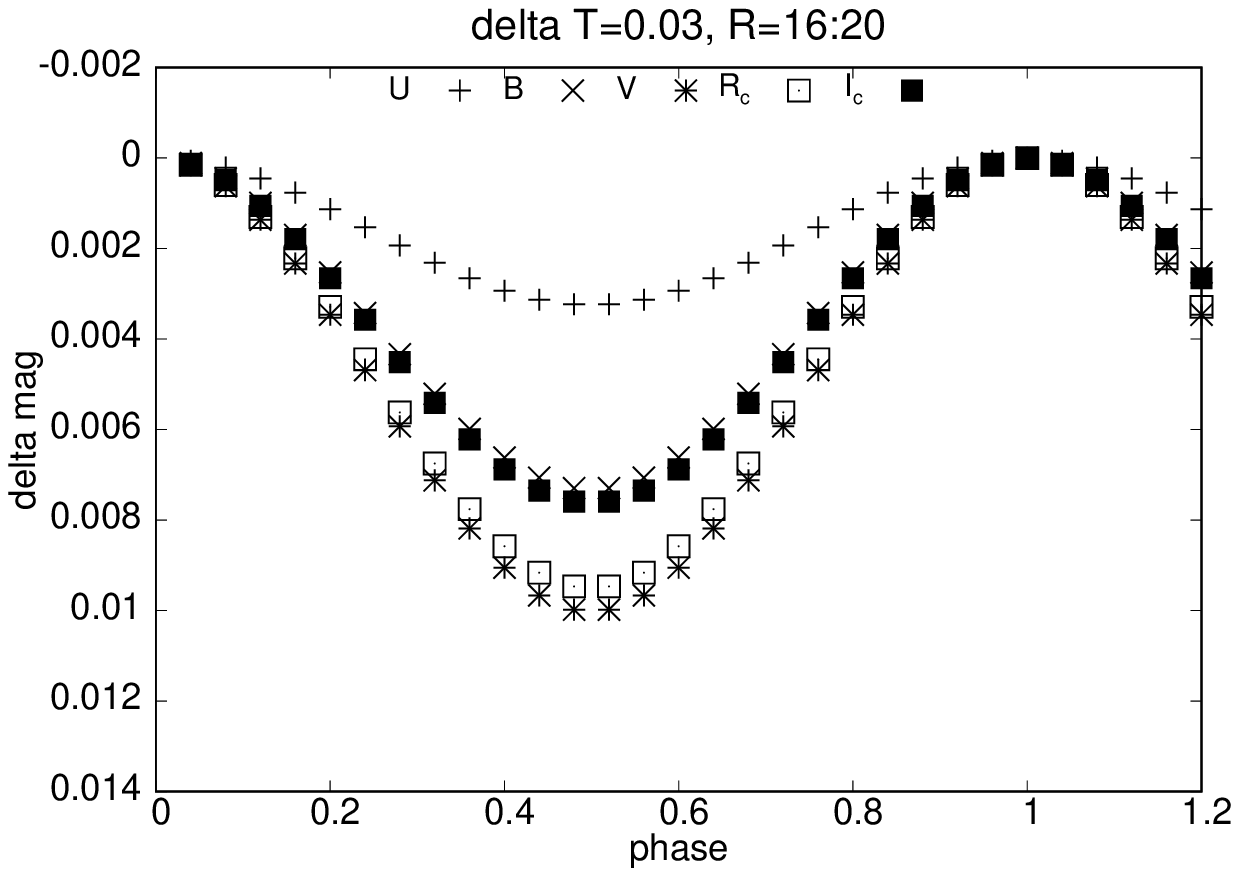}

\caption{Sample of synthetic light curves in $UBVR_cI_c$ filters used during 
the localisation process of {\it Segment~II} light variations. Most similar 
amplitudes in $BVR_cI_c$ filters are obtained for disc inhomogeneities located 
between 13-20~R$_{\sun}$, but the slightly higher amplitude in the $U$ filter may suggest 
somewhat closer localisation, from 12 to 15~R$_{\sun}$. 
The values used in a given model ($\Delta T$, $R_{pert}^{inn}$ 
and $R_{pert}^{out}$) are given at the top of each panel.}

\label{Fig.rez8}

\end{figure*}

We searched the parameter space manually with step of 0.05 in $\Delta T$ for {\it Segments~I} and {\it III}, 
and 0.01 for {\it Segment~II}. The same $\Delta T$ was always assumed for all filters. A step of $\Delta R=1$~R$_{\sun}$ was used to estimate $R_{pert}^{inn}$, $R_{pert}^{out}$ as well as an optimal width of the disc inhomogeneity $R_{pert}^{out}-R_{pert}^{inn}$. We obtained the following results for the three pre-defined light curve segments:

\begin{enumerate}

\item To reproduce the largest 0.07~mag variations and colour-magnitude diagrams for {\it Segment~I}, 
we found that effective temperatures in disc annuli 
between 16-20~R$_{\sun}$ must deviate by $\Delta T\approx0.2$. 

This distance is similar to the preliminary mid-inhomogeneity radius of 20~R$_{\sun}$, which was obtained using blackbody approximation instead of model atmospheres, as briefly stated in \citet{siwak17}. Degeneracy between $\Delta T$ and the position and size of the perturbed area did not turn out to be significant. An increase of the disc inhomogeneity size simultaneously with decrease of $\Delta T$ (and vice versa) results in colour-magnitude diagrams that do not match those observed. 
Similarly, attempts to set the inhomogeneous disc area either very close to the star (6-8~R$_{\sun}$) 
or at a greater distance (25-30~R$_{\sun}$) were also completely unsuccessful: they resulted in all positive 
or all negative values of slopes in synthetic colour-magnitude diagrams, respectively.\newline 
We present synthetic colour-magnitude diagrams and their comparison with the best-defined {\it SAAO} observations in Figure~\ref{Fig.rez7}. We stress that they also match well most of the {\it MSO}  and the combined {\it SAAO} and {\it MSO} colour-magnitude diagrams. We note that for this solution,  synthetic amplitudes of $V$-filter light variations 
are also almost identical to those observed. 
The numerical values of slopes of synthetic colour-magnitude diagrams shown in Figure~\ref{Fig.rez7} are as follows: $-0.40$ for $V-(U-V)$, $-0.16$ for $V-(B-V)$, $+0.03$ for $V-(V-R_c),$ and $+0.23$ for $V-(V-I_c)$.  The theoretical values of the colour indices are also similar to those observed. The discrepancies, i.e. constant shifts in colour indicies (e.g. -0.10~mag for the $V-(U-V)$ diagram), applied manually 
to match observations, are indicated in all four panels in Fig.~\ref{Fig.rez7}. 
These discrepancies result from imperfections of the model such as the choice of spectral intensities for ordinary supergiants, zero-point calibration and interstellar extinction estimate errors, and limited availability of the stellar models below 2300~K.

\item The similarity of amplitudes of {\it MOST} and ground-based $UBVR_cI_c$ light curves in {\it Segment~II} may also suggest their disc origin. Unfortunately, the lack of any trends in the colour-magnitude diagrams severely limits precision of localisation of the inhomogeneous plasma parcels by means of the light synthesis model. Therefore, we can rely on the coarse constancy of their amplitudes, as inferred from Figures~\ref{Fig.rez6}b-e. Even though most similar amplitudes in $BVR_cI_c$ filters are obtained from our model for disc inhomogeneities located between 13-20~R$_{\sun}$, the higher observed amplitude in $U$ filter (Fig.~\ref{Fig.rez6}a,f) may suggest a slightly closer location of between 12-15~R$_{\sun}$. 
The observed amplitudes were roughly reproduced by our model for $\Delta T=0.03$ (Figure~\ref{Fig.rez8}). Although the lack of precise $U$-filter data seriously limits precision of this estimate, the location of the inhomogeneity at the inner disc rim, as expected from the short 3-1.38~d period of this wave train, can be excluded within the disc model of \citet{zhu07}. Otherwise we would observe almost constant light in the $I_c$ filter, and the amplitude would gradually increase with decreasing effective wavelengths of the remaining filters, as suggested by the first panel 
in Fig.~\ref{Fig.rez8}.

\item The same conclusion as for {\it Segment~I} can also be true for {\it Segment~III}. This is due to similarity of trends observed in respective colour-magnitude diagrams for Johnson-Cousins filters, as shown by the $s$ values (Fig.~\ref{Fig.rez5}g,h,i). The large scatter of $U$-filter data makes it impossible to find any firm relation from the $V-(U-V)$ diagram (Fig.~\ref{Fig.rez4}a), but the slope appears to be negative on the auxiliary $\Delta y-\Delta(v-y)$ diagram (Fig.~\ref{Fig.rez4}c). Therefore we conclude that {\it Segment~III} light variations could arise somewhere at the distance of 14-19~R$_{\sun}$ from the star (for $\Delta T\approx0.15$). 

\end{enumerate}

We did not consider disc inclinations other than 55~deg and temperature distributions corresponding to a range of $M {\dot M}$ values. For instance, the variability shape observed in {\it Segments~I} and {\it III} could be fairly well reconstructed for disc inclinations closer to 70~deg, as derived for FU~Ori by \citet{gramajo14}; such extended computations may be meaningful when more precise multi-colour observations are available from future space telescopes.

The main conclusion of this model is not to pay too much attention to the exact numbers
obtained, but rather to point out that long-periodic (10-11~d) families do not arise either very close ($\sim 5-10$~R$_{\sun}$) or very far from the star ($\sim 30-40$~R$_{\sun}$), but near 15-20~R$_{\sun}$. More accurate modelling, including reliable three-dimensional approximation of disc inhomogeneity 
instead of its crude parameterisation by $\Delta T$ factor, will allow us to refine these results in the future. 
In this paper we assume that $\Delta T$ parameter automatically takes into account all phenomena 
related to the fact that real disc inhomogeneities probably have the form of waves or warps. 
Once illuminated by the inner disc, they may cast shadows 
on more distant parts of the disc.


\begin{figure*}

\centering

\includegraphics[width=0.33\linewidth]{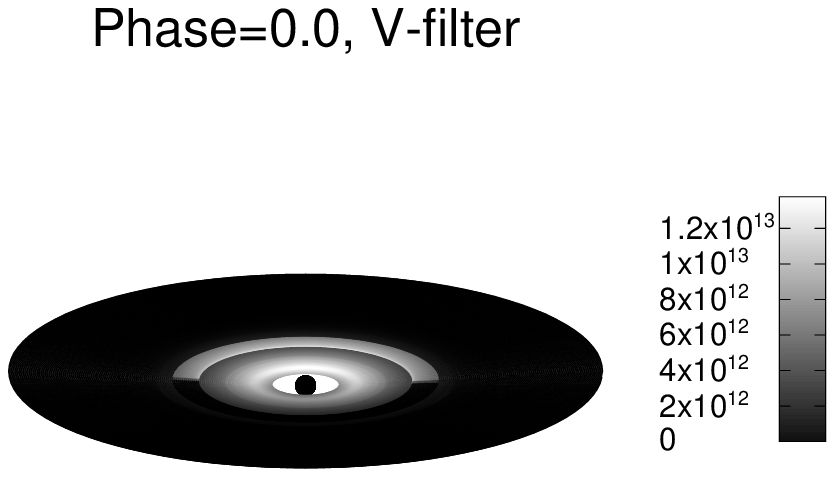}
\includegraphics[width=0.33\linewidth]{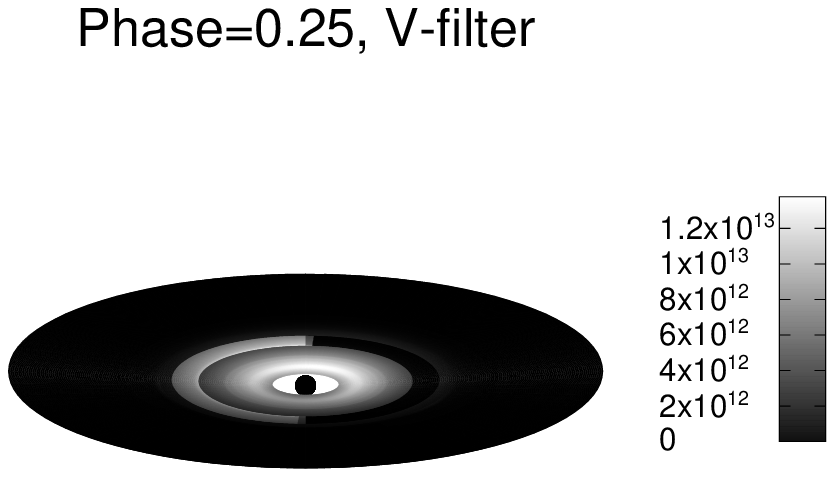}
\includegraphics[width=0.33\linewidth]{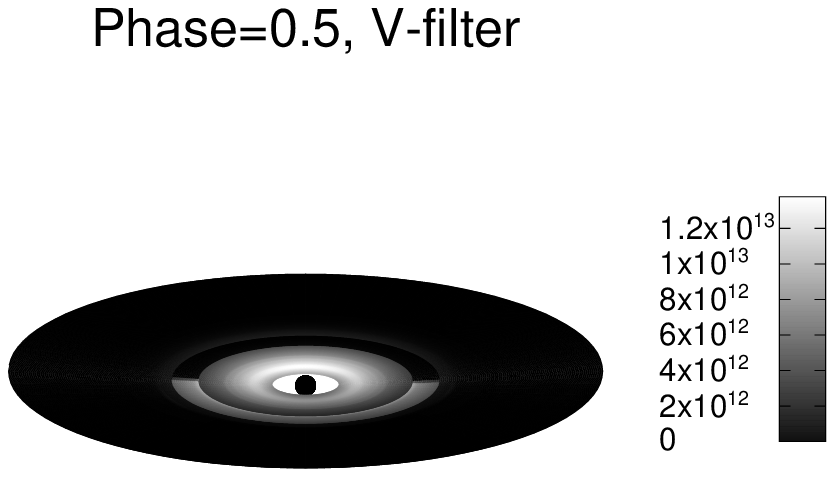}

\caption{Picture proposed for a qualitative explanation of the {\it Segment~I} variability. 
The lighter semi-ring represents the 16-20~R$_{\sun}$ disc inhomogeneity in three rotational phases.
The disc fluxes calculated for $V$ filter are expressed in greyscales (as defined on respective bars) 
and are left in temporary model units; they are also slightly affected by limitations of our plotting software. 
The stellar flux is expressed arbitrarily; the stellar radius was set to 1.5~R$_{\sun}$.}

\label{Fig.rez10}

\end{figure*}


\section{Discussion} 

\label{discussion}

The new 2013-2014 {\it MOST} light curve of FU~Ori was collected over a twice as long interval as the first, gathered in the 2010-2011 season. This enabled us to identify three families of light variations characterised by different quasi-periods and variability patterns: namely {\it Segments~I, II,} and {\it III}, as defined in Section~\ref{general_descr}. The colour-magnitude diagrams for respective segments constructed from the $uvyUBVR_cI_cRI$ data taken simultaneously from the ground were used to pinpoint mechanisms leading to the observed variability.

\subsection{Results for long-periodic light variations}
\label{disc_long}

The light variations observed in {\it Segment~I} and {\it Segment~III} are probably driven by the same mechanism, as inferred from analysis of their colour-magnitude diagrams (Fig.~\ref{Fig.rez5}, Fig.~\ref{Fig.rez7}). Using the disc and star light synthesis model, we ruled out the possibility that the QPOs seen in {\it Segment~I} could be due to changing visibility of accretion hotspots on the star (Sec.~\ref{spotmodel}). Instead, we argue that they could arise owing to revolution of the disc inhomogeneity located 
between 16-20~R$_{\sun}$ (see in Sec.~\ref{discmodel} and in Figure~\ref{Fig.rez10}). 
Our observations indicate that its lifetime does not exceed several revolutions around the star. We note that the average radius 18~R$_{\sun}$ of the 4~R$_{\sun}$ wide inhomogeneity obtained from the light synthesis model for the 10.75~d quasi-period is by 4.3~R$_{\sun}$ larger than the Keplerian radius of 13.7~R$_{\sun}$, expected for the stellar mass of 0.3~M$_{\sun}$. This small disagreement disappears if we assume 0.7~M$_{\sun}$ for the central star, in accordance with \citet{gramajo14}. Assuming that {\it Segment~III} was also due to Keplerian revolution, the stellar mass derived for average radius (16.5~R$_{\sun}$) of the inhomogeneity (14-19~R$_{\sun}$) would be smaller, i.e. about 0.45~M$_{\sun}$. Nevertheless, no one can consider this particular result as significant owing to the doubts concerning 
the quasi-periodic nature of these light variations, as described in point~3 of Section~\ref{general_descr}.

The formation mechanism of disc inhomogeneities at the distance of 0.05-0.1~AU responsible for the $\sim10-11$~d oscillations is not clear to us. These inhomogeneities could perhaps be related to a mechanism leading to FUor outbursts themselves. Recently, \citet{liu16} presented {\it Subaru Hi-CIAO} differential linear-polarisation imaging observations, showing large-scale asymmetrical structures around FU~Ori, Z~CMa, V1735~Cyg, and V1057~Cyg. Their result supports the gravitational instability of the disc as a mechanism creating spirals and clumps falling towards the star and leading to accumulation of mass near the inner disc and to enhanced accretion (see also in \citealt{vorobyov05, vorobyov06, vorobyov15}). Once the matter is slowly accumulated owing to gravitational instabilities, it may trigger thermal instabilities in the inner disc. This also activates magneto-rotational instabilities leading to the FUor-type outburst \citep{zhu09}. These authors calculated light curves showing small-amplitude quasi-periodic light variations after the outburst, which might be caused by convective eddies formed in the {\it high state-low state} transition region. The convection is especially strong and even penetrates the mid-plane of the disc in the regions of 0.15-0.35~AU, but is still present although confined only to the disc surface for smaller disc radii.\newline
It is not excluded that convection eddies may be responsible for the longer family of light variations observed 
by {\it MOST} and earlier from the ground, as mentioned in Section~\ref{intro}.
Assuming this scenario is correct, our {\it MSO} observations, lasting 194~days,  also had the potential 
to detect quasi-periods up to 65~days long and probe light variations occurring 
at 50-60~R$_{\sun}$ (0.25~AU).  Although periodograms presented in the first panel of Figure~\ref{Fig.rez2} show two wide peaks at 13-17 and 42-48~d, their significance is very low. If these longest variations really arise owing to convection eddies that are persistent for a few revolutions 
around the star, then their amplitudes seen in infrared filters should be larger than in visual bands.

The FUor phenomenon could also be initiated by tidal disruption of a few Jupiter mass planet, as obtained by \citet{lodato04} and \citet{nayakshin12}. These authors showed that before the disc gap is opened a planet can easily migrate to the distance of $\sim0.1$~AU from the star; this is similar to the inhomogeneity average radius estimated for {\it Segments~I} and {\it III}. The material from the planet may feed the disc through Roche lobe overflow, leading to an enhanced mass accretion and a major disc brightness increase. It is not excluded that tidal disruption of a planet may create disc inhomogeneities in close vicinity of the star. However, in these circumstances permanent rather than quasi-periodic light variations should be observed. We note that also \citet{powell12} considered this scenario to explain the persistent 3.6~d periodic modulation found in cross-correlation function profiles of FU~Ori; this periodicity is not visible as a constant feature in the {\it MOST} light curves.

The very interesting possibility is also offered by \citet{romanova13}, who found waves induced in the disc plasma structure from interactions with a rotating tilted magnetosphere of a star. The authors obtained two major types of solutions:
\begin{enumerate} 

\item The first is for typical CTTS, where a magnetospheric radius is similar to the co-rotation radius (see Sec.~3.2 of their paper).  In this case a strong warp is formed that rotates with the stellar frequency. This scenario can be applied to AA~Tau-like stars, periodically obscured by a warped disc \citep{bouvier99, bouvier07, mcginnis15},

\item The second solution is for the case in which the magnetospheric radius is much smaller than the co-rotation radius (see Sec.~3.3 of \citealt{romanova13}). Such a situation occurs in CTTS with enhanced mass accretion, when the disc plasma pressure compresses the magnetosphere. 
Our guess is that the latter solution could also apply to FU~Ori, as suggested by \citet{audard14}. Although \citet{romanova13} claimed in Sec.~4.1 of their paper that waves created in CTTS discs cannot be directly observed because of the large brightness contrast with the dominant star, it does not apply to FU~Ori, where the disc overwhelms the stellar luminosity by a hundred times\footnote{We also note that \citet{flaherty16} found a few dozen examples of such interactions in infrared {\it SPITZER} long-term observations of young stellar objects in Chamaleon~I star forming region.}.
The authors found {\it high-frequency inner bending waves,} i.e. inhomogeneities whose rotational frequency around the star is almost equal to, or slightly lower than, the Keplerian velocity of the inner disc. The waves originate only during periods of unstable accretion and are located at the inner edge of the disc. In addition, the authors also found {\it lower frequency waves} that propagate to larger distances and are sometimes enhanced at the disc co-rotation radius. The major warp, causing AA~Tau-type occultations, does not appear.

\end{enumerate}

If the latter solution is applicable to FU~Ori, then the 10-11~d light variations observed in {\it Segments I} and {\it III} could represent modulation in the visibility of the inhomogeneity caused by a locally enhanced {\it lower frequency wave} at (or near) the disc co-rotation radius and perhaps also the stellar rotational period. We note that stable over three seasons 14.8~d periodic modulation of P Cygni profiles 
by the disc wind found in FU~Ori spectra by \citet{herbig03}, and later confirmed (at 13.48~d) by \citet{powell12}, 
was proposed by the first authors to be the rotational period of the star. 
So far, our preliminary attempts to explain {\it Segment~I} and {\it III} variability 
by axially non-symmetrical dusty disc wind parameterised by means of single $(1-\Delta T)$ term 
in Equation~\ref{inhom} resulted in a non-physical solution; we obtained 
that the light emerging from the disc semi-ring between 11-16~R$_{\sun}$ must be almost completely absorbed. 
This solution maintains both the colour-period relation for the stellar mass of 0.3~M$_{\sun}$, 
observed amplitudes, and the same values of negative and positive slopes in consecutive 
colour-magnitude diagrams. However, such a mechanism would lead to significant modulations of the disc rotational profiles, 
which would certainly have been noticed by previous authors.
Coordinated space-based photometric and ground-based high-resolution spectroscopic observations may enable the study of possible relationships in the future. 

\subsection{Results for short-periodic light variations}

{\it Segment~II} of the light curve is composed of a short-period, sine-like wave train of much smaller ($\sim 0.01$~mag) amplitude. A period shortening of each successive oscillation is seen directly in the light curve. This wave train became visible at the end of {\it Segment~I} as the $\sim3.5$~d signal,  and ceased after 8-9~days, when its period shortened down to $1.38\pm0.04$~d. If this light variability is driven by the Keplerian motion of disc inhomogeneities drifting to the inner disc parts, as previously deduced from its continuously decreasing period \citep{siwak13}, then for the stellar mass of 0.3~M$_{\sun}$ we obtain the value of the inner disc radius of 3.5~R$_{\sun}$ or 4.6~R$_{\sun}$ for 0.7~M$_{\sun}$. These values are also in accord with interferometric observations of \citet{malbet05}, who obtained $5.5^{+2.9}_{-1.8}$~R$_{\sun}$, and with the value of the stellar radius of 3.6~R$_{\sun}$ derived by \citet{konigl11}.

The short-periodic sine-like variability pattern revealed by {\it MOST} in the 2010-2011 light curve, was re-analysed in Section~\ref{MOSTwav}. The conclusions are somewhat different from the preliminary findings by \citet{siwak13}. The lower values of periods obtained in Section~\ref{MOSTwav} may be used for refinement of the inner disc radius value obtained in \citet{siwak13}: if 2.1~d is the lower period limit then the change in inner disc radius value is small, from 4.8 to 4.5~R$_{\sun}$ for 0.3~M$_{\sun}$, or 6~R$_{\sun}$ for 0.7~M$_{\sun}$.  Assuming that the 1.08~d value was due to the revolution of a plasma parcel with a local Keplerian speed at the inner disc radius, the respective radii would be equal to 3~R$_{\sun}$ or 3.9~R$_{\sun}$.

Unexpectedly, the above interpretation regarding the origin of (at least) the 2013-2014 
short-periodic light variations was questioned by the disc  and star light synthesis model. We found that similarity of the amplitudes observed in $UBVR_cI_c$ filters (Fig.~\ref{Fig.rez6}a-e) can only be explained by the changing visibility of the disc inhomogeneities parameterised by $\Delta T=0.03-0.04$ and located between 12-15~R$_{\sun}$.
This is in strong conflict with the location predicted with the assumption of purely Keplerian motion of the plasma parcels in the innermost disc region ($\sim3-8$~R$_{\sun}$), as discussed above. To avoid this conflict, we attempted to explain these light variations by assuming that they 
are caused by modulations of the innermost disc flux. 
For example, we made an attempt to explain these variations by a dusty disc wind, approximated by means of a single term in Equation~\ref{inhom}, i.e. $(1 - \Delta T)\times T_{eff}(R)$, and moderate-to-large values of $\Delta T$;  
however this was also unsuccessful.\newline 
We conclude that these short-periodic light variations cannot be assigned to {\it high-frequency waves} 
(see in in point 2 of Section~\ref{disc_long}) assuming that the temperature distribution in the disc 
follows the model of \citet{zhu07}.
If  these QPOs really arise between 12-15~R$_{\sun}$,
then they cannot be driven by Keplerian revolution of the disc inhomogeneities. In these circumstances our model obviously should not be used to describe short-periodic oscillations. Maybe a clue to the real mechanism is hidden in the fact that light variations observed in {\it Segment~I} smoothly transfer into the {\it Segment~II} sine-like wave with decreasing amplitude and period. This may suggest that these two light curve segments were in fact physically linked. It is not excluded that the first ({\it Segment~I}) was due to the Keplerian revolution of a disc inhomogeneity around the star, while the second ({\it Segment~II}) was the signature of some hypothetical disc plasma oscillations, excited during dissipation of the previously dominating major  disc inhomogeneity.

We suppose, however, that a physically more consistent explanation can be offered by the assumption that the magnetospheric gap in FUors is not always devoid of visible light sources. The inclusion of this possibility would require proper modification of our light synthesis model in the future.\newline
As mentioned in Section~\ref{intro}, to explain the observed colour-magnitude relations in {\it UBVR} filters, \citet{kenyon2000} proposed that the light variations in FU~Ori mostly arise 
in the narrow zone between the radius, where the disc temperature reaches its maximum, 
and the stellar photosphere (i.e. at 1.1-1.2~$R^{\star}$ in their model units).
This was in accordance with their Monte Carlo computations indicating that random fluctuations 
of a characteristic timescale no longer than 1~d dominate in the light curve. However, our {\it MOST} observations do not necessarily confirm this view. Quasi-periods of 1-3~days are seen only during very limited time intervals and these variations appear to be time coherent. 
Moreover, the existence of a typical boundary layer zone in FU~Ori was later questioned by \citet{zhu07}. 
Instead, it turned out that in spite of enhanced mass transfer, FU~Ori may possess a small magnetosphere. 
Assuming that the observationally determined inner disc radius of 5~R$_{\sun}$ \citep{malbet05,zhu07} 
is equal to the magnetospheric radius $r_m$, the lower limit of $r_m/R^{\star}\approx1.4$ was derived 
by \citet{konigl11}. 
According to the authors this size is in accordance with the result of \cite{donati05}, 
who measured the poloidal component of the inner disc magnetic field at 1~kG. 
In these circumstances short-lived unstable accretion tongues rotating with the inner disc rotational frequency can be formed \citep{kulkarni08,kulkarni09, blinova16} 
and are expected to transfer disc plasma towards the star. 
If plasma carried in these tongues would be cooler by 1500-2000~K than the maximum disc temperature (6420~K), then changing visibility of these tongues could lead 
to the short-periodic, small-amplitude light variations of similar amplitudes in Johnson filters, 
as observed in {\it Segment~II} (Fig.~\ref{Fig.rez6}).

FU~Ori is not the only FUor, where short-periodic QPOs were observed. 
The shortest detected 1.28~d period in FU~Ori-type star V2493~Cyg was also 
attributed to Keplerian rotation of plasma parcels emerging at the disc magnetospheric radius by \citet{green13}. Similarly, about 1~d flux modulation due to the changing visibility of two antipodal accretion hotspots on the star was found in X-ray observations of the EXor/FUor star V1647~Ori \citep{hamaguchi12}. 
The hot X-ray component in FU~Ori is also variable (at 0.8~d) and viewed through heavy absorption from a disc wind or accretion stream \citep{skinner10}. These results appear to be in agreement with the result of \cite{blinova16}, who found that for small 
magnetospheres as in FU~Ori, an {\it ordered unstable regime} may create one or two tongues and related 
hotspots that are not fixed on the star, but rotate with the inner disc rotational frequency. 
This scenario would also explain period shortening observed by {\it MOST} in {\it Segment~II}, 
by assuming that the accretion rate inside a tongue increases, as predicted by \citet{kulkarni09}.

It is a matter of debate, whether about twice greater amplitude observed in $U$ filter 
during {\it Segment~II} definitely speaks for the so-called hotspot mechanism, 
at least occasionally operating in FU~Ori. 
Accurate flux-calibrated spectra obtained simultaneously with space-based photometric (and ideally also X-ray) observations may be helpful to catch signatures of these additional hot radiation sources at short wavelengths during future occurrences of $\sim 1-3$~d light variations.

\section{Summary}

\label{summary}

We observed FU~Ori simultaneously from space and the ground in winter 2013-2014 with the aim to determine the mechanisms leading to light variations discovered by {\it MOST} during the first run in 2010-2011. Comparison of ground-based and synthetic colour-magnitude diagrams specifically prepared for each of three distinct oscillatory patterns identified in the new light curve indicates that the longer, $\sim$10-11~d QPOs are most likely due to the changing visibility of disc inhomogeneities localised at a distance of about 16-20~R$_{\sun}$. These inhomogeneities could represent convection eddies in the disc \citep{zhu09} and/or {\it low-frequency waves} caused by interactions of a tilted stellar magnetosphere with the disc plasma and enhanced at the disc co-rotation radius \citep{romanova13}.

The local Keplerian periodicity in the middle ($18$~R$_{\sun}$) of the major inhomogeneity is 11~days if we assume a stellar mass of 0.7~M$_{\sun}$. This result is in reasonable agreement with the colour-period relationship claimed 
in Section~\ref{intro}.  However, no similar agreement was obtained for the short-periodic 3-1.38~d variability. According to our light synthesis model, this variability appears to arise somewhere between 12-15~R$_{\sun}$. The mechanism engaging Keplerian revolution of a disc inhomogeneity on a spiral orbit, however suggests a gradual approach of the inhomogeneous plasma parcel towards 
the star from 5.9 to 3.5~R$_{\sun}$ or from 7.8 to 4.6~R$_{\sun}$ assuming reasonable stellar masses of 0.3 or 0.7~M$_{\sun}$, respectively. This disagreement might be temporarily resolved by assuming that one or two unstable tongues, 
in which disc plasma of the temperature of about 4500-5000~K is transmitted towards the star, 
appear in the small magnetospheric gap at least for a short time. 
Our $U$-filter observations also indicate the possibility that these short-periodic 
light variations may also be driven by related hotspot(s), 
revolving on stellar surface with the local Keplerian velocity of related tongue(s).

Further accurate broadband simultaneous photometric and spectroscopic 
observations are needed to clarify the issues left with a question mark in this paper. With the end of the {\it MOST} satellite activity, the next possibility to observe 
FU~Ori should appear during the {\it TESS} mission. However, {\it TESS} will still provide single-band observations only. 
Unfortunately, the apertures of the {\it BRITE} satellites fleet \citep{weiss14}, which provides six-month-long, blue- and red-band light curves, are too small to provide data on FU~Ori. Two-colour, high-precision, space-based data may be provided by {\it UVSat} \citep{pigulski17}. This would eliminate the problems arising from the limited accuracy of ground-based observations in the u band, and would be very suitable for exploration of disc dynamics of the brightest FUors.

\section*{Acknowledgments}

\label{thanks}

This study was based on
(1)~data from the {\it MOST\/} satellite, a Canadian Space Agency mission jointly operated by Dynacon Inc., the University of Toronto Institut of Aerospace Studies, and the University of British Columbia, with the assistance of the University of Vienna; 
(2)~observations made at the Mount Suhora Astronomical Observatory, Cracow Pedagogical University; and
(3)~observations made at the South African Astronomical Observatory. 
This paper also made use of NASA's Astrophysics Data System (ADS) Bibliographic Services.\newline
MS, MW, MD, and WO are grateful to the Polish National Science Centre for the grant 2012/05/E/ST9/03915. GS is grateful for the Polish National Science Centre for the grant 2011/03/D/ST9/01808. 
Polish participation in SALT is funded by grant No. MNiSW DIR/WK/2016/07. 
The Natural Sciences and Engineering Research Council of Canada supports the research of DBG,JMM, AFJM, and SMR. Additional support for AFJM was provided by FQRNT (Qu{\'e}bec). CC was supported by the Canadian Space Agency. RK and WWW are supported by the Austrian Science Funding Agency (P22691-N16). MS acknowledges Dr. Francois van Wyk and the entire {\it {\it SAAO}} staff for their hospitality, as well as the observers, who obtained observations at the {\it MSO} during single nights, i.e. Dr. hab. Andrzej Baran, mgr. Micha{\l} {\.Z}ejmo, and Dr. Jan Janik.\newline
Special thanks are also due to an anonymous referee for highly useful suggestions and comments on the previous version of the paper.

\begin{appendix}
\section{Description of the model}
\label{model_app}

The disc geometry of the model is assumed as in Section~3 of \citet{zhu07}, i.e.\ the disc vertical height $H$ is a function of the disc radius $R$, measured from the central star, given by 
\begin{equation}
H(R)=H_0 \left(\frac{R}{R_{inn}}\right)^{\frac{9}{8}},
\end{equation}
where the thickness of the inner disc, presumably truncated by the stellar magnetosphere at $R_{inn}=5$~R$_{\sun}$, is assumed to be $H_0=0.1~R_{inn}$. The distribution of the disc effective temperature $T_{eff}$ as a function of disc  radius $R$ is assumed for the stationary accretion case \citep{pringle81}
\begin{equation}
\label{teff-r} 
T_{eff}^{4}(R)= \frac{3GM{\dot M}}{8{\pi}{\sigma}R^{3}} \left[1-\left(\frac{R_{inn}}{R}\right)^{\frac{1}{2}}\right],
\end{equation}
where $G$ is the universal gravitation constant, $M$ is the stellar mass, $\dot M$ is the mass accretion rate transferred from the disc onto the central star, and $\sigma$ is the Boltzmann constant. In accordance with \citet{zhu07}, we assume $M {\dot M} = 7.2\times10^{-5}$~M$_{\sun}^{2}$~yr$^{-1}$ and that the effective temperature of disc annuli located closer than $1.36~R_{inn}$ is equal to the maximum disc temperature value, calculated by 
\citet{zhu07} to be 6420~K.

The surfaces of consecutive disc annuli were approximated by surface areas of truncated cones. Subsequently, each disc annulus was divided into smaller elements, evenly distributed in azimuthal angle $\varphi$.  The flux emerging from each fine element was computed taking into account its effective surface area $dS=d{\varphi}dR$, which is a function of inclination, the orientation angle of a particular annulus with respect to the disc mid-plane, and the azimuthal angle ($\varphi$) of the centre of a surface element $dS$. In accord with previous works \citep{kenyon88, zhu07}, we assumed that atmospheres of consecutive disc annuli radiate in the same way as atmospheres of  supergiant stars with the effective temperatures given by Equation~3. Hence, for the flux calculations we used the emerging intensities (a full radiation field) of supergiant stars from the {\it PHOENIX} library \citep{husser13}.The theoretical intensities $I_{\lambda}$ in this library are calculated for 78 values of $\mu = cos \gamma$, where $\gamma$ is the angle of view between the normal vector of the infinitely small element of the photosphere and the observer. The intensities are calculated with a resolution of 1 \AA~in wavelength and 100~K in effective temperature, starting from 500 \AA~and 2300~K, respectively. Special attention was given to the choice of intensities with a proper $\log g$ (in cgs units), which changed from 1.5 (6500-5300~K) through 1.0 (5200-4200~K), 0.5 (4100-3700~K), 0.0 (3600-3200~K), to -0.5 (3100-2300~K). Because of the lack of proper intensities for $\log g=-0.5$ for the last temperature range in the library, we were forced to extrapolate the intensities for $\log g=0.0$ using coefficients estimated by comparison of intensities calculated for $\log g=-0.5$ and 0.0 in the range 3200-3600~K; luckily the above operation has negligible significance for the final result. To obtain emergent intensities $I_{\lambda}(\mu)$ for each disc element $dS$, we interpolated the library intensities in $\mu$ and $T_{eff}$ to account for the full range of visibility angles $\gamma = 46.5-63.6$~deg and temperatures $T_{eff} = 2300-6420$~K appropriate for our case. Finally we calculated the flux $F_{\lambda}^{d}$ emitted in a given direction from the disc by integration,
\begin{equation}
F_{\lambda}^{d}=\int_{S}{I_{\lambda}(\mu)}{dS}=\int_{R}\int_{\varphi}{I_{\lambda}(\mu)}{d\varphi dR},
\end{equation}
where $0 \leqslant \varphi < 2\pi$, $5 \leqslant R \leqslant 44.6$~R$_{\sun}$. The upper limit of 44.6~R$_{\sun}$ is imposed by the lack of theoretical models for temperatures lower than 2300~K in the {\it PHOENIX} library. This is fortunately not a serious limitation of our results, as the contribution of disc light emerging beyond this radius should be small. This amounted to only $2\%$ for the $I$ filter in the very first FU~Ori disc model of \cite{kenyon88} and was practically equal to zero for other Johnson filters.

For the observed disc inclination of 55~deg and the inner disc radius of 5~R$_{\sun}$ \citep{malbet05,zhu07}, the star can be slightly obscured by the front disc rim only for $R^{\star}$ larger than 2~R$_{\sun}$. 
In spite that \citet{konigl11} proposed $R^{\star}=3.6$~R$_{\sun}$, we considered a range of smaller values between 1.5-2~R$_{\sun}$ and effective temperatures T$_{eff}^{\star}$ of 3500-4000~K, which are more 
typical of CTTS. The results turned out to be very weakly dependent on the parameters within the above ranges, and we finally chose the temperature of 4000~K and  radius of 2~R$_{\sun}$ for the central star in further calculations.To integrate the star flux ($F_{\lambda}^{\star}$), we used $I_{\lambda}(\mu=1)$, calculated for solar metallicity $[Fe/H]=0$ and $\log\,g=4.0$ from the respective {\it PHOENIX} model and then we applied the linear limb-darkening law using the coefficients from the \citet{diaz-cordoves95} and the \citet{claret95} tables for {\it UBVRI} filters.

The emitted disc and star summed fluxes $F_{ft}$ in individual photometric pass bands $ft\in\{U,B,V,R_c,I_c\}$ were determined using the transmission function profiles $\tau_{\lambda_{ft}}$ from \citet{bessel90}, i.e.
\begin{equation}
F_{ft}=\frac{\int_{\lambda}{(F_{\lambda}^{d}+F_{\lambda}^{\star})}{\tau_{\lambda_{ft}}}{d\lambda}}{\int_{\lambda}{\tau_{\lambda_{ft}}}{d\lambda}},
\end{equation}
where $\lambda$ varied within the wavelength range appropriate for the respective bandpass $\tau_{\lambda_{ft}}$.The fluxes $F_{ft}$ were later transformed to a magnitude scale. The published $\tau_{\lambda_{ft}}$ profiles may not accurately represent the actual {\it MSO} $RI$-filter transmissions but are sufficient for our qualitative analyses.

\section{Model calibration and validation}
\label{valid}

Before we can carry out a serious application of this model we first need to calibrate and validate it. First, we estimated the zero points necessary for proper calculation of colour indices. For this purpose we used {\it PHOENIX}'s emerging intensity of a main sequence A0V star (9600~K, $\log g = 4.0$, $\mu=1$) and the linear limb-darkening law. Obviously, the disc light was ignored during these calculations. The synthetic fluxes of the star in Johnson-Cousins filters were then properly added to meet the well-known requirement that colour indices of an A0V unreddened star are equal to almost zero.

In the second step, we made an estimate of the interstellar extinction to check whether the results obtained with zero points calculated above are in compliance with the historical results. The unreddened colour indices given by our model
are as follows:\newline
$(U-V)_m^0\approx1.21$, $(B-V)_m^0\approx0.72$, $(V-R_c)_m^0\approx0.43$, and $(V-I_c)_m^0\approx0.93$.\newline
Using the \citet{cardelli89} equations~1, 3a, and 3b for the mean $R_V$-dependent extinction law, 
and assuming $E(B-V)\approx0.57$, estimated on the basis of results obtained from our model and our observations ($B-V\approx1.29$), we obtained $A_V=1.77$ for $R_V=3.1$. Integration of the stellar extinction curve over $\tau_{\lambda_{ft}}$ in the respective wavelength ranges leads to the following values:\newline $E(U-V)\approx1.00$, $E(V-R_c)\approx0.34$, $E(V-I_c)\approx0.73$.\newline 
This in turn leads to the following, model-dependent colour indices:\newline $(U-V)_m\approx2.20$, $(B-V)_m\approx1.29$, $(V-R_c)_m\approx0.74$ and $(V-I_c)_m\approx1.66$~mag.\newline
 All values are in good agreement with the mean values calculated from all multi-colour observations obtained during the {\it MOST} run, i.e. 2.121, 1.291, 0.770, and 1.594~mag, respectively (see also in Tab.~\ref{Tab.comp} for one particular night). The value $A_V\approx1.8$ obtained in this work agrees well with estimates of \citet{zhu07} at $A_V=1.5^{+0.7}_{-0.2}$ and \citet{pueyo12} at $A_V=2.0^{+0.5}_{-0.5}$.

\begin{figure}

\includegraphics[width=1\linewidth]{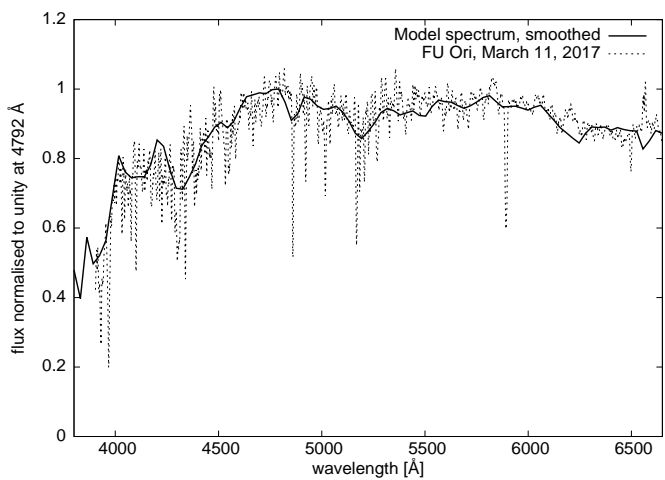}

\caption{Comparison of the smoothed model spectrum (the continuous line) and the flux-calibrated 
and de-reddened FU~Ori spectrum (the broken line) obtained on March 11, 2017, three years after 
our photometric campaign.}

\label{Fig.rez11}

\end{figure}

In Figure~\ref{Fig.rez11} we show the final test of our model. The observed (Sec.~\ref{spec-saao}) and model-synthesised FU~Ori spectra were normalised to unity at the maximum, which for both spectra appear to exist near 4792~\AA. Because of the large number of spectral lines in our model spectrum we decided to smooth it for easier comparison. The same $E(B-V)$ value as above was assumed during the de-reddening process of the real FU~Ori spectrum within the {\small \sc IRAF} task {\it deredden}. Similarity of both spectra appears to be satisfactory and it practically does not depend on a particular FU~Ori spectrum nor the standard star used for the flux calibration. However, one can note systematically lower observed flux in the region of ultraviolet and blue wavelengths. The observed flux appears to be slightly higher at red wavelengths. This suggests that the maximum disc temperature is currently slightly lower than one or two decades ago, when \citet{zhu07} et al. performed their observations and should perhaps be updated. This is not an unexpected finding as the disc brightness, as it is now 81 years after the outburst, is constantly decreasing. Nevertheless, we decided to refrain from this task as this small discrepancy may also be caused by systematic errors in our flux calibration procedure.


One of our model imperfections 
is the inability to reproduce the shapes of the oscillations as observed in {\it Segments I} and {\it III} ; only reproduction of sine-like wave trains as in {\it Segment~II} is possible. This is likely due to simplified assumptions about the disc inhomogeneity structure. Additional weakness comes with the assumption that stellar atmosphere models closely match properties of the disc atmosphere as well as with the lack of $I_{\lambda}$ for temperatures lower than 2300~K, which slightly influences our results obtained from colour-magnitude diagrams utilising near-infrared filters. We have also omitted the rotational broadening of the spectrum due to the Keplerian rotation of the disc as this effect is negligible for spectrum synthesis results in broadband filters.
\end{appendix}


\begin{thebibliography}{}

\bibitem[Audard et al.(2014)]{audard14}   Audard, M., Abraham, P., Dunham, M. M., Green, J. D., Grosso, N., et al., 2014,   Protostars and Planets VI, edited by Henrik Beuther, Ralf S. Klessen, Cornelis P. Dullemond,   and Thomas Henning, University of Arizona Press, Tucson, 914 pp., p.387-410



\bibitem[Avenhaus et al.(2018)]{avenhaus18}  Avenhaus, H., Quanz, S. P., Garufi, A., Perez, S., Casassus, S., et al., 2018, arXiv:1803.10882

\bibitem[Benisty et al.(2015)]{benisty15}    Benisty, M., Juhasz, A., Boccaletti, A., Avenhaus, H., Milli, J., et al., 2015, A\&A, 578, 6   



\bibitem[Bessel(1990)]{bessel90}     
Bessel, M. S., 1990, PASP, 102, 1181 

    

\bibitem[Blinova et al.(2016)]{blinova16}    Blinova, A. A., Romanova, M. M., \& Lovelace R. V. E., 2016, MNRAS, 459, 2354    



\bibitem[Bouvier et al.(1999)]{bouvier99}    Bouvier, J., Chelli, A., Allain, S., Carrasco, L., Costero, R., et al., 1999, A\&A, 349, 619



\bibitem[Bouvier et al.(2007)]{bouvier07}    Bouvier, J., Alencar, S. H. P., Boutelier, T., Dougados, C., \& Balog, Z., 2007, A\&A, 463, 1017



\bibitem[Cardelli et al.(1989)]{cardelli89}   Cardelli, J. A., Clayton, G. C., \& Mathis, J. S., 1989, AJ, 345, 245



\bibitem[Claret et al.(1995)]{claret95}   Claret, A., Diaz-Cordoves, J., \& Gimenez, A., 1995, A\&AS, 114, 247



\bibitem[Clarke et al.(2005)]{clarke05}   Clarke, C., Lodato, G., Melnikov, S. Y., \& Ibrahimov, M. A., 2005, MNRAS, 361, 942

   

\bibitem[Crause et al.(2016)]{crause16}   Crause, L. A., Carter, D., Daniels, A., Evans, G., Fourie, P., et al., 2016, SPIE, 9908E, 27 



\bibitem[Diaz-Cordoves  et al.(1995)]{diaz-cordoves95}    Diaz-Cordoves, J., Claret, A., \& Gimenez A., 1995, A\&AS, 110, 329



\bibitem[Dodin(2018)]{dodin18}   Dodin, A., 2018, MNRAS, 475, 4367



\bibitem[Donati et al.(2005)]{donati05}   Donati, J.-F., Paletou, F., Bouvier, J., \& Ferreira, J., 2005, Nature, 438, 466



\bibitem[Flaherty et al.(2016)]{flaherty16}   Flaherty, K. M., DeMarchi, L., Muzerolle, J., Balog, Z., Herbst, W., Megeath, S. T., Furlan, E., \&    Gutermuth, R., 2016, ApJ, 883, 104



\bibitem[Follette et al.(2017)]{follette17}   Follette, K. B., Rameau, J., Dong, R., Pueyo, L., Close, L. M., et al., 2017, AJ, 153, 264

   

\bibitem[Fukugita et al.(1996)]{fukugita96}   Fukugita, M., Ichikawa, T., Gunn, J. E., Doi, M., Shimasaku, K., \& Schneider, D. P.,    1996, AJ, 111, 1748



\bibitem[Gramajo et al.(2014)]{gramajo14}     Gramajo, L. V., Rod{\'o}n, J. A., \& G{\'o}mez, M., 2014, AJ, 147, 140



\bibitem[Green et al.(2013)]{green13}     Green, J. D., Robertson, P., Baek, G., Pooley, D., Pak, S., et al., 2013, AJ, 764, 22



\bibitem[Hamaguchi et al.(2012)]{hamaguchi12}    Hamaguchi, K., Grosso, N., Kastner, J. H., Weintraub, D. A., Richmond, M., Petre, R.,      Teets, W. K., \& Principe D., 2012, ApJ, 754, 32 



\bibitem[Hartmann \& Kenyon(1985)]{hartmann85}    Hartmann, L., \& Kenyon, S. J., 1985, ApJ, 299, 462 



\bibitem[Hartmann \& Kenyon(1996)]{hartmann96}    Hartmann, L., \& Kenyon, S. J., 1996, ARA\&A, 34, 207



\bibitem[Hartmann(1998)]{hartmann98}    Hartmann, L., Accretion processes in Star Formation, 1998, Cambridge, 34, 207

 

\bibitem[Herbig(1977)]{herbig77}    Herbig, G. H., 1977, ApJ, 217, 693



\bibitem[Herbig et al.(2003)]{herbig03}    Herbig, G. H., Petrov, P. P., \& Duemmler, R., 2003, ApJ, 595, 384



\bibitem[Husser et al.(2013)]{husser13}    Husser, T.-O., Wende-von Berg, S., Dreizler, S., Homeier, D., Reiners, A., Barman, T., \&    Hauschildt, P. H., 2013, A\&A, 553, A6 



\bibitem[Ibragimov(1993)]{ibragimov93}     Ibragimov, M. A., 1993, Ap, 35, 257 



\bibitem[Kenyon et al.(1988)]{kenyon88}    Kenyon, S. J., Hartmann, L., \& Hewett, R., 1988, ApJ, 325, 231



\bibitem[Kenyon et al.(2000)]{kenyon2000}    Kenyon, S. J., Kolotilov, E. A., Ibragimov, M. A., \& Mattei, J. A.,     2000, ApJ, 531, 1028



\bibitem[Kolotilov \& Petrov(1985)]{kolotilov85}    Kolotilov, E. A., \& Petrov, P. P., 1985, Sov. Astron. Lett., 11, 385 


\bibitem[K{\"o}nigl et al.(2011)]{konigl11}    K{\"o}nigl, A., Romanova, M. M., \& Lovelace, R. V. E., 2011, MNRAS, 416, 757


\bibitem[Kulkarni \& Romanova(2008)]{kulkarni08}       Kulkarni, A. K., \& Romanova, M. M., 2008, MNRAS, 386, 673 



\bibitem[Kulkarni \& Romanova(2009)]{kulkarni09}       Kulkarni, A. K., \& Romanova, M. M., 2009, MNRAS, 398, 701 



\bibitem[Kulkarni \& Romanova(2013)]{kulkarni13}       Kulkarni, A. K., \& Romanova, M. M., 2013, MNRAS, 433, 3048

     

\bibitem[Liu et al.(2016)]{liu16}    Liu, H. B., Takami, M., Kudo, T., Hashimoto, J., Dong, R., et al., 2016, Science Advances, vol. 2, no. 2, e1500875  

    

\bibitem[Lodato \& Clarke(2004)]{lodato04}        Lodato, G., \& Clarke, C. J., 2004, MNRAS, 353, 841

    

\bibitem[Luybarskii(1997)]{luybarskii1997}        Luybarskii, Y. E., 1997, MNRAS, 292, 679    

 

\bibitem[Matthews et al.(2004)]{M2004}    Matthews, J. M., Kusching, R., Guenther, D. B., Walker, G. A. H.,    Moffat, A. F. J., Rucinski, S. M., Sasselov, D., \& Weiss, W. W.,     2004, Nature, 430, 51



\bibitem[Malbet et al.(2005)]{malbet05}    Malbet, F., Lachaume, R., Berger, J.-P., Colavita, M. M., Folco, E.Di,     et al., 2005, A\&A, 437, 627



\bibitem[McGinnis et al.(2015)]{mcginnis15}     McGinnis, P. T., Alencar, S. H. P., Guimaraes, M. M., Sousa, A. P., Stauffer, J., et al., 2015, A\&A, 577, A11

   

\bibitem[Menzies et al.(1989)]{menzies89}   Menzies, J. W., Cousins, A. W. J., Banfield, R. M., \& Laing, J. D., 1989, {\it SAAO} Circulars, 13, 1-13



\bibitem[Motl(2011)]{Motl11}   Motl, D., 2011, http://c-munipack.sourceforge.net



\bibitem[Nayakshin \& Lodato(2012)]{nayakshin12}    Nayakshin, S., \& Lodato, G., 2012, MNRAS, 426, 70



\bibitem[P{\'e}rez et al.(2016)]{perez16}        P{\'e}rez, L. M., Carpenter, J. M., Andrews, S. M., Ricci, L., Isella, A., et al., 2016, Sci, 353, 1519



\bibitem[Pigulski et al.(2017)]{pigulski17}                 Pigulski, A., Baran, A., Bzowski, M., Cugier, H., Czerny, B., et al., 2017,     Proceedings of the PAS (Proc. of the 2nd BRITE Science conference, Innsbruck), vol. 5, p. 76

    

\bibitem[Powell et al.(2012)]{powell12}        Powell, S. L., Irwin, M., Bouvier, J., \& Clarke, C. J.,2012, MNRAS, 426, 3315

    

\bibitem[Pueyo et al.(2012)]{pueyo12}         Pueyo, L., Hillenbrand, L., Vasisht, G., Oppenheimer, B. R., Monnier, J. D., et al., 2012, ApJ, 757, 57 

    

\bibitem[Pringle(1981)]{pringle81}        Pringle, J. E., 1981, ARA\&A, 19, 137     



\bibitem[Romanova et al.(2004)]{romanowa04}     Romanova, M. M., Ustyugova, G. V., Koldoba, A. V., \& Lovelace, R. V. E., 2004, ApJ, 610, 920



\bibitem[Romanova et al.(2013)]{romanova13}        Romanova, M. M., Ustyugova, G. V., Koldoba, A. V., \& Lovelace, R. V. E., 2013, MNRAS, 430, 699



\bibitem[Rowe et al.(2006)]{rowe06}    Rowe, J.F., Matthews, J.M., Seager, S., et al., 2006, ApJ, 646, 1241



\bibitem[Rucinski et al.(2008)]{ruc08}          Rucinski, S. M., Matthews, J. M., Kuschnig, R., Pojmanski, G.,         Rowe, J., et al., 2008 MNRAS, 391, 1913

    

\bibitem[Sicilia-Aguilar et al.(2015)]{sicilia-aguilar15}        Sicilia-Aguilar, A., Fang, M., Roccatagliata, V., Collier Cameron, A., K{\'o}sp{\'a}l, {\'A}., Henning, T.,     {\'A}brah{\'a}m, P., \& Sipos, N., 2015, A\&A, 580, 82



\bibitem[Sipos et al.(2009)]{sipos09}        Sipos, N., {\'A}brah{\'a}m, P., Acosta-Pulido, J., Juh{\'a}sz, A., K{\'o}sp{\'a}l, {\'A}., Kun, M.,     Mo{\'o}r, A., \& Setiawan, J., 2009, A\&A, 507, 881



\bibitem[Siwak et al.(2011)]{siwak11}    Siwak, M., Rucinski, S. M., Matthews, J. M., Pojmanski, G., Kuschnig, R., et al., 2011, MNRAS, 410, 2725



\bibitem[Siwak et al.(2013)]{siwak13}    Siwak, M., Rucinski, S. M., Matthews, J. M., Kuschnig, R., Guenther, D. B., et al., 2013, MNRAS, 432, 194



\bibitem[Siwak et al.(2014)]{siwak14}    Siwak, M., Rucinski, S. M., Matthews, J. M., Kuschnig, R., Guenther, D. B., et al., 2014, MNRAS, 444, 327



\bibitem[Siwak et al.(2016)]{siwak16}    Siwak, M., Ogloza, W., Rucinski, S. M., Moffat, A. F. J., Matthews, J. M., et al., 2016, MNRAS, 456, 3972 



\bibitem[Siwak et al.(2017)]{siwak17}   Siwak, M., Rucinski, S. M., Matthews, J. M., Cameron, C., Guenther, D. B, et al., 2017,     Proceedings of the PAS (Proc. of the 2nd BRITE Science conference, Innsbruck) vol.5, p. 214 


\bibitem[Siwak et al.(2018)]{siwak18}   Siwak, M., Ogloza, W., Moffat, A. F. J., Matthews, J. M., Rucinski, S. M., et al., 2018, MNRAS, 478, 758


\bibitem[Skinner et al.(2010)]{skinner10}    Skinner, S. L., G{\"u}del, M., Briggs, K. R., \& Lamzin, S. A., 2010, ApJ, 722, 1654

    

\bibitem[Stetson(1987)]{stet87}    Stetson, P. B., 1987 PASP, 99, 191

    

\bibitem[Stolker et al.(2016)]{stolker16}        Stolker, T., Dominik, C., Avenhaus, H., Min, M., de Boer, J., et al., 2016, A\&A, 595, 113



\bibitem[Vorobyov \& Basu(2005)]{vorobyov05}    Vorobyov, E. I., \& Basu, S., 2005, ApJ, 633, L137



\bibitem[Vorobyov \& Basu(2006)]{vorobyov06}    Vorobyov, E. I., \& Basu, S., 2006, ApJ, 650, 956



\bibitem[Vorobyov et al.(2015)]{vorobyov15}    Vorobyov, E. I., \& Basu, S., 2015, ApJ, 805, 115



\bibitem[Walker et al.(2003)]{WM2003}    Walker, G. A. H., Matthews, J. M., Kuschnig, R., Johnson, R., Rucinski, S. M., et al., 2003, PASP, 115, 1023



\bibitem[Warmels(1991)]{W1991}        Warmels, R. H., 1991, PASP Conf. Series, 25, 115



\bibitem[Weiss et al.(2014)]{weiss14}    Weiss, W. W., Rucinski, S. M., Moffat, A. F. J., Schwarzenberg-Czerny, A.,     Koudelka, O. F., et al., 2014, PASP, 126, 573



\bibitem[Zhu et al.(2007)]{zhu07}    Zhu, Z., Hartmann, L., Calvet, N., Hernandez, J., Muzerolle, J., \& Tannirkulam, A.-K., 2007, ApJ, 669, 483



\bibitem[Zhu et al.(2009)]{zhu09}        Zhu, Z., Hartmann, L., Gammie, C., \& McKinney, J. C., 2009, ApJ, 701, 620

     

\end{thebibliography}
\end{document}